\documentclass[twocolumn]{aastex631}
\usepackage{graphicx}	% Including figure files
\usepackage{amsmath}	% Advanced maths commands
\usepackage{amssymb}	% Extra maths symbols
\usepackage{natbib}
\usepackage{float}
\usepackage{placeins}

\usepackage{subfigure}

\usepackage{newtxtext,newtxmath}

\usepackage[T1]{fontenc}
\usepackage{ae,aecompl}

\begin{document}

\title{High significance detection of the dark substructure in gravitational lens SDSSJ0946+1006 by image pixel supersampling}

\author[0000-0002-4270-6453]{
Quinn Minor}
%Quinn Minor\thanks{E-mail: qminor@bmcc.cuny.edu}
\affiliation{Department of Science, Borough of Manhattan Community College, City University of New York, New York, NY 10007, USA}
\affiliation{Department of Astrophysics, American Museum of Natural History, New York, NY 10024, USA}
\affiliation{Graduate Center, City University of New York, 365 5th Avenue, New York, NY 10016, USA}

%\pubyear{2024}

%\label{firstpage}
%\maketitle

\begin{abstract}
Recent studies have shown that the dark substructure reported in the gravitational lens SDSSJ0946+1006 has a high central density which is in apparent tension with the $\Lambda$CDM paradigm. However, its detection significance has been found in \cite{ballard2024} to be sensitive to prior assumptions about the smoothness of the source galaxy. Here we show that the detection significance of the substructure is higher than previously reported (log-Bayes factor $\Delta\ln{\cal E} \approx 143$, equivalent to a $\sim17\sigma$ detection) by approximating the integration of light over each pixel via ray tracing and averaging over many subpixels---a technique known as supersampling---and this result is insensitive to the assumed prior on the source galaxy smoothness. Assuming a dark matter subhalo, the combination of supersampling and modeling both sets of lensed arcs also tightens the subhalo constraints: we find the subhalo's projected mass within 1 kpc lies in the range $(2.2-3.4)\times10^9M_\odot$ at 95\% confidence in our highest evidence model, while the log-slope of the subhalo's projected density at 1 kpc is steeper than $-1.75$ at the 95\% confidence level, further establishing it as an outlier compared to expectations from CDM. We also identify a systematic that has biased the slope of the primary lensing galaxy's density profile in prior studies, which we speculate might be due to the presence of dust or an imperfect foreground subtraction. Our analysis places the existence of the substructure on firmer ground, and should motivate deeper follow-up observations to better constrain its properties and clarify its apparent tension with $\Lambda$CDM.
\end{abstract}


\section{Introduction}\label{sec:intro}

Among the hundreds of currently known galaxies that act as strong gravitational lenses, precious few have been reported to contain an observable perturbation to the lensed images by a dark substructure with high significance \citep{vegetti2010,vegetti2012,hezaveh2016,nierenberg2014}. Of these, the system J0946+1006 (hereafter J0946) stands out in that its perturbing substructure has been reported by \cite{minor2021} (and more recently \citealt{despali2024}) to have a high central density which appears to be in tension with the Cold Dark Matter (CDM) paradigm in view of its lack of an observed stellar luminosity. It also happens to be a rare ``double Einstein ring'' system with two lensed sources at different redshifts \citep{gavazzi2008,collett2014}, earning it the nickname ``the Jackpot''; later, a third faint lensed source was also discovered \citep{collett2020}.

Recently, \cite{ballard2024} (hereafter B24) have used more flexible modeling of all three lensed sources in multiple bands and confirmed this tension, although it is somewhat alleviated in their models. Surprisingly, B24 also found that the detection significance of the substructure varied dramatically depending on the assumed prior for the smoothness of the source galaxy. If the source galaxy is allowed to have greater fluctuations in surface brightness (using gradient-based regularization; \citealt{suyu2006}), noise-level residuals are achieved even without the presence of a substructure, whereas the fit is much poorer if the source is required to be very smooth (as in curvature regularization). By contrast, when a substructure is included, the source is allowed to be smoother regardless of the source regularization scheme used; this trend was also pointed out in the original discovery paper of \cite{vegetti2010}. Hence the detection significance using gradient regularization, reported as 5.9$\sigma$ in B24, is entirely driven by the difference in the smoothness of the source, rather than by any significant improvement in the quality of the fit. Using curvature regularization, which imposes more smoothness, results in a higher detection significance ($\sim 11.3\sigma$). Taken at face value, it seems that our inference of the very existence of the substructure is dependent on our assumptions about how smooth the source galaxy should be, a rather uncomfortable result.

In the aforementioned studies that modeled a perturbing substructure in J0946 using adaptive grid methods (with the exception of \citealt{nightingale2024}), the surface brightness in each image pixel produced by the model is determined by ray tracing the pixel centers to the source plane. In reality, however, the photons striking a camera pixel are equally likely to strike anywhere within that pixel, not just its center point. This can be approximated through the supersampling technique: dividing each pixel into subpixels, ray tracing the subpixel centers, and subsequently averaging the surface brightnesses of all the subpixels. Besides more closely approximating the integration of light over an entire pixel, supersampling is well known as a spatial anti-aliasing technique, often used in computer graphics to minimize the aliasing effects (``jaggies'') that result from sampling an image at low resolution \citep{goss1999}. In astronomy this is especially important when modeling cuspy or pointlike sources \citep{galan2024}. In the context of multiply imaged sources, however, we will see that supersampling accomplishes much more than that: it enforces the requirement that image pixels that overlap when ray traced to the source plane have consistent surface brightnesses, regardless of prior assumptions about the intrinsic smoothness of the source galaxy.

In this paper, we show that the detection significance of the dark substructure in J0946 increases dramatically when supersampling of the image pixels is performed during lens modeling. With even a modest level of supersampling (2$\times$2 splitting of pixels), significant residuals appear in the vicinity of the purported substructure and its corresponding counterimage when a subhalo is not included in the model; this is a consequence of the fact that pixels that map to the same region of the source are required to have consistent surface brightnesses. These residuals are reduced to noise level when a subhalo is included, resulting in a detection significance $\sim17\sigma$ which is quite insensitive to the assumed source prior. In addition, by enforcing consistency of multiply imaged pixels, supersampling also reduces the allowed parameter space of the substructure (in particular its central density and slope). Hence the presence of the dark substructure and its inferred central density and slope are placed on firmer ground, which should motivate additional follow-up observations to elucidate its apparent tension with CDM. 

The paper is organized as follows. The lens modeling procedure is described in Section \ref{sec:lensmodeling}. The substructure detection significance without supersampling is discussed in Section \ref{sec:results_no_ss}. In Section \ref{sec:results_with_ss} we discuss the detection significance with supersampling and show why supersampling makes a dramatic difference when a substructure is not included in the model. In Section \ref{sec:galaxyslope} we investigate an important bias on the host galaxy's density slope as revealed by supersampling. The effect of supersampling on the subhalo parameter constraints is presented in Section \ref{sec:subhalo_constraints}. In Section \ref{sec:discussion} we discuss the possible origin of the relative dark spot in the \textit{HST} image which is biasing the primary galaxy's density slope, and finally we conclude in Section \ref{sec:conclusions}.

\begin{figure*}
	\centering
 		\subfigure[\textit{HST} image]
	{
		\includegraphics[height=0.35\hsize,width=0.32\hsize]{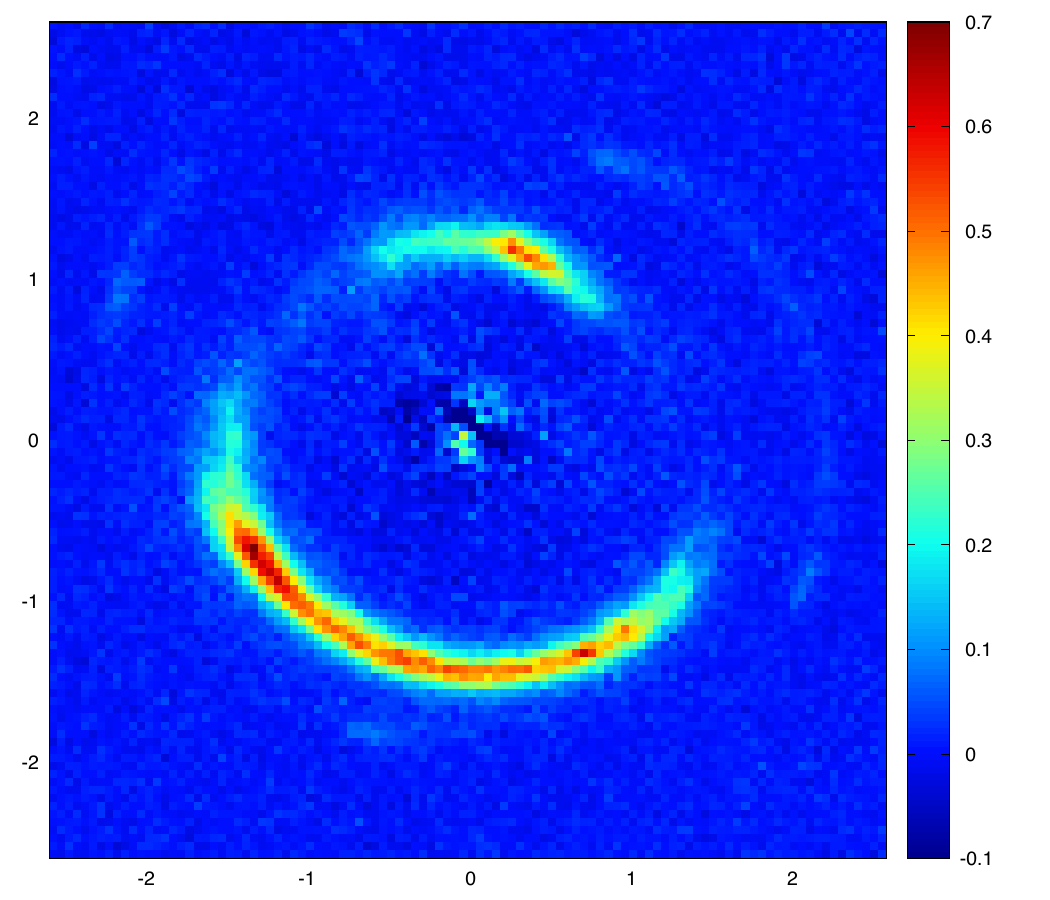}
		\label{fig:data}
	}
		\subfigure[$s1$ mask]
	{
		\includegraphics[height=0.35\hsize,width=0.32\hsize]{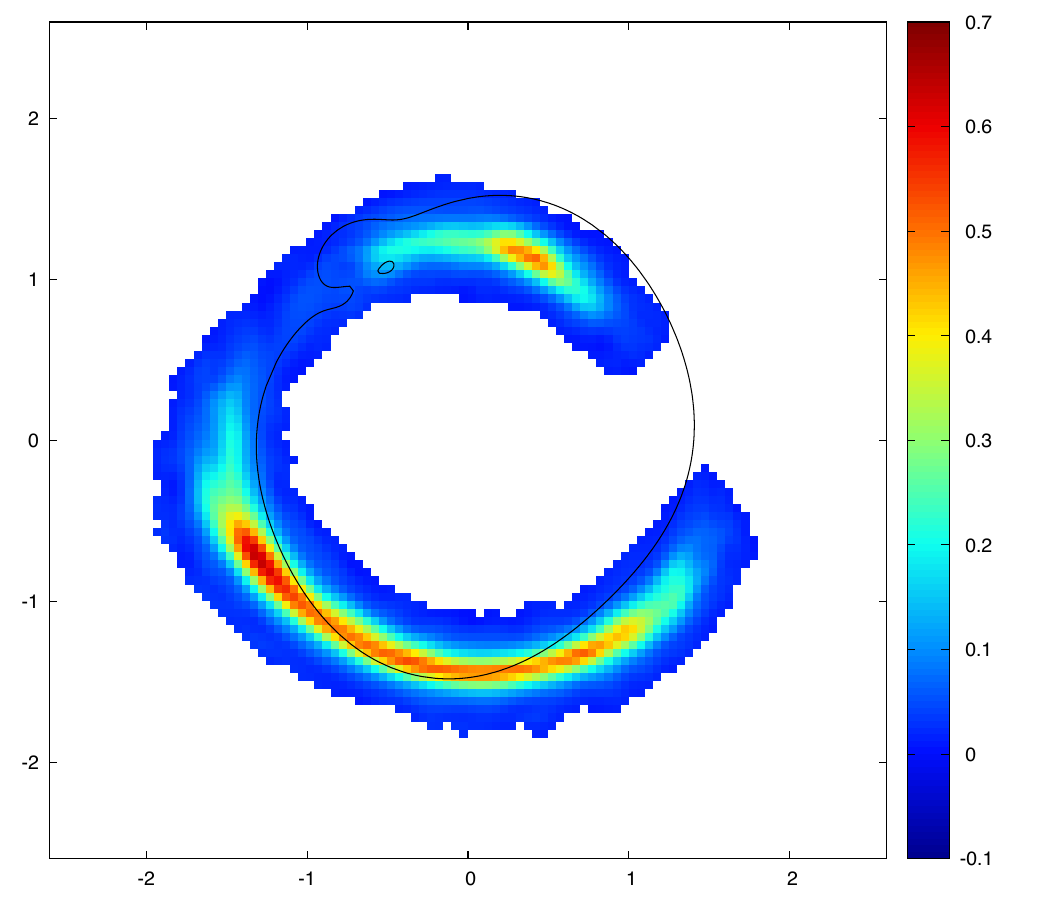}
		\label{fig:mask1}
	}
		\subfigure[$s2$ mask]
	{
		\includegraphics[height=0.35\hsize,width=0.32\hsize]{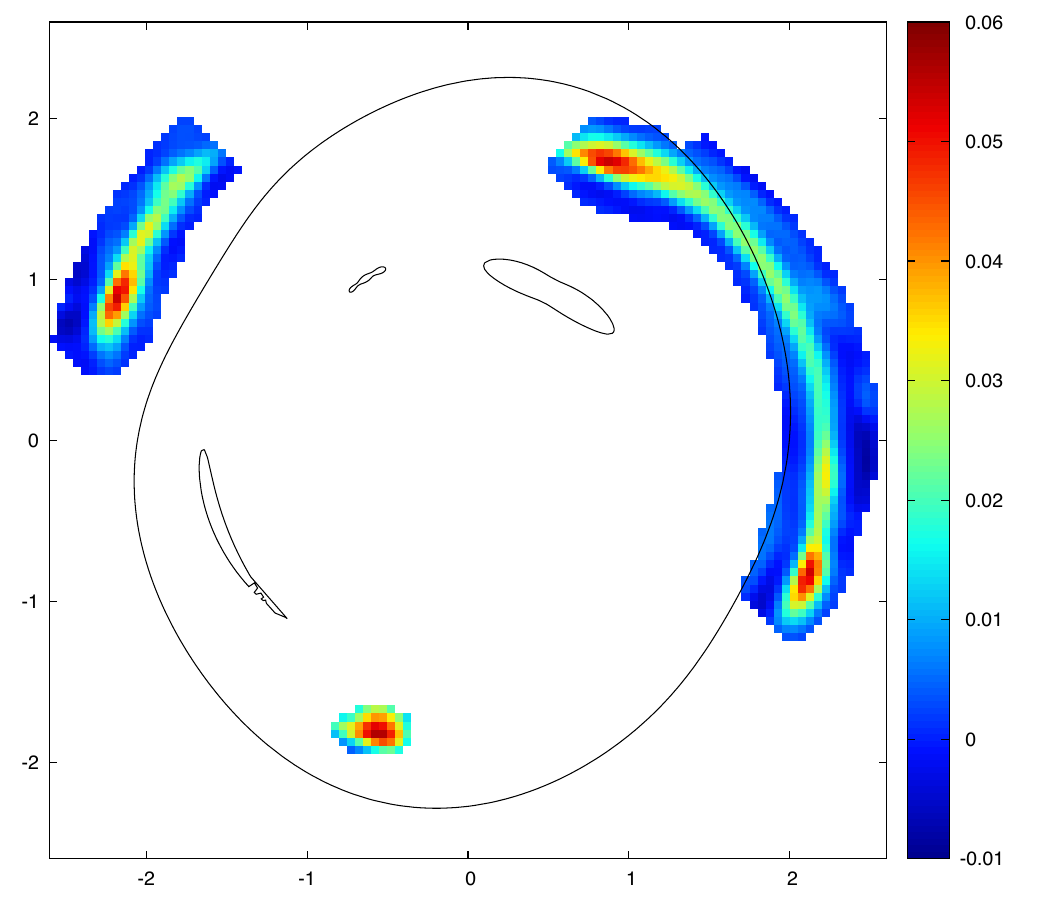}
		\label{fig:mask2}
	}
	\caption{\textit{HST} image of J0946 in the F814W band, with foreground galaxy subtracted out. In (b) and (c) we show the masks used to reconstruct the images of the source galaxies $s1$ and $s2$ respectively; care has been taken to make sure the masks do not overlap, although they include pixels that neighbor each other in the lower half of the image. Our best-fit image reconstruction from our highest evidence model (which includes both a subhalo and additional angular structure in the projected density in the form of multipoles and will be explored in Section \ref{sec:subhalo_constraints}) is shown, along with the critical curves in black.}
\label{fig:data_and_masks}
\end{figure*}

\section{Lens modeling}\label{sec:lensmodeling}

For this analysis, we model both sets of lensed images in the I-band (F814W) Hubble Space Telescope (\textit{HST}) image of J0946 \citep{gavazzi2008}. This requires reconstructing two sources at different redshifts: the inner arcs from source $s1$ at redshift $z_{s1}=0.609$, and the much fainter outer arcs from source $s2$ at redshift $z_{s2}=2.035$ \citep{smith2021}. These are shown in the foreground-subtracted image in Figure \ref{fig:data}. We use the same foreground subtraction, point spread function (PSF) and noise map as employed in B24. However, in contrast to B24, we do not include any additional bands in our modeling and also omit the faint third source which has been identified by VLT/MUSE observations due to its relatively low resolution compared to the other two lensed sources \citep{collett2020}. 

\begin{figure*}
	\centering
	\subfigure[residuals (without subhalo)]
	{
		\includegraphics[height=0.35\hsize,width=0.32\hsize]{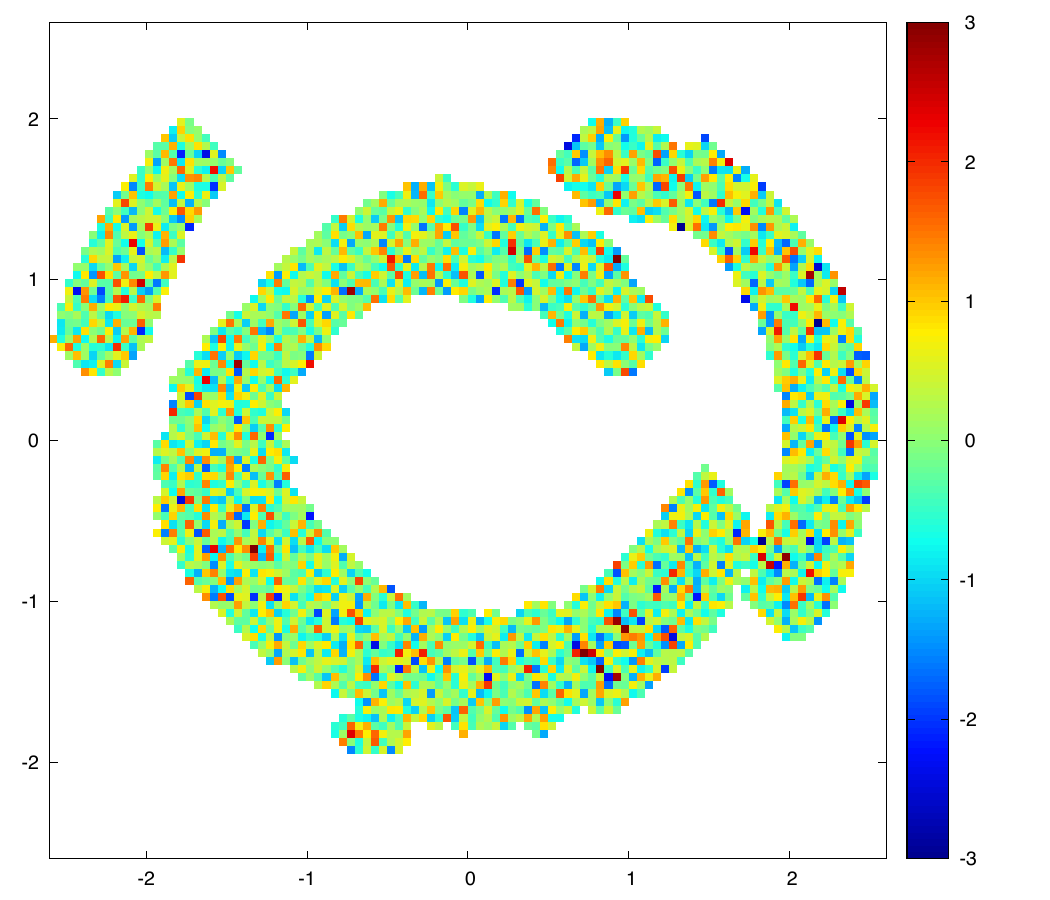}
		\label{nospl0_nosub}
	}
	\subfigure[reconstructed source $s1$ ($z_s$=0.609)]
	{
		\includegraphics[height=0.35\hsize,width=0.32\hsize]{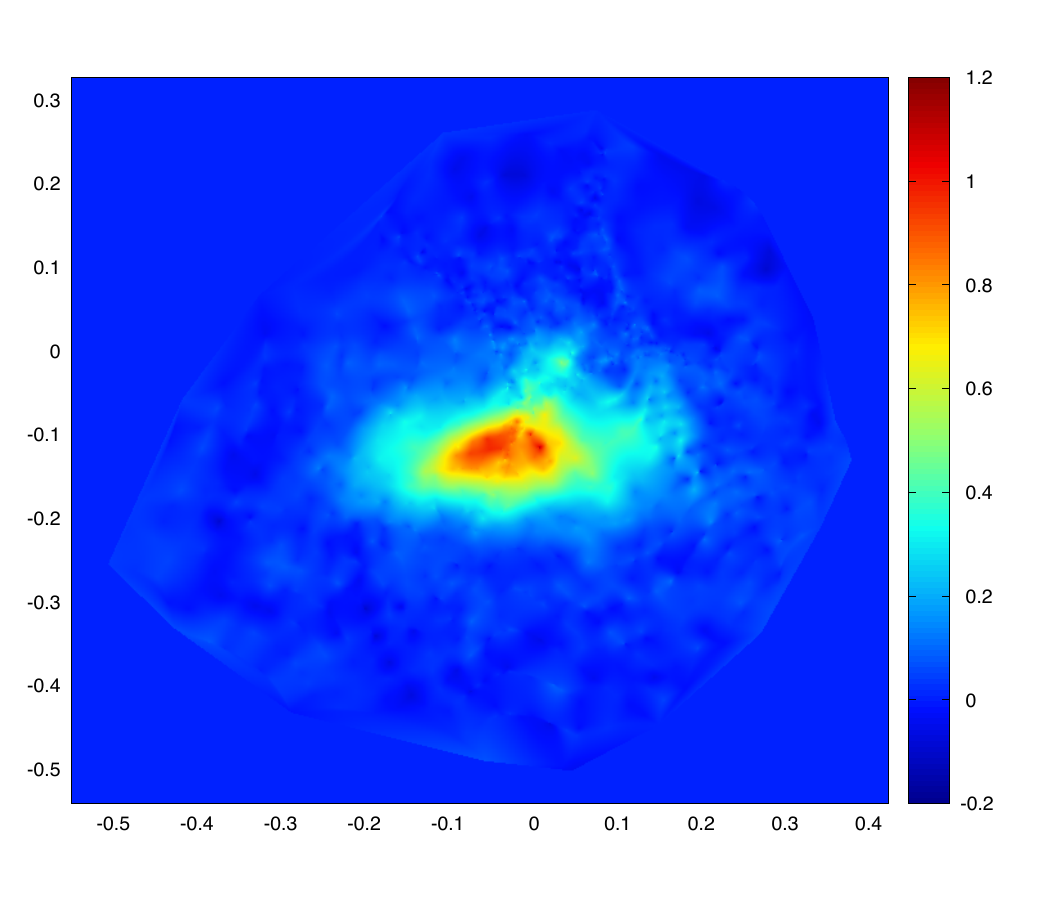}
		\label{nospl0_nosub_src1}
	}
		\subfigure[reconstructed source $s2$ ($z_s$=2.035)]
	{
		\includegraphics[height=0.35\hsize,width=0.32\hsize]{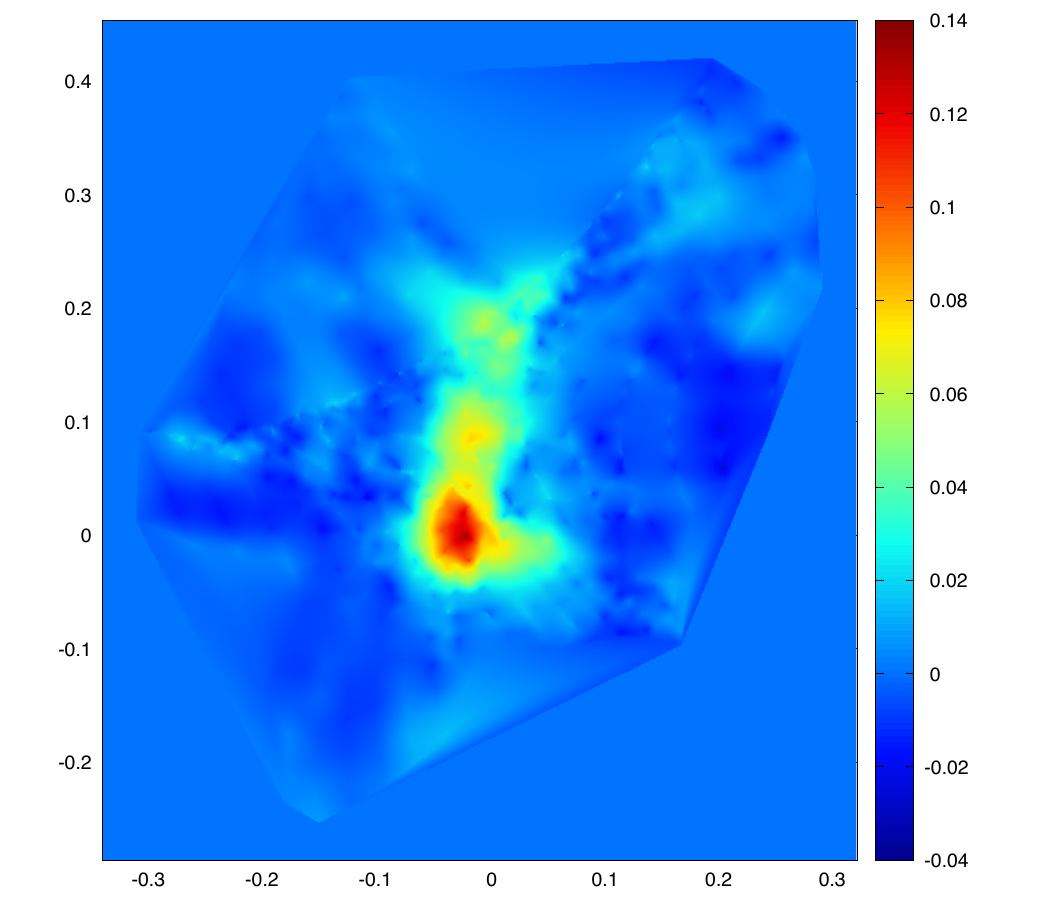}
		\label{nospl0_nosub_src2}
	}
	\subfigure[residuals (with subhalo)]
	{
		\includegraphics[height=0.35\hsize,width=0.32\hsize]{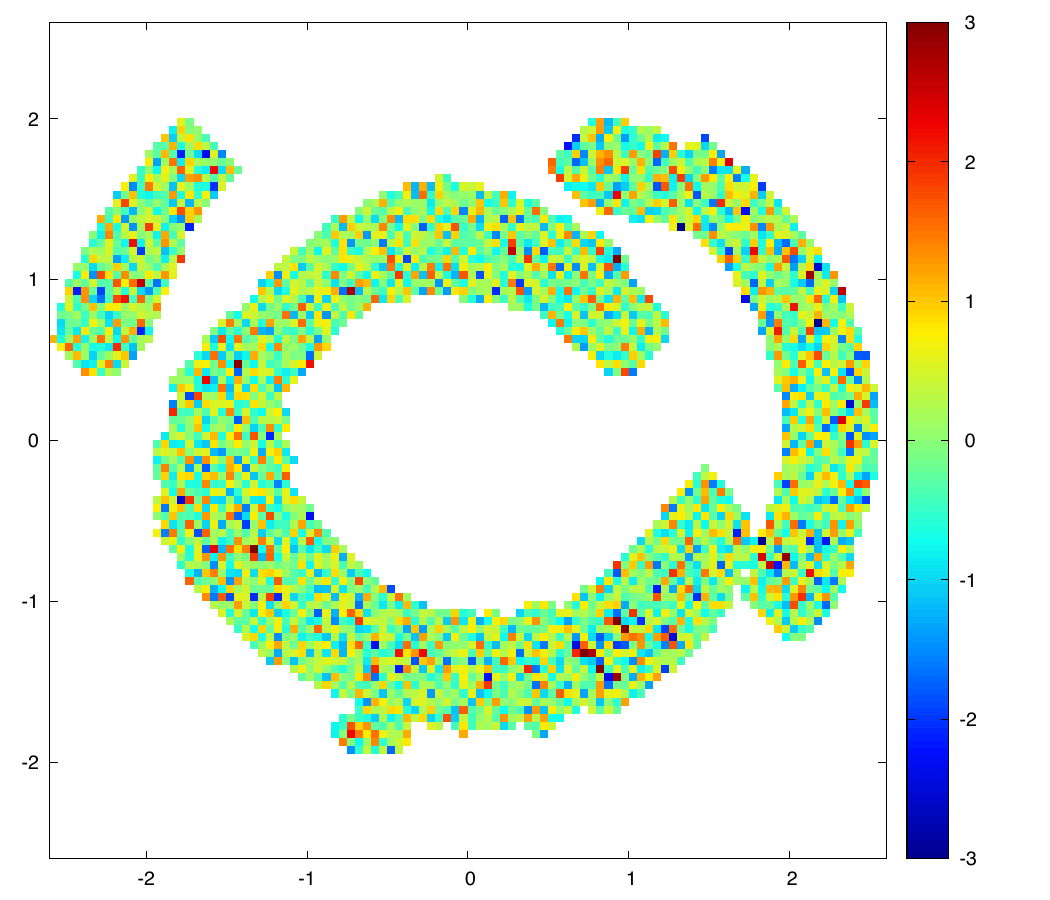}
		\label{nospl0_tnfwsub}
	}
	\subfigure[reconstructed source $s1$ ($z_s$=0.609)]
	{
		\includegraphics[height=0.35\hsize,width=0.32\hsize]{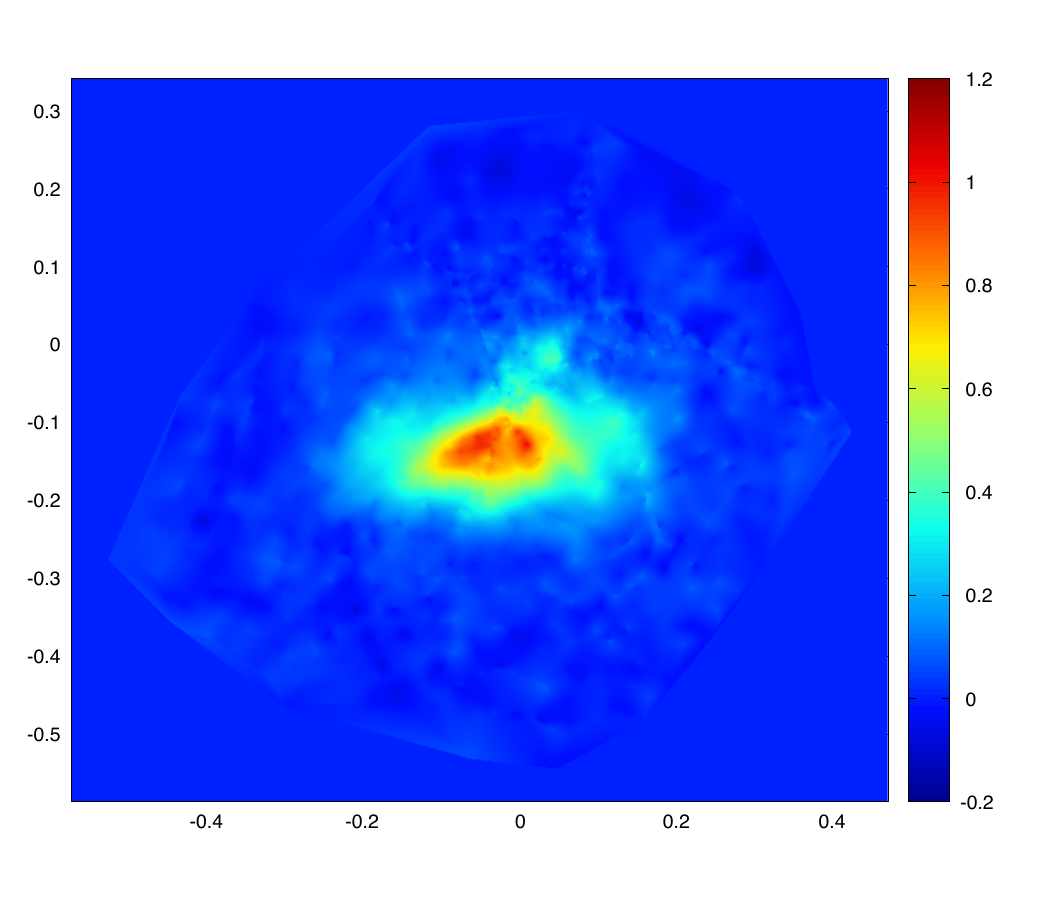}
		\label{nospl0_tnfwsub_src1}
	}
		\subfigure[reconstructed source $s2$ ($z_s$=2.035)]
	{
		\includegraphics[height=0.35\hsize,width=0.32\hsize]{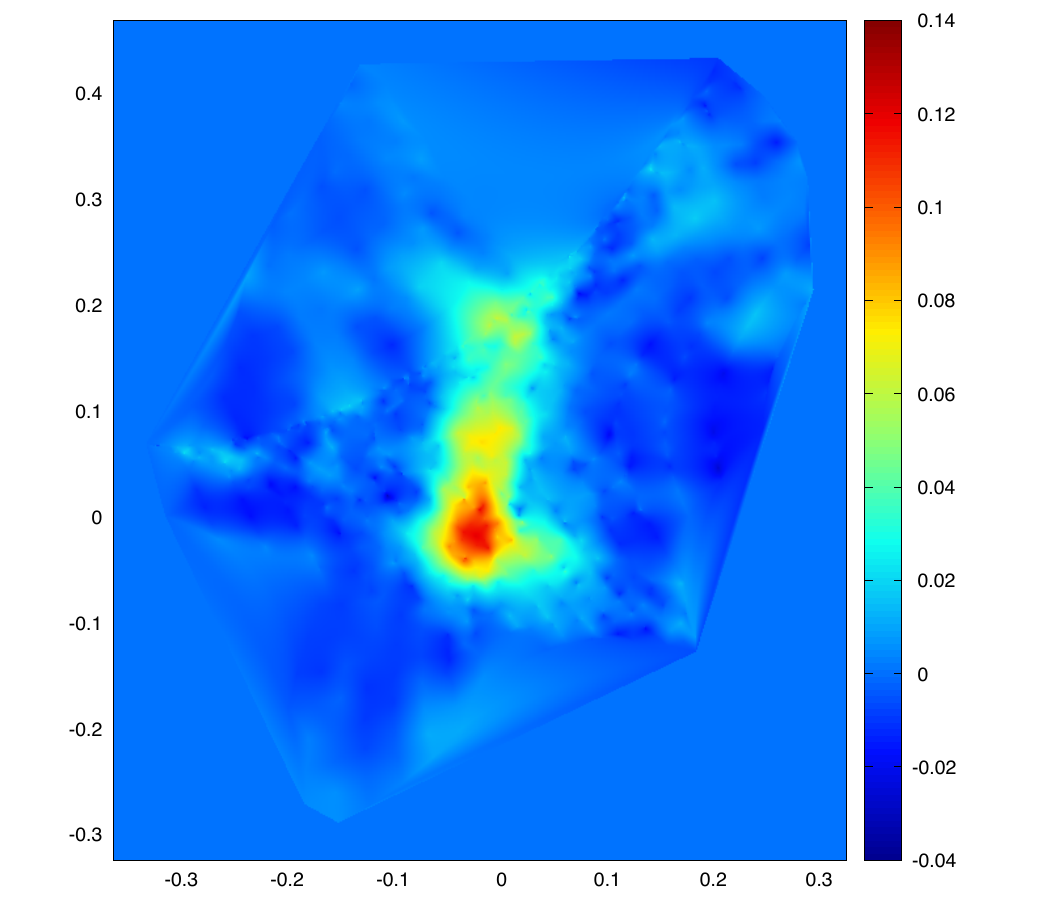}
		\label{nospl0_tnfwsub_src2}
	}
	\caption{Normalized residuals and reconstructed sources $s1$ and $s2$ for the best-fit models without supersampling of image pixels, first without a subhalo (top row) and then with a subhalo included in the model (bottom row). Note that when plotting our sources, we are interpolating in the source pixels of the adaptive grid using natural neighbor interpolation, rather than showing the Voronoi pixels themselves.  }
\label{fig:bestfit_nosup}
\end{figure*}

\subsection{Mass model}
Our model for the lens galaxy's density profile and subhalo are identical to those of \cite{minor2021} and B24, albeit with different parameterizations: first we choose a power law ellipsoid for the projected mass density profile of the primary lensing galaxy \citep{tessore2015}, which we parameterize as

\begin{equation}
\Sigma(\xi) = \frac{2-\alpha}{2}\left(\frac{R_e}{\xi}\right)^\alpha\Sigma_{\rm cr}(z_l,z_{s}),
\label{eq:power_law}
\end{equation}
where $\Sigma_{\rm cr}(z_l,z_s)$ is the critical lensing density for a given lens and source redshift \citep{schneider1992}, $R_e$ is the Einstein radius (defined as the radius within which the average density $\bar\Sigma = \Sigma_{\rm cr}$), $\alpha$ is the (negative) log-slope of the projected density, and the elliptical radius is defined as $\xi = \sqrt{qx^2+y^2/q}$ with $q$ being the axis ratio. Note that since we are modeling two source galaxies with different redshifts $z_{s1}$ and $z_{s2}$, in principle we could choose to normalize the density profile according to the Einstein radius at either source plane. Since the low-redshift source $s1$ has much higher signal-to-noise and the purported substructure is close to a lensed image from $s1$, we choose $R_e$ to be the Einstein radius of source $s1$, i.e. $\bar\Sigma(\xi = R_e) = \Sigma_{\rm cr}(z_l,z_{s1})$.
Rather than varying the axis ratio $q$ and position angle $\theta$ as free parameters, we vary the ellipticity components, defined as $e_1 = (1-q)\cos{2\theta}$ and $e_2 = (1-q)\sin{2\theta}$; this has the advantage of not having a coordinate singularity when the ellipticity is very small. An external shear term is added to the lensing potential to capture perturbations from distant galaxies, while in practice also capturing (to some extent) possible departures from an ellipsoidal mass distribution \citep{etherington2024}. We likewise parameterize the external shear using shear components $\Gamma_1$ and $\Gamma_2$.

For the perturbing subhalo, we choose a smoothly truncated Navarro-Frenk-White (tNFW) profile \citep{baltz2009} whose density profile declines steeply as $\sim r^{-5}$ well beyond the truncation radius. The parameters of the subhalo's density profile are the virial mass and concentration $m_{200}$ and $c_{200}$ and truncation radius $r_t$. (Note that while $m_{200}$ approximates the virial radius, it is defined as the mass within the radius $r_{200}$ whose average density is 200 times the critical density of the Universe.) Although we are modeling a subhalo, we can formally define $m_{200}$ and $c_{200}$ as the virial mass and concentration the halo \emph{would} have if it were a field halo with the given density profile, in the absence of tidal stripping. (Note that even in the absence of tidal stripping, $r_t$ could represent the virial radius at infall, which would have been smaller than the virial radius in the field at the given redshift.) In addition, because the lower redshift source \emph{s1} lies along the line of sight to the higher redshift source, it contributes to the lensing effect by perturbing the deflections generated by the host galaxy. We approximate the mass distribution of the perturber \emph{s1} as a singular isothermal sphere (whose density is given by Eq.~\ref{eq:power_law} with $\alpha=1$), and in addition we take into account the recursive lensing that occurs with multiple lens planes by solving the recursive lensing equation \citep{schneider1992,petkova2014}, as is done in \cite{collett2014} and the fiduciary model in B24. Since the lensing mass corresponding to $s1$ only impacts the images of the more distant source $s_2$, we normalize its density profile by its Einstein radius with respect to $z_{s2}$; in other words, we define $R_{e,s1}$ to be the Einstein radius it \emph{would} have if it were lensing a source at $z_{s2}$, assuming there were no foreground lens present.

Finally, following \cite{minor2021} and B24, we will also consider a model with additional angular structure in the primary lensing galaxy, in the form of $m=3$ and $m=4$ multipole terms added to the projected density of the lens (to be explored in Sections \ref{sec:multipoles} and \ref{sec:subhalo_constraints}). These take the form

\begin{equation}
\Delta\Sigma_m = R^{-\alpha}\left[A_m\cos{m\phi}+B_m\sin{m\phi}\right]\Sigma_{\rm cr}(z_l,z_s),
\end{equation}
where $\alpha$ is taken to be the same power-law index as in the primary lensing galaxy's density profile (Eq.~\ref{eq:power_law}). Thus our additional free parameters in the multipole fits will be $A_3,B_3$ and $A_4,B_4$.

\subsection{Source reconstruction}\label{sec:src_reconstruction}

We model the lensed sources using pixellated source reconstruction with an adaptive source pixel grid \citep{vegetti2009}. This was accomplished using the \textsc{QLens} software, which has been employed in previous works \citep{galan2024,andrade2022,minor2021,minor2021b,andrade2019,minor2017,minor2008} and is planned for public release in the coming year (Minor et al., in prep). During each likelihood evaluation, a lensing matrix is constructed to generate the model image from the source pixel amplitudes, along with a source regularization matrix, after with the source pixels are solved for by a linear inversion to produce the pixel-level Bayesian evidence \citep{suyu2006}.

In the models without supersampling, the center of all the pixels within the mask are ray traced to the source plane, after which we construct a Delaunay triangulation to generate the source pixel grid in a manner identical to B24 and \cite{vegetti2010}. In the case where supersampling is performed, each image pixel is split into $N_{\rm sp}\times N_{\rm sp}$ subpixels, where $N_{\rm sp}$ is referred to as the supersampling factor. We choose $N_{\rm sp}=4$ for our base set of runs, but in Section \ref{sec:num_splittings} we will investigate whether our inferences are significantly affected by the number of subpixels.

For the supersampled models, we define our source pixels in a manner similar to \cite{nightingale2015}: we ray trace the center of each subpixel to the source plane, after which a K-means clustering algorithm is used to find positions for the source pixels, such that there is a greater density of source pixels in regions where there is greater clustering of ray-traced points. (This form of pixellation is shown in Figure \ref{fig:src_pixcomp} in Appendix \ref{sec:pixellation}.) This has the advantage of ensuring good coverage of the source plane, whereas if the image pixel centers are used (as in the models without supersampling), pixels from separate images may end up in almost identical spots when ray traced to the source plane, leaving considerable gaps between source points in neighboring regions. To accomplish this procedure in a reasonable time, we use a dual-tree K-means algorithm implemented in the \textsc{MLPACK} software package \citep{curtin2023} which is specifically designed to handle a large number of clusters. From the resulting centroids, we construct a Delaunay triangulation to generate the source pixel grid. To keep the computation time reasonable, we choose the number of source pixels to be equal to half the number of image pixels within the mask (i.e. $N_s = N_d/2$); although this is half the number of source pixels compared to our models without supersampling, the constraints are only slightly affected by increasing the number of source pixels to $N_s = N_d$ (see Appendix \ref{sec:pixellation} for a detailed comparison). To generate  the lensed images, the surface brightnesses of the subpixels are found using natural neighbor interpolation \citep{sibson1981}, which uses the Voronoi grid that is dual to the Delaunay triangulation we have constructed. (Note that for the unsupersampled models, no interpolation is required at all, since each source pixel is associated with a single image pixel.) After interpolating to find surface brightness values for each subpixel, the subpixels are averaged to find the (unconvolved) surface brightness of each image pixel. The resulting image is then convolved with the point spread function (PSF). Note that we do not supersample the PSF itself, which would make the convolutions much more computationally expensive.

To discourage pathological solutions, we impose two additional priors specific to the lens modeling: first, we implement a prior that discourages producing lensed images outside the mask. We accomplish this by temporarily unmasking after the source pixel surface brightnesses are found, generating the lensed images, and imposing a steep penalty if any surface brightness is found outside the original mask whose value is greater than 0.2 times the maximum surface brightness of the images. Second, we place a prior on the number of lensed images produced. This is accomplished by creating a $60\times60$ Cartesian grid in the source plane whose dimensions are set by the extent of the ray-traced points. We then find the overlap area of all the ray-traced image pixels for each Cartesian grid cell; by dividing the total overlap area by the area of each grid cell, we obtain the number of images produced by that cell. To avoid counting highly demagnified images (which can be problematic in multi-component lenses), we only include the overlap area from pixels whose magnification is greater than 0.1. We then find the average number of images over all the cells, and impose a steep penalty if the average number of images is less than 1.7. This especially discourages solutions that are not multiply imaged, where the source looks similar to the observed configuration of lensed images. With these priors in place, we can obtain a good solution with a single nested sampling run, provided the parameter priors are broad enough.

Since this lensed system has two lensed sources at different redshifts ($z_{s1}=0.609$ and $z_{s2}=2.035$) whose lensed images do not overlap, we draw two separate masks around each set of lensed images (shown in Figure \ref{fig:data_and_masks}) and perform the lensing inversions for each source separately. For each source, we regularize the reconstructed source with gradient regularization using the method of \cite{nightingale2024}, which uses natural neighbor interpolation to calculate the gradient in the neighborhood of each source pixel. Rather than include the regularization strengths $\lambda_1$ and $\lambda_2$ as free parameters during sampling, we optimize the regularization strengths for the two sources $\lambda_1$ and $\lambda_2$ during each likelihood evaluation using Brent's method. This has the advantage of reducing the burden on the sampler, and can be done quickly since the ray tracing and construction of the source grid do not need to be redone while the regularization strength is being optimized, and the posterior is generally well-peaked around the optimal regularization strength.

Since the reconstructed source $s1$ has two central peaks \citep{vegetti2010}, the optimal placement of the lensing mass of $s1$  may differ slightly from the centroid of the source light. To account for this, we vary the displacement of the lensing mass from the centroid of the source light as free parameters $\Delta x_{s1}$ and $\Delta y_{s1}$. Thus the lens $s1$ has three free parameters: $\Delta x_{s1}$ and $\Delta y_{s1}$, and its Einstein radius parameter $R_{e,s1}$ (which we define with respect to the second source redshift, i.e. the radius of an Einstein ring it would produce from a source at the redshift of $s2$ if there were no other lens present). We find an approximate centroid of the source before the inversion is performed, by ray tracing the data pixels to the source plane $s1$ and calculating the centroid and radial dispersion $\sigma_{\rm src}$ of the ray-traced points, weighted by their observed surface brightnesses; we include only the pixels whose surface brightnesses are greater than 5 times the estimated noise level for that pixel, and we further remove points that lie beyond $3\sigma_{\rm src}$ and repeat the procedure iteratively until no outlier points remain.

\begin{figure*}
	\centering
 		\subfigure[residuals (without subhalo)]
	{
		\includegraphics[height=0.48\hsize,width=0.48\hsize]{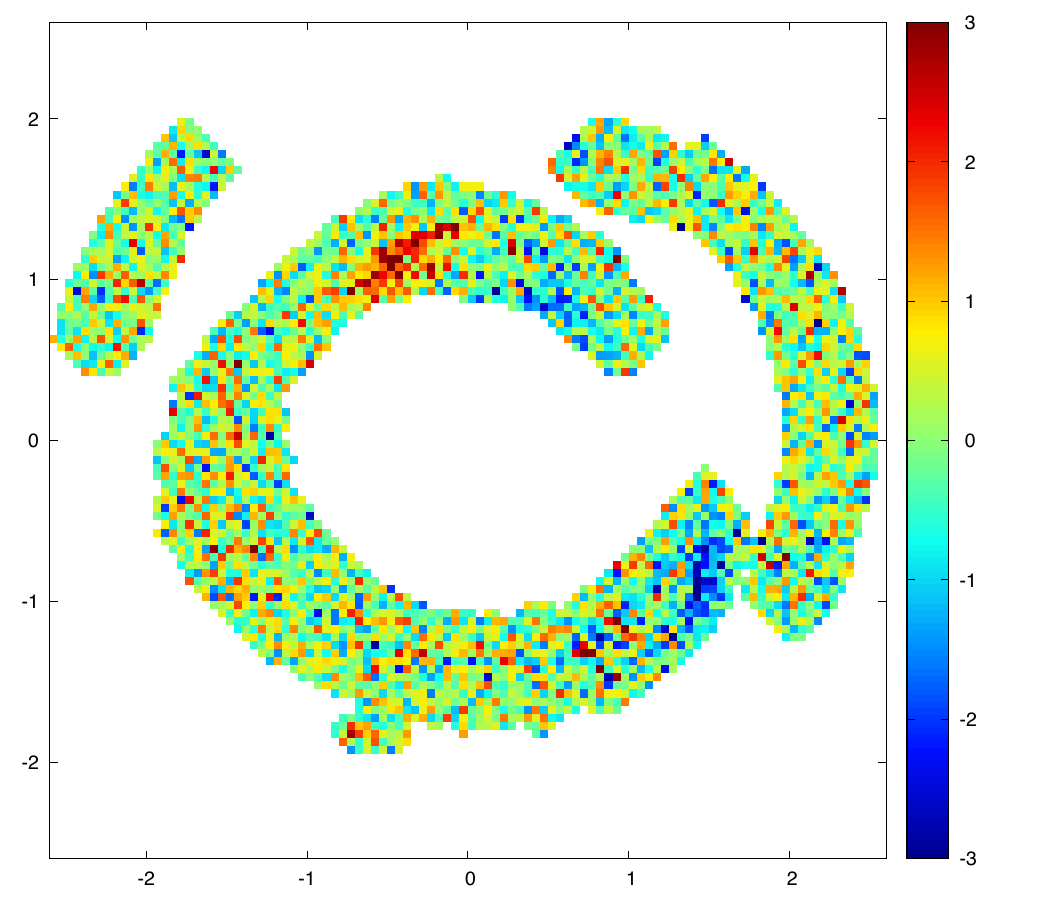}
		\label{nospl0_nosub_spl4_resid}
	}
		\subfigure[residuals (with subhalo)]
	{
		\includegraphics[height=0.48\hsize,width=0.48\hsize]{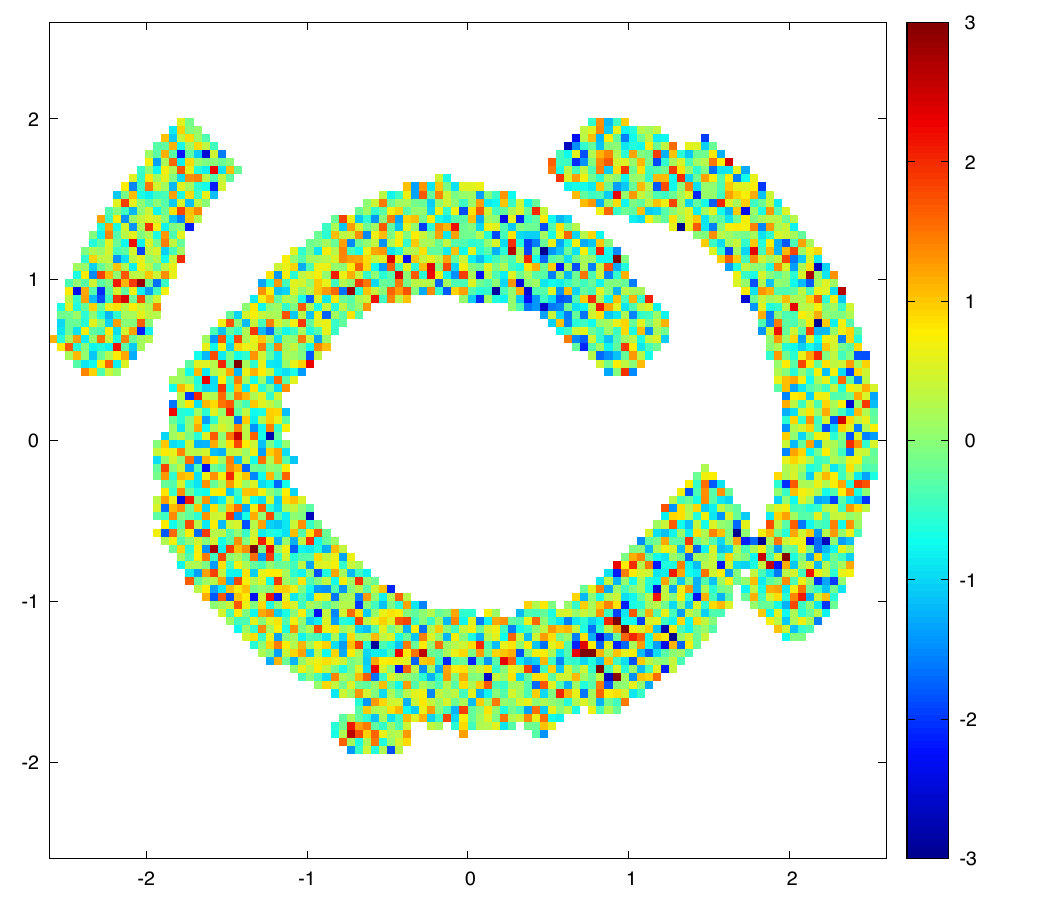}
		\label{nospl0_tnfwsub_spl4_resid}
	}
	\caption{Normalized residuals for the same best-fit models as in Figure \ref{fig:bestfit_nosup}, except with $4\times 4$ supersampling turned on. Figure on the left shows the model without a subhalo, figure on the right shows model with subhalo included in the model. Note that in this case, supersampling was \emph{not} used when constraining the lens model parameters. After adopting the best-fit model, we turn on supersampling and reoptimize the regularization parameters before reconstructing the source galaxies and lensed images.}
\label{fig:bestfit_sup_turned_on}
\end{figure*}

\begin{figure*}
	\centering
		\subfigure[before supersampling turned on]
	{
		\includegraphics[height=0.48\hsize,width=0.48\hsize]{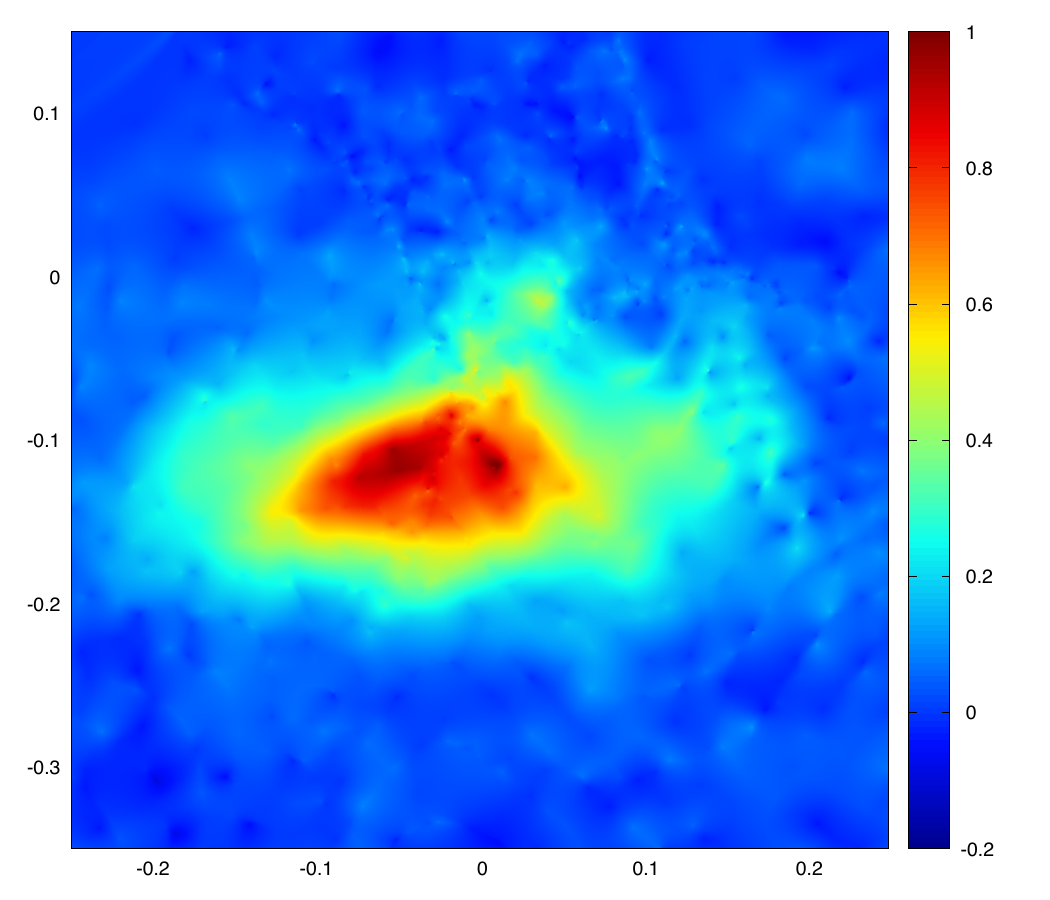}
		\label{zoomed_src_noss}
	}
		\subfigure[after supersampling turned on]
	{
		\includegraphics[height=0.48\hsize,width=0.48\hsize]{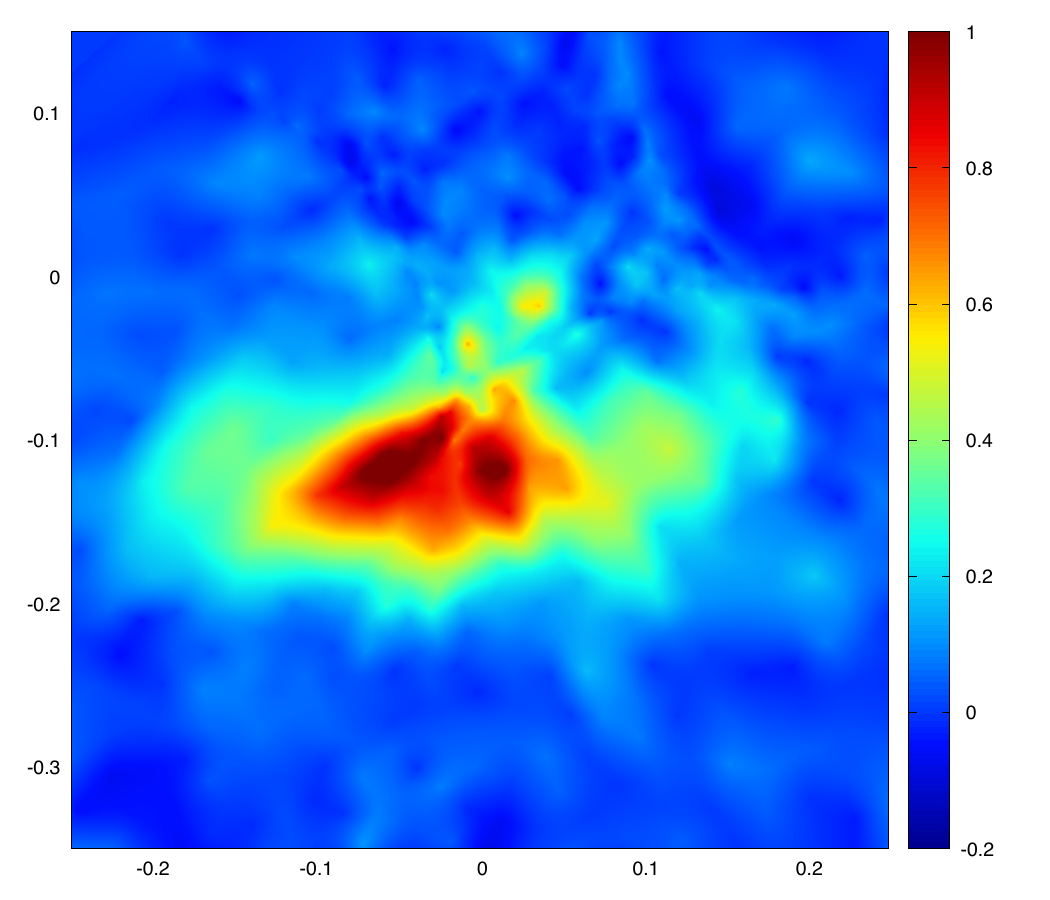}
		\label{zoomed_src_ss}
	}
	\caption{Zoomed-in view of the reconstructed source $s1$ for the best-fit model without a subhalo, obtained without supersampling, (a) before and (b) after supersampling is turned on. As in Figure \ref{fig:bestfit_nosup}, we plot the source using natural neighbor interpolation in the source pixels, rather than showing the Voronoi pixels themselves. Note that the noisy region of the source in the upper right is smoothed out when supersampling is used (which directly leads to the residuals seen in Figure \ref{nospl0_nosub_spl4_resid}).}
\label{fig:sources_sup_turned_on}
\end{figure*}

\section{Results of lensing analysis}\label{sec:results}

\subsection{Substructure detection significance without supersampling of the image pixels}\label{sec:results_no_ss}

We first model both lensed sources $s1$ and $s2$ without supersampling the image plane; in other words, we map each image pixel to a single point in the source plane. The posterior is sampled using the MultiNest sampler \citep{feroz2009} with 1000 live points, with 16 nonlinear parameters. The resulting reconstructed sources and residuals for the model without a subhalo included are shown in the top row of Figure \ref{fig:bestfit_nosup}, with inferred parameters shown in Table \ref{tab:posterior_inferences}. We find the best-fit parameters for the host galaxy and the lensing mass corresponding to $s1$ are broadly consistent with those in B24, although the inferences are not expected to be exactly alike since B24 also included U-band data and VLT/MUSE imaging of a third lensed source at higher redshift. As in B24, we obtain noise-level residuals except in the lower-right region which maps to the brightest part of the source $s1$. The reconstructed source in \cite{minor2021} shows that the brightest region in $s1$ is actually quite cuspy; since the regularization does not allow sufficiently rapid fluctuations in the source surface brightness, this central cusp is not well-reproduced, leaving residuals in the lower-right region of the mask \citep{galan2024}. Since this is far from where the purported subhalo is located, we do not expect this to significantly affect the detection significance of the subhalo, although whether these residuals could indirectly affect the inferred subhalo properties (e.g. by changing the source regularization parameter) is an interesting question which is outside the scope of this work.

Now we repeat the procedure, but this time including a dark matter subhalo in the model as in \cite{minor2021} The resulting residuals and reconstructed sources are shown in the bottom row of Figure \ref{fig:bestfit_nosup}. Here again we find similarly good residuals compared to the model without a subhalo (top row). We defer a discussion of the inferred subhalo parameters to Section \ref{sec:subhalo_constraints}, focusing here only on the detection significance. Comparing the Bayesian evidence to the model without a subhalo, we find an increase in log-evidence $\Delta\ln{\cal E} = 16.3$, which corresponds to a detection significance $\sim5\sigma$, similar to what is found in B24. However, since the residuals are similar in either solution, the improved Bayesian evidence in the model with the subhalo is entirely due to the fact that the reconstructed source $s1$ is smoother when a subhalo is included. If curvature regularization is used to impose more smoothness on the source, we find the detection significance is much higher ($\sim 11.3\sigma$ in B24, and $\sim 16\sigma$ in \citealt{vegetti2010}). The detection significance of the subhalo can thus vary dramatically depending on how noisy the reconstructed source is allowed to be, e.g. using gradient versus curvature regularization. As noted in B24, this by itself is troubling if we are claiming a substructure detection, since the ``noisiness'' of the reconstructed source is heavily influenced by our regularization prior, which was not astrophysically motivated in the first place. It is not difficult to imagine that perhaps some region of the source might have significantly more fine structure than the rest, perhaps due to star-forming clumps or merger activity, and that this perturbation in the source might be mistaken for the presence of a dark matter substructure. We will see shortly, however, that this result does not hold up when supersampling is used during lens modeling.

\begin{figure*}
	\centering
	\subfigure[residuals (without subhalo)]
	{
		\includegraphics[height=0.35\hsize,width=0.32\hsize]{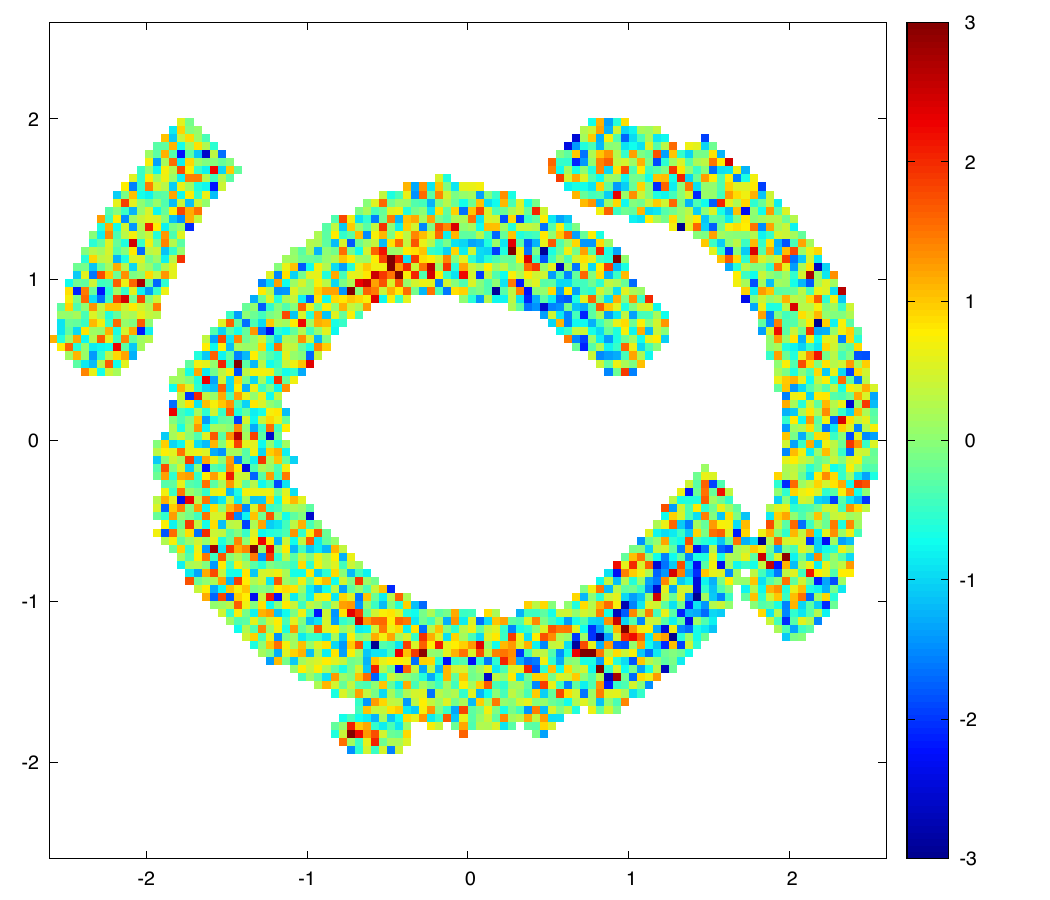}
		\label{spl4_nosub}
	}
	\subfigure[reconstructed source $s1$ ($z_s$=0.609)]
	{
		\includegraphics[height=0.35\hsize,width=0.32\hsize]{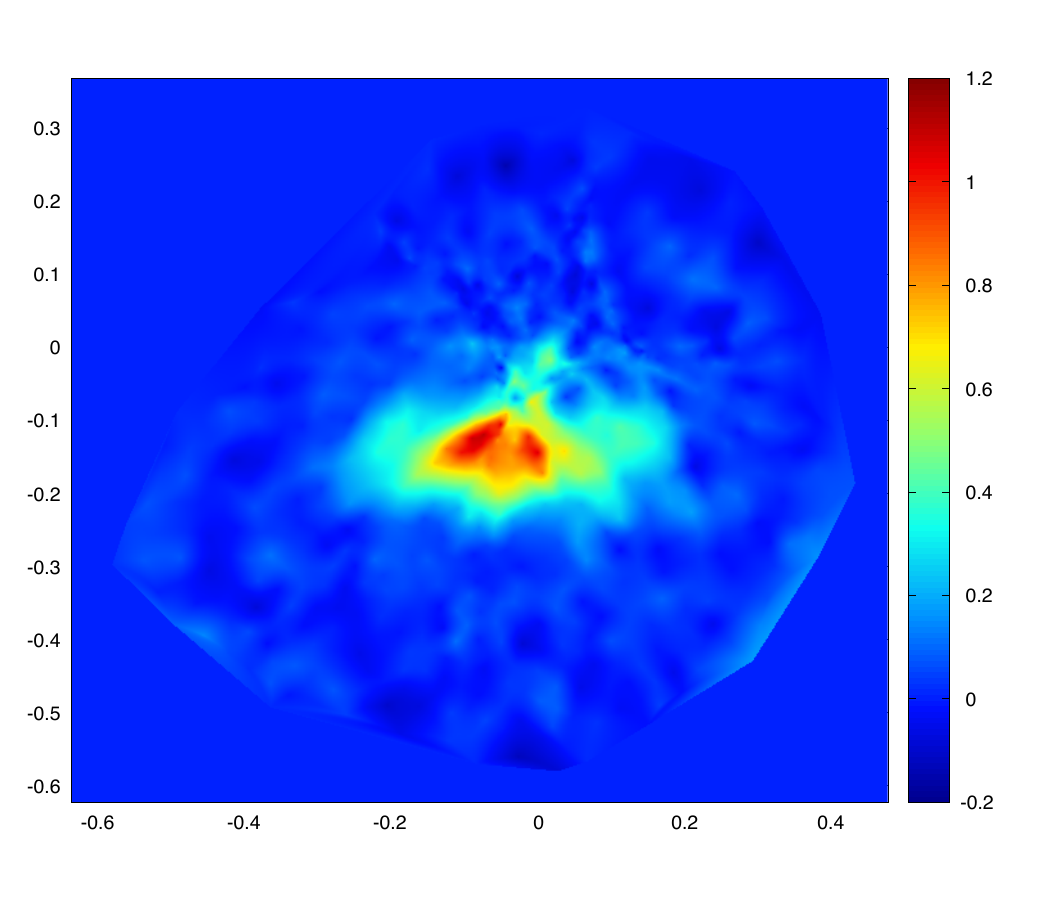}
		\label{spl4_nosub_src1}
	}
		\subfigure[reconstructed source $s2$ ($z_s$=2.035)]
	{
		\includegraphics[height=0.35\hsize,width=0.32\hsize]{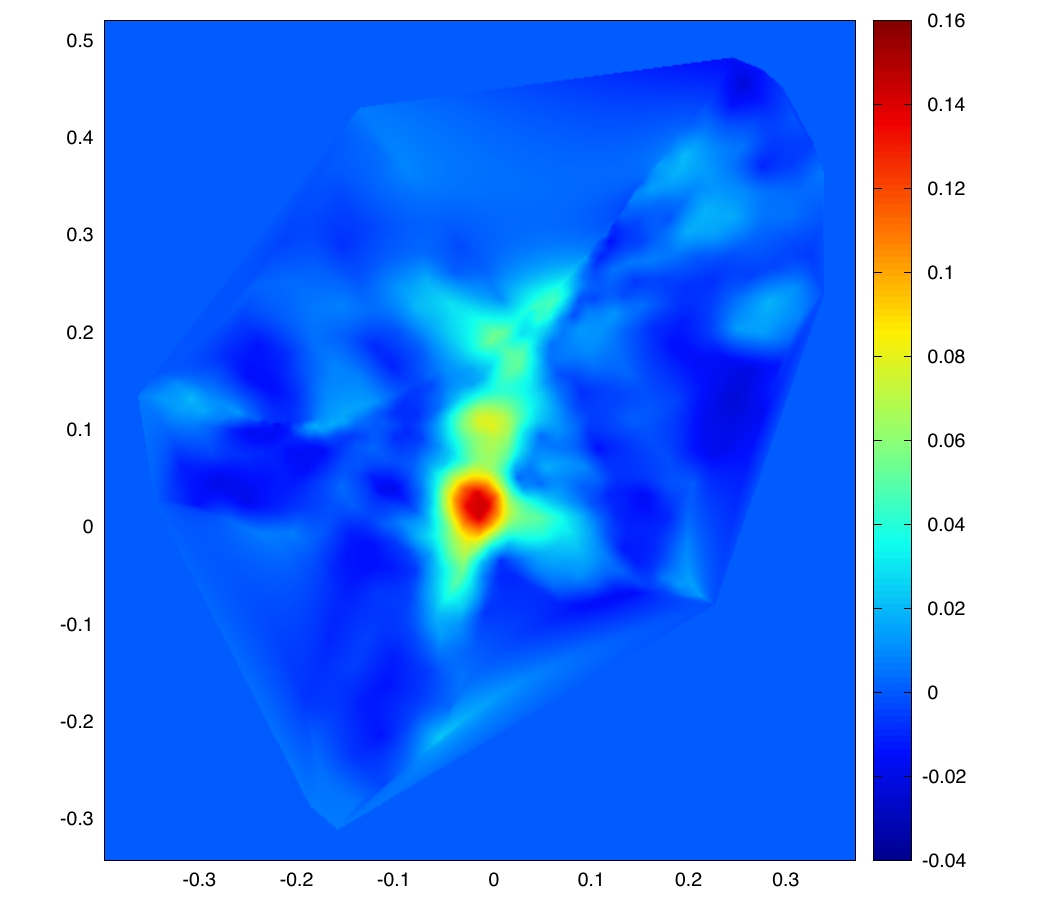}
		\label{spl4_nosub_src2}
	}
		\subfigure[residuals (with subhalo)]
	{
		\includegraphics[height=0.35\hsize,width=0.32\hsize]{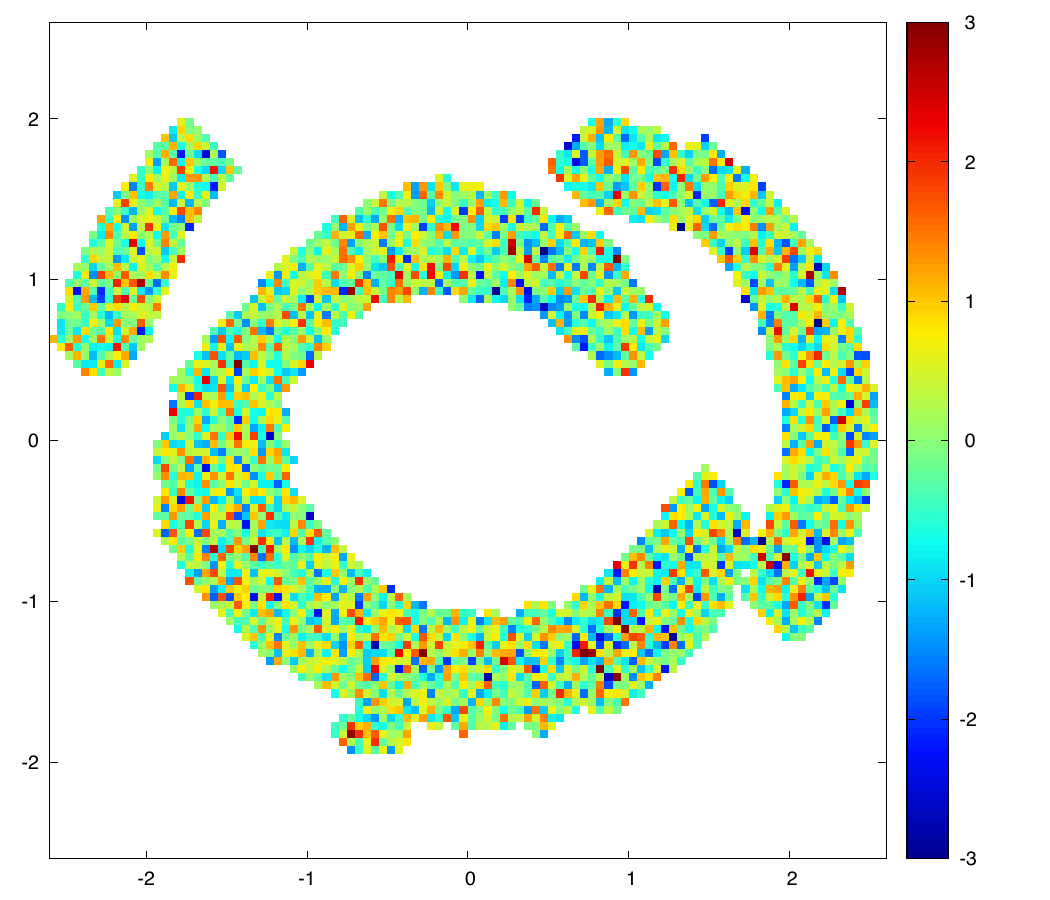}
		\label{spl4_tnfwsub_resid}
	}
		\subfigure[reconstructed source $s1$ ($z_s$=0.609)]
	{
		\includegraphics[height=0.35\hsize,width=0.32\hsize]{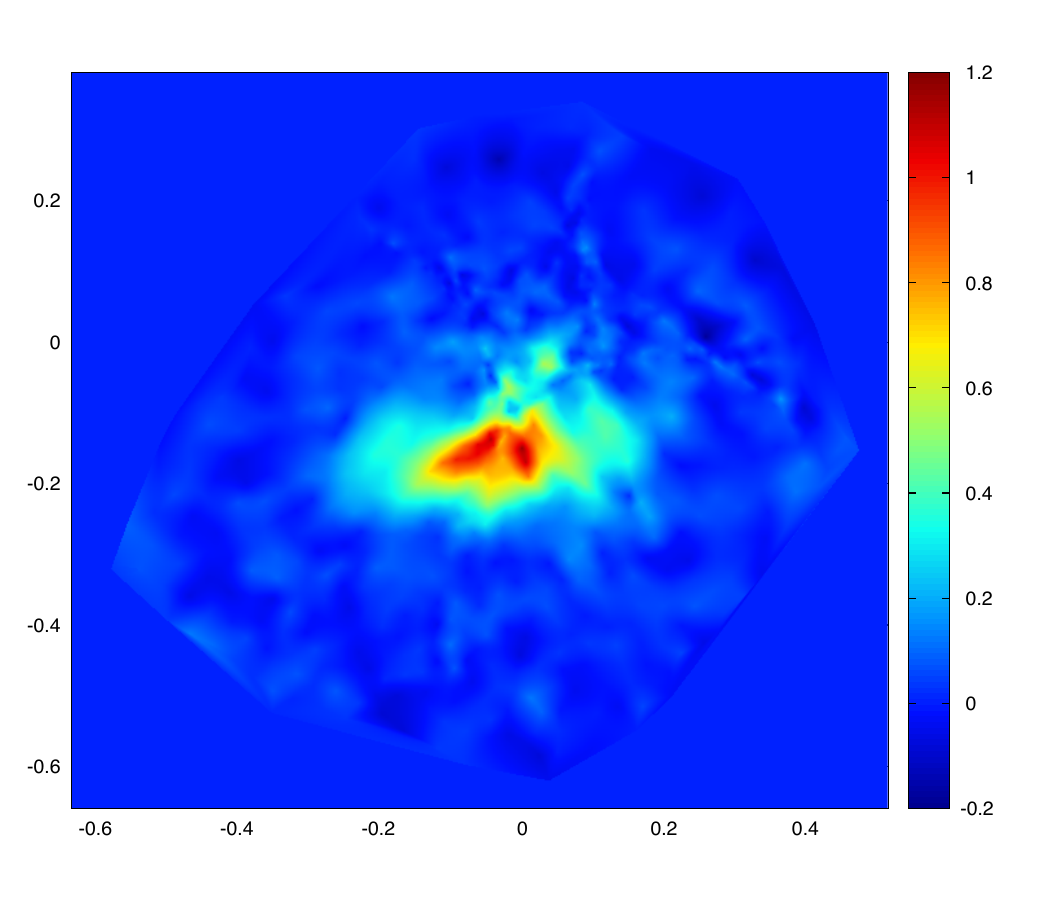}
		\label{spl4_tnfwsub_src1}
	}
		\subfigure[reconstructed source $s2$ ($z_s$=2.035)]
	{
		\includegraphics[height=0.35\hsize,width=0.32\hsize]{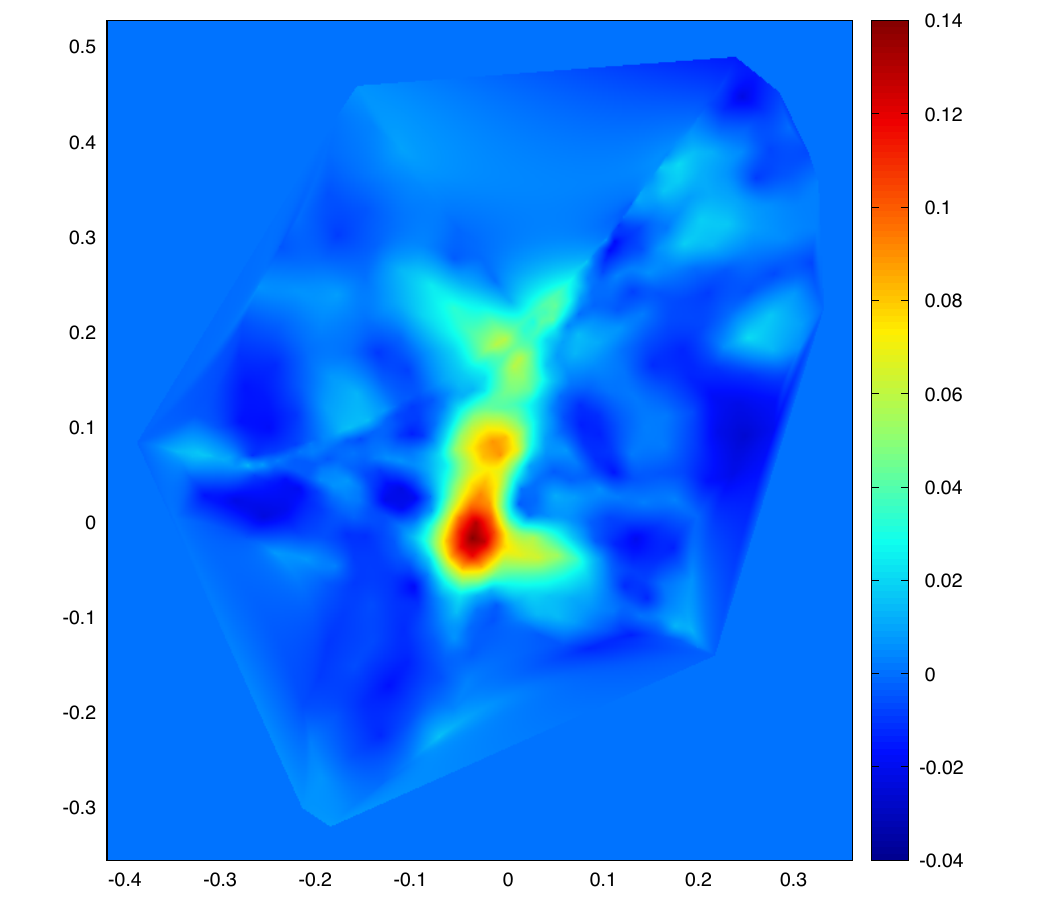}
		\label{spl4_tnfwsub_src2}
	}
 
		\subfigure[residuals (subhalo, multipoles)]
	{
 		\includegraphics[height=0.35\hsize,width=0.32\hsize]{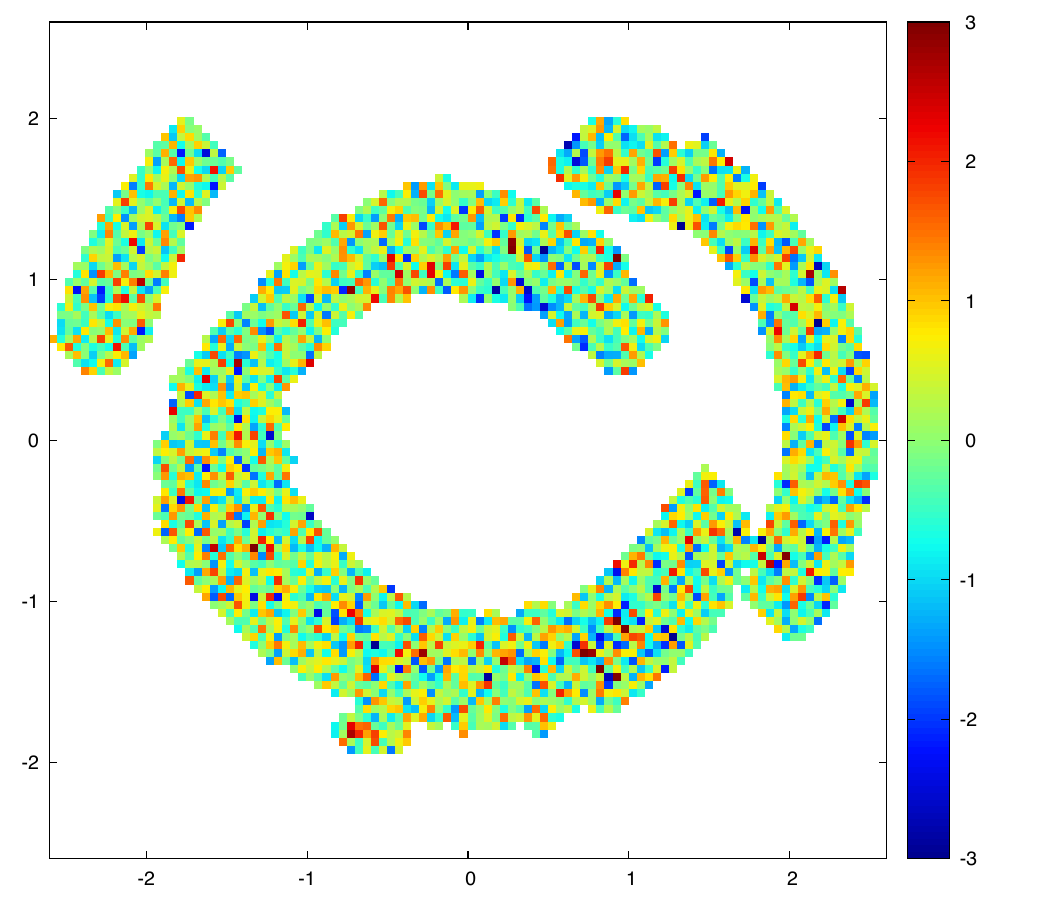}
		\label{spl4mult_tnfwsub_resid}
	}
		\subfigure[reconstructed source $s1$ ($z_s$=0.609)]
	{
		\includegraphics[height=0.35\hsize,width=0.32\hsize]{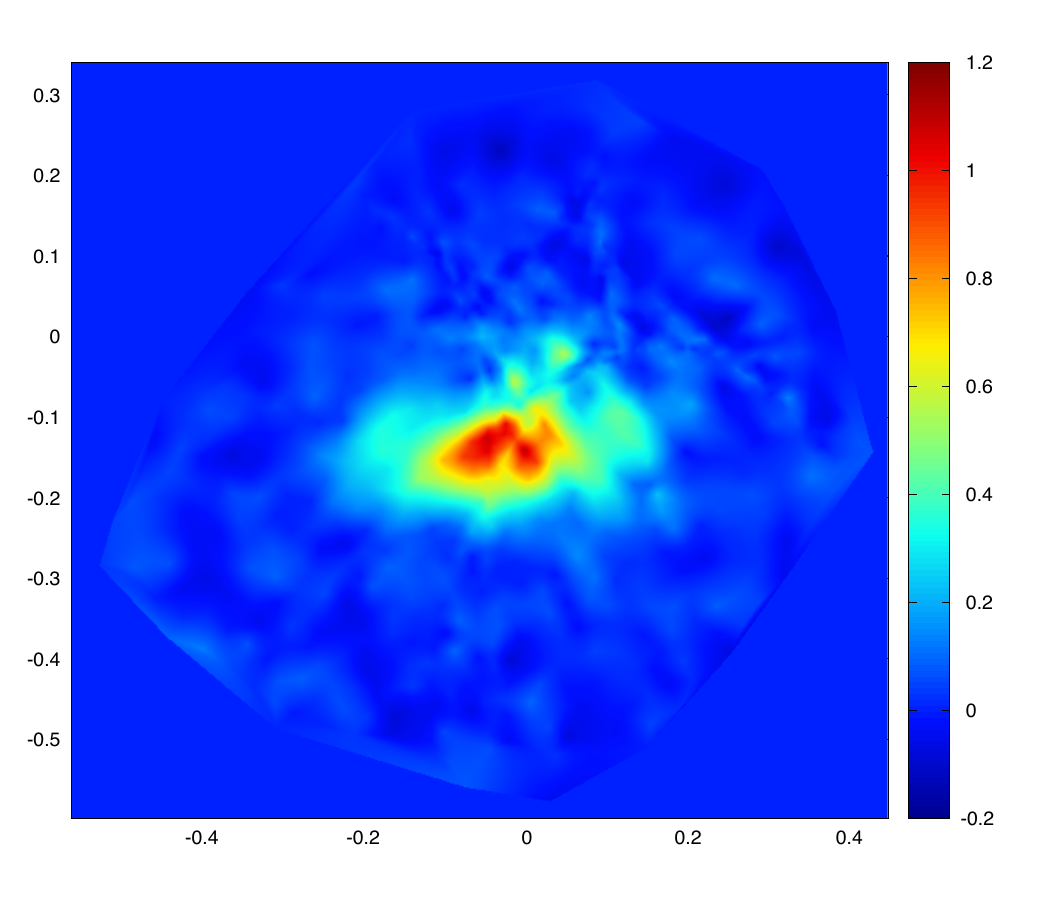}
		\label{spl4mult_tnfwsub_src1}
	}
		\subfigure[reconstructed source $s2$ ($z_s$=2.035)]
	{
		\includegraphics[height=0.35\hsize,width=0.32\hsize]{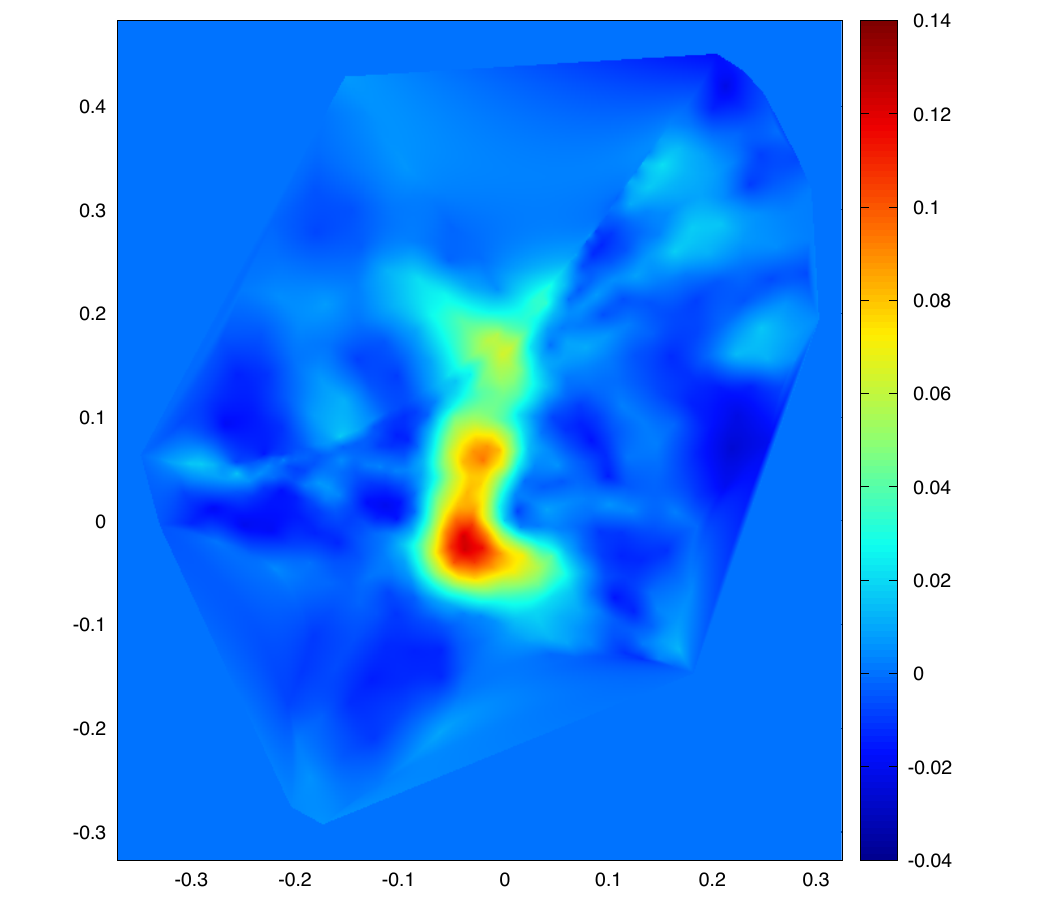}
		\label{spl4mult_tnfwsub_src2}
	}
	\caption{Normalized residuals and reconstructed sources for models that use $4x4$ supersampling of image pixels during model fitting. Top row shows the best-fit model that does not include a subhalo; middle row shows the same plots for the best-fit model that includes a subhalo, and bottom row shows plots for the best-fit model that includes both a subhalo and $m=3$, $m=4$ multipoles in the projected density of the lens.}
\label{fig:bestfit_supersampled}
\end{figure*}
\subsection{Substructure detection significance with supersampling}\label{sec:results_with_ss}

\begin{figure*}
	\centering
	\subfigure[b][reconstructed source (without subhalo)]
	{
		\includegraphics[height=0.47\hsize,width=0.47\hsize]{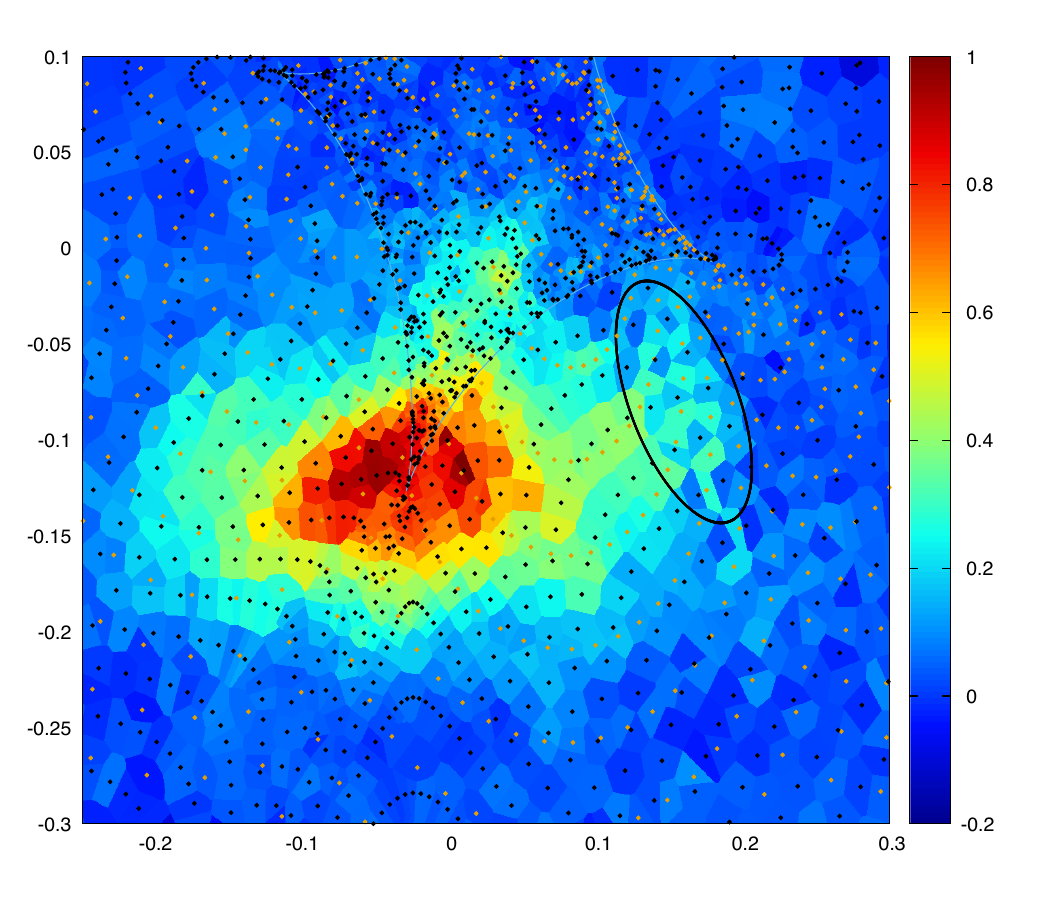}
		\label{checkerboard_nospl_nosub}
	}
	\subfigure[b][model image (without subhalo)]
	{
		\includegraphics[height=0.47\hsize,width=0.47\hsize]{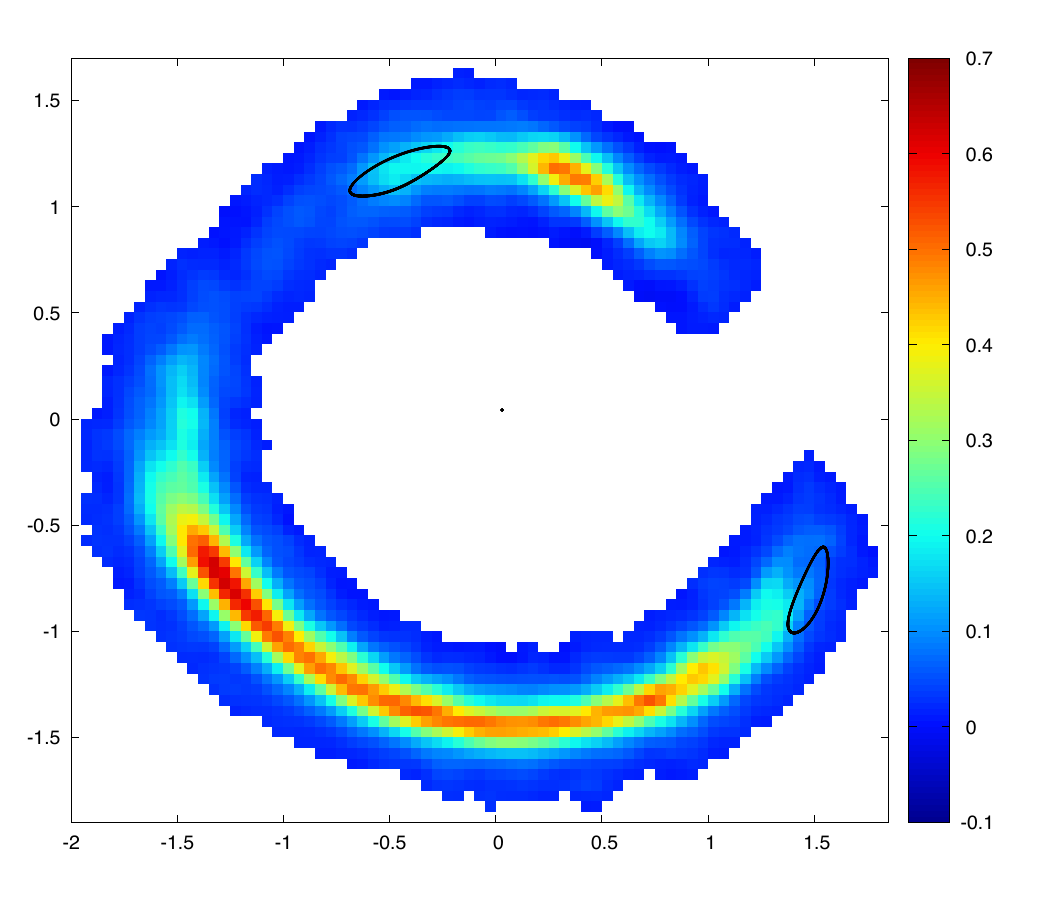}
		\label{imgpixels_tagged}
	}
	\caption{Inferred source pixel surface brightnesses for reconstructed source $s1$ (without supersampling) for the model without a subhalo, with the region of interest circled (black ellipse). Figure (b) shows the corresponding lensed images along with the images of the black ellipse (black curves). The points in (a) that correspond to ray-traced image pixels from the upper arc in (b) are colored orange, while the remaining points that are ray-traced from the lower arc are colored black. Note that the source pixels coming from the upper arc (orange points) require a significantly higher surface brightness than those coming from their counterpart image in the lower arc (black points), resulting in a ``checkerboard'' pattern of source pixels within the circled region. }
\label{fig:bestfit_sourcegrid}
\end{figure*}
\begin{figure*}
	\centering
 	\subfigure[reconstructed source (with subhalo)]
	{
		\includegraphics[height=0.47\hsize,width=0.47\hsize]{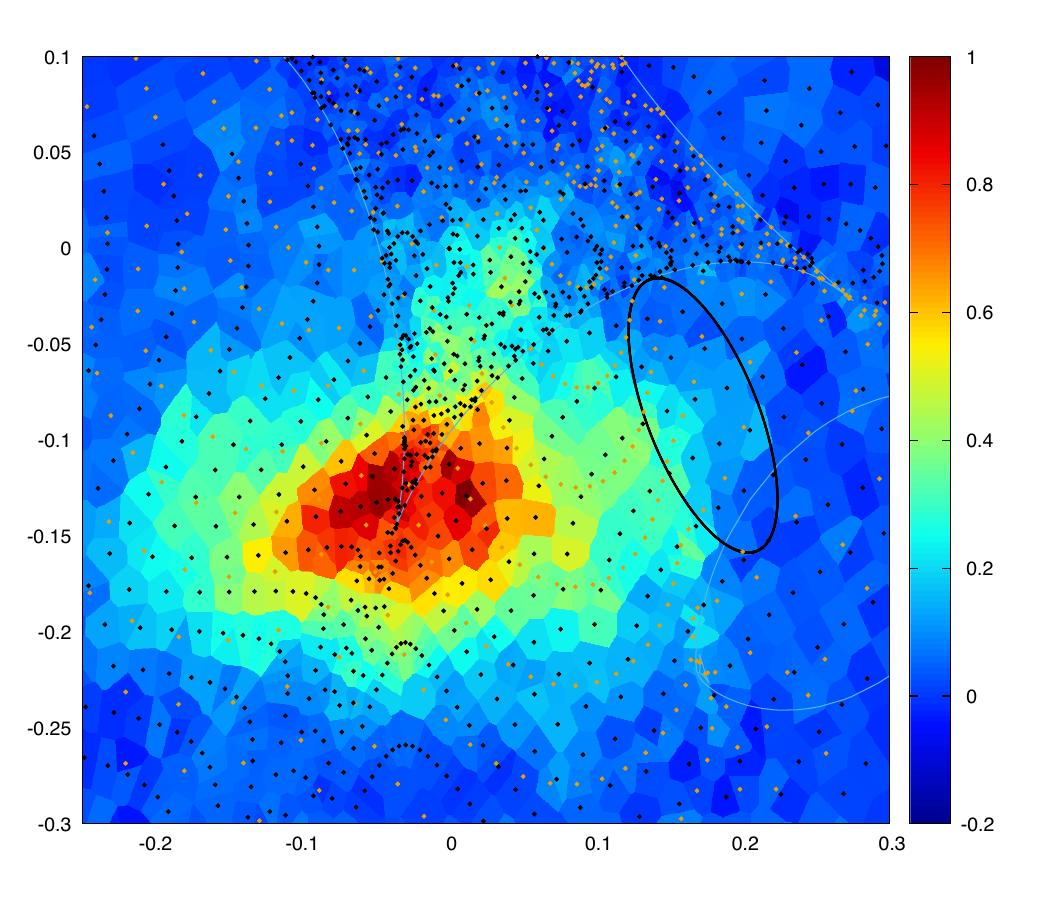}
		\label{checkerboard_nospl_tnfwsub}
	}
		\subfigure[model image (with subhalo)]
	{
		\includegraphics[height=0.47\hsize,width=0.47\hsize]{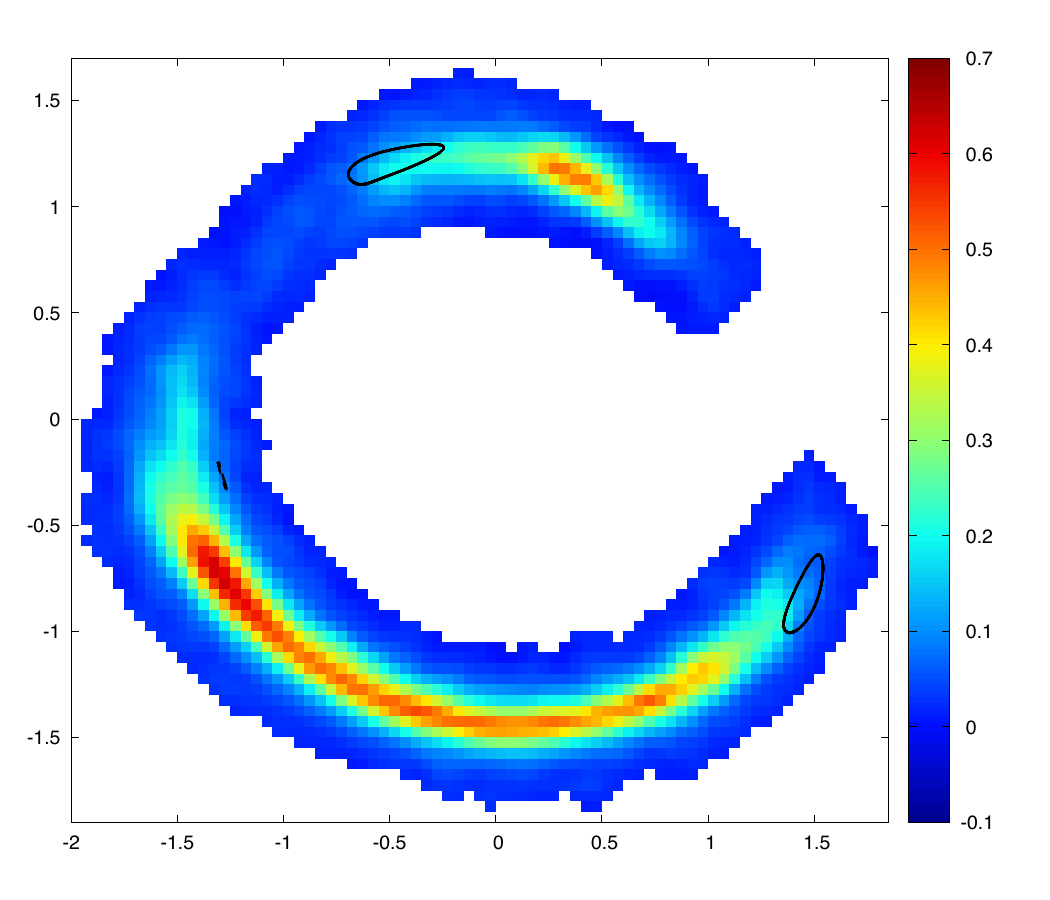}
		\label{imgpixels_tagged_tnfwsub}
	}
	\caption{Same as in Figure \ref{fig:bestfit_sourcegrid}, except this time the model includes a subhalo. The lensed images (b) show that this model yields more consistent surface brightness in the circled regions in the upper and lower arcs, and thus no checkerboard pattern is evident in (a); for this reason, this model yields greatly improved residuals when supersampling is used.}
\label{fig:bestfit_sourcegrid_sub}
\end{figure*}

In actual observations, the light detected by each camera pixel does not simply come from a single point in the sky, but rather is integrated over the entire area taken up by the pixel. We now approximate this by performing $4\times 4$ supersampling of the image pixels, as discussed in Section \ref{sec:lensmodeling}. First, we investigate whether the solution without a subhalo (top row of Figure \ref{fig:bestfit_nosup}) achieves good residuals if supersampling is now turned on. We turn on supersampling and then reoptimize the regularization parameters $\lambda_1$ and $\lambda_2$ followed by a final inversion of the lensing matrix. The resulting residuals are shown in Figure \ref{fig:bestfit_sup_turned_on}. For the model without a subhalo (Figure \ref{nospl0_nosub_spl4_resid}), we see strong positive residuals in the location of the purported subhalo in \cite{vegetti2010}, beyond 3$\sigma$ in some cases, with corresponding negative residuals in the counterimage in the lower right. Indeed, we have found that the fit is poor regardless of the splitting used; e.g. a $2\times 2$ splitting results in residuals of similar magnitude. To see the effect on the reconstructed source $s1$ for this model, in Figure \ref{fig:sources_sup_turned_on} we show a zoomed-in view of the inferred source $s1$ before and after supersampling is turned on. After supersampling is turned on (Figure \ref{zoomed_src_ss}), we see that the upper-right region of the source is considerably smoother than the original without supersampling (Figure \ref{zoomed_src_noss}), and the optimal regularization strength is higher resulting in a smoother source overall. 

For the model that includes a subhalo, however, the story is very different when supersampling is activated (Figure \ref{nospl0_tnfwsub_spl4_resid}): in contrast to the model without a subhalo, we do not find a significant change in the residuals beyond noise-level in the inferred location of the subhalo or its counterimage in the lower right. We do notice significant residuals appearing in a region on the right side of the smaller lensed arc from $s1$ where negative residuals are evident; these residuals are also present in the model without a subhalo, and play an important role in the inferred host galaxy slope which we will investigate in Section \ref{sec:galaxyslope}. Overall, however, we find no significant residuals beyond 2$\sigma$ in the purported location of the subhalo and its counterimage.

Although the model without a subhalo explored thus far provides a poor fit when supersampling is activated, it is reasonable to ask whether the lens model parameters could be adjusted to eliminate the residuals. We investigate this by performing a new nested sampling run with supersampling used during each likelihood evaluation. The resulting residuals are shown in the top row of Figure \ref{fig:bestfit_supersampled}. Indeed the residuals are reduced with adjusted lens model parameters, but significant residuals remain. Repeating the analysis with subhalo included in the model now results in significantly improved residuals (middle row). The log-Bayes factor between the two models is 142.6, resulting in a detection significance of $\sim16.7\sigma$, far greater than the detection significance in B24.\footnote{This result is also supported by \cite{nightingale2024} who used supersampling in a similar manner as we have done here, but modeling only the lower redshift source $s1$. They find a Bayes factor $\Delta\ln {\cal E} = 75.8$, also a high significance detection although somewhat lower than ours. The reason for the differing Bayes factor compared to ours may be because the authors employed an NFW subhalo without truncation and imposed a concentration-mass relation on the subhalo, resulting in a much lower concentration than is required to achieve noise-level residuals in the vicinity of the substructure.} This detection significance is far less sensitive to the source prior compared to when supersampling is not used; for example, when curvature regularization is used instead of gradient regularization, we find a detection significance of $\sim17.1\sigma$, a relatively minor increase in significance.

\subsubsection{Why is the fit without a substructure so poor when supersampling is used?}\label{sec:checkerboarding}

To understand why the solution without a subhalo fares so poorly when supersampling is used, it is useful to consider where the two lensed arcs map to in the source plane. In Figure \ref{fig:bestfit_sourcegrid} we plot the reconstructed source $s1$ without supersampling in the form of a Voronoi grid, so the source pixels can be clearly distinguished. The source pixels that come from ray tracing pixels from the upper arc (defined by $y>0.4$ arcsec in the image plane) are shown by the orange points, while the remaining pixels are shown by the black points. Note that in the solution without a subhalo (Figure \ref{checkerboard_nospl_nosub}), in the upper right region of the source (circled by the black ellipse) there is a ``checkerboard'' effect where the source pixels have alternating surface brightness. As the lensed images show (Figure \ref{imgpixels_tagged}), this is because the pixels from the upper arc prefer a higher surface brightness, while the pixels from the lower arc prefer a lower surface brightness. When gradient regularization is used, the source is allowed to fluctuate sufficiently from pixel to pixel that this ``checkerboarding'' is allowed, despite the fact that this noisiness of the source is not apparent in any \emph{individual} lensed image. Unfortunately, there is nothing in the source prior that would specifically penalize this solution based on the enormous coincidence it would require, namely that the pixels from the top arc all happen to map to brighter spots while the pixels from the bottom arc all happen to map to darker spots.\footnote{Note that the checkerboard effect should still be present even if one is modeling multiple bands, as done in B24, since the ray tracing is still the same for each band. Indeed, close inspection shows the checkerboard pattern to be present in their reconstructions of $s1$ that do not include a subhalo, similar to our reconstruction when no supersampling is performed. A similar pattern is evident in the more recent reconstructions of \cite{despali2024} who also used gradient regularization.}

When supersampling is used, however, the checkerboard effect is no longer possible: the subpixels within each image pixel map to different regions around the Voronoi cells, and interpolation is used to calculate the surface brightness of each subpixel, after which subpixels are averaged together. Thus, the darker and brighter subpixels are both included in the interpolation and averaging that produces the image pixel surface brightnesses, as would be expected in reality since we observe the integrated light over the area of each pixel, not just the light striking the center of each pixel. As a result, the reconstructed source is naturally smoother when supersampling is used, since it gets little benefit from the checkerboard effect, resulting in strong residuals. In contrast, when a subhalo is included in the model, the reconstructed source without supersampling (Figure \ref{checkerboard_nospl_tnfwsub}) does not require checkerboarding to achieve a good fit; the arc near the subhalo and its counterimage now both have surface brightnesses that are more consistent with each other (Figure \ref{imgpixels_tagged_tnfwsub}), resulting in a naturally smoother source. Hence, when supersampling is turned on, the morphology of the reconstructed source is virtually unchanged, and a good fit is still achieved.

\begin{figure*}
	\centering
 	\subfigure[model with interpolation, but no supersampling]
	{
		\includegraphics[height=0.49\hsize,width=0.48\hsize]{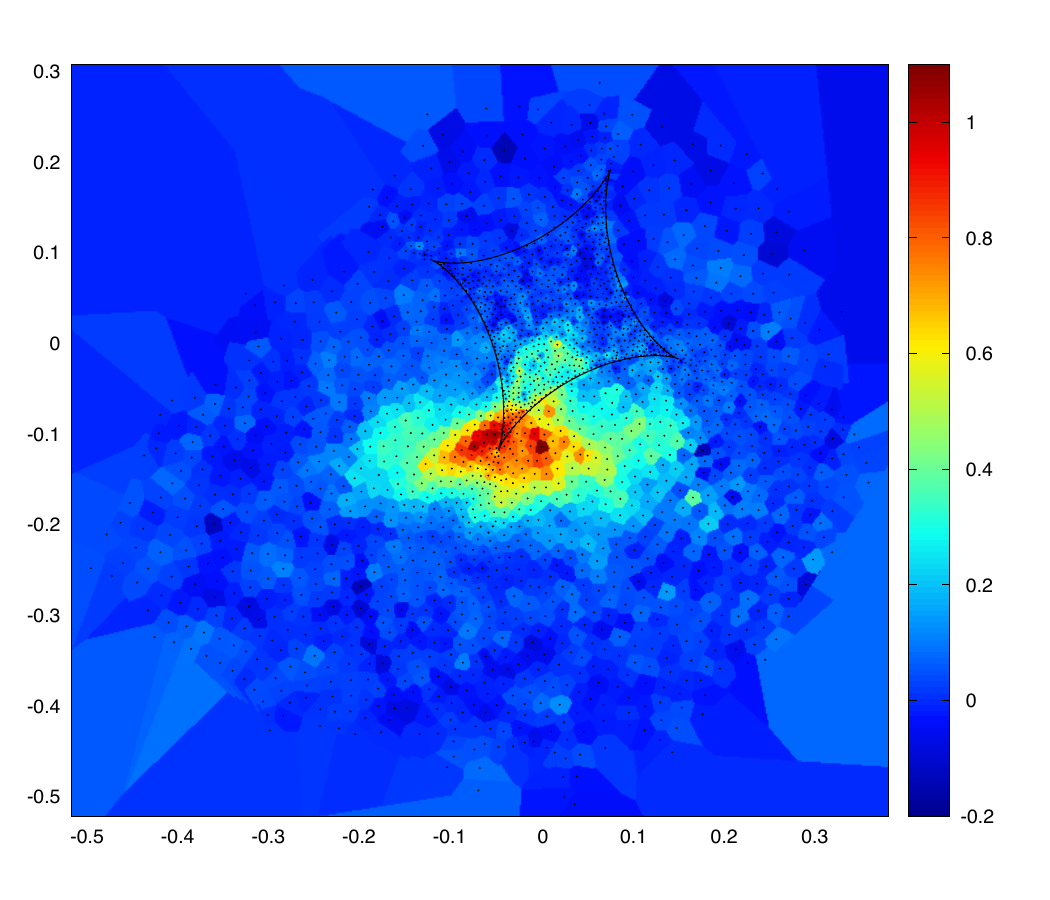}
		\label{fig:interpolation_noss_nosub}
	}
	\subfigure[same as (a), but with supersampling turned on]
	{
		\includegraphics[height=0.49\hsize,width=0.48\hsize]{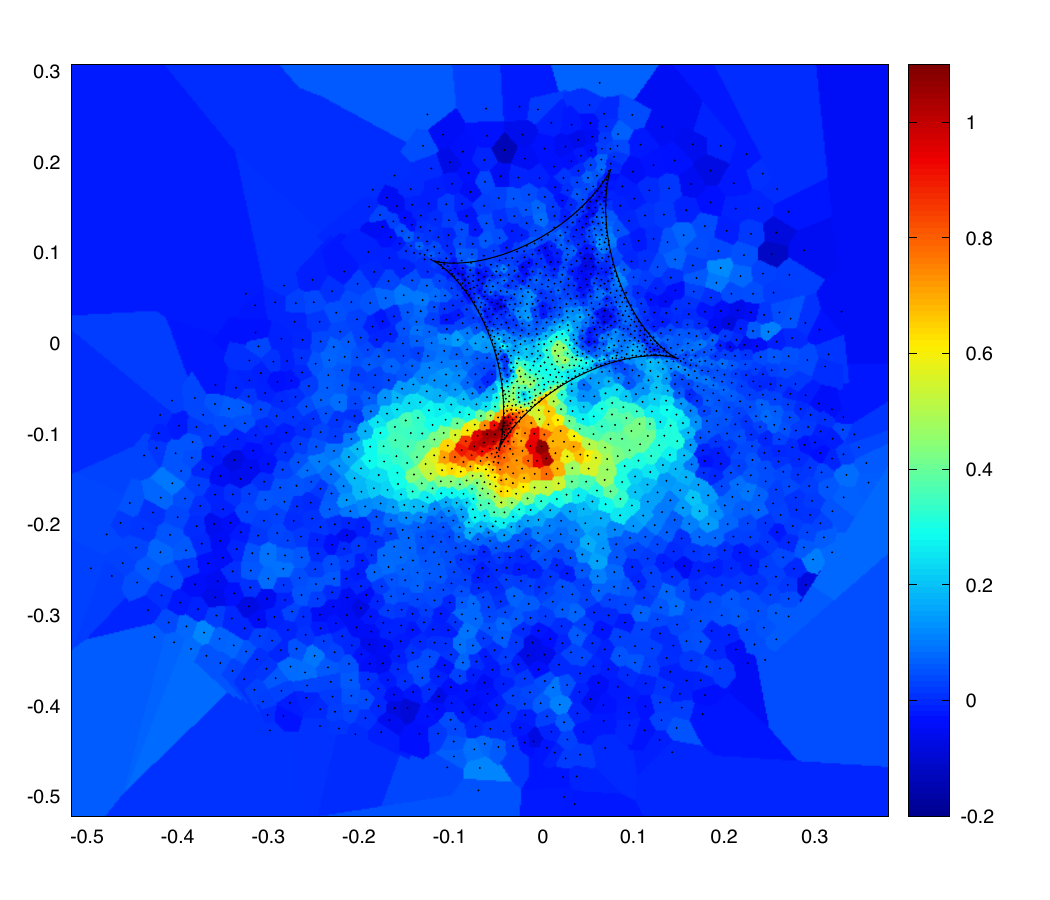}
		\label{fig:interpolation_sson_nosub}
	}
	\caption{The reconstructed source $s1$ for the model without a subhalo, obtained without supersampling during model fitting, but the source pixel grid is generated using the $k$-means clustering algorithm described in Section \ref{sec:src_reconstruction}. Hence, all image pixel surface brightness values are found by ray tracing the pixel centers and interpolating in nearby source pixels. (a) shows the best-fit reconstructed source from this procedure, while (b) shows the same solution but with $N_{sp}=4$ supersampling turned on and the regularization parmeter reoptimized. Note that in (a), there is still a checkerboard effect, although it is reduced by the interpolation; by contrast, the source is far smoother when supersampling is used.}
\label{fig:interpolation_noss}
\end{figure*}

\subsubsection{The role of source pixel interpolation versus supersampling}\label{sec:supersampling_versus_interpolation}

It is noteworthy that even in the absence of supersampling, the checkerboard effect described above can be diminished if the number of source pixels is chosen to be smaller than the number of image pixels $N_d$ within the mask. For example, if we choose $N_s = N_d/2$, then half of the image pixels do not map directly to a source pixel, hence their surface brightness values must be determined by interpolating in nearby source pixels. This reduces the checkerboarding effect and produces significant residuals if a subhalo is not included. If we redo the analysis using $N_d/2$ source pixels and no supersampling, we find a subhalo detection significance of $\sim 9\sigma$; this is not nearly as high compared to when supersampling is used, but more significant nonetheless.

This raises the question whether our subhalo detection significance might be just as high without supersampling, if the surface brightness of \emph{all} the ray-traced image pixels are found by interpolation. \footnote{Although we have focused on pixellated sources, we should mention that for lens modeling methods that reconstruct the source using basis functions such as shapelets \citep{birrer2015}, interpolation may not be performed since the basis functions can be evaluated directly. Thus, if a large number of basis functions are used to fit high magnification regions, supersampling \emph{must} be performed to reliably avoid the checkerboard effect.} To test this, we fit a model where the source pixels are found via the source pixel clustering method (as outlined in Section \ref{sec:src_reconstruction}), but the inferred surface brightness values are obtained by ray tracing the image pixel centers, i.e. without supersampling. (Note that the clustering algorithm does require splitting the pixels and ray-tracing the subpixels, but this is only to generate the source pixel Voronoi grid.) We redo our analysis with and without a subhalo and find a subhalo detection significance of $\sim 12\sigma$, which is still not as high as in the supersampled case (which was $\sim 17\sigma$ when supersampling with $N_{sp}=4$), but a significant detection nonetheless. As an interesting side note, we found the posterior to be noisier in the model without supersampling; this is likely due to the less complete coverage of the source plane that results from ray-tracing only the image pixel centers. By contrast, the supersampled model averages over a much larger number of interpolated source points, reducing the noisiness of the likelihood.

To see why the detection significance is not as strong as in the supersampled case, in Figure \ref{fig:interpolation_noss_nosub} we show the best-fit source rectonstruction for the model without a subhalo. Although the source is smoother than the unsupersampled model that did not use interpolation (Figure \ref{checkerboard_nospl_nosub}), there is still a noticeable checkerboard pattern present, particularly in lower magnification regions. To see the difference, we turn on supersampling and reoptimize the regularization parameter, resulting in the reconstructed source shown in Figure \ref{fig:interpolation_sson_nosub}. Here the checkerboard effect is eliminated, resulting in a smoother source reconstruction overall. Likewise, we find the image residuals from the unsupersampled (but interpolated) source are stronger compared to the model without interpolation, but nevertheless weaker than in the supersampled version, resulting in a somewhat smaller detection significance. 

\begin{figure*}
	\centering
	\subfigure[$\alpha=0.96$]
	{
		\includegraphics[height=0.35\hsize,width=0.32\hsize]{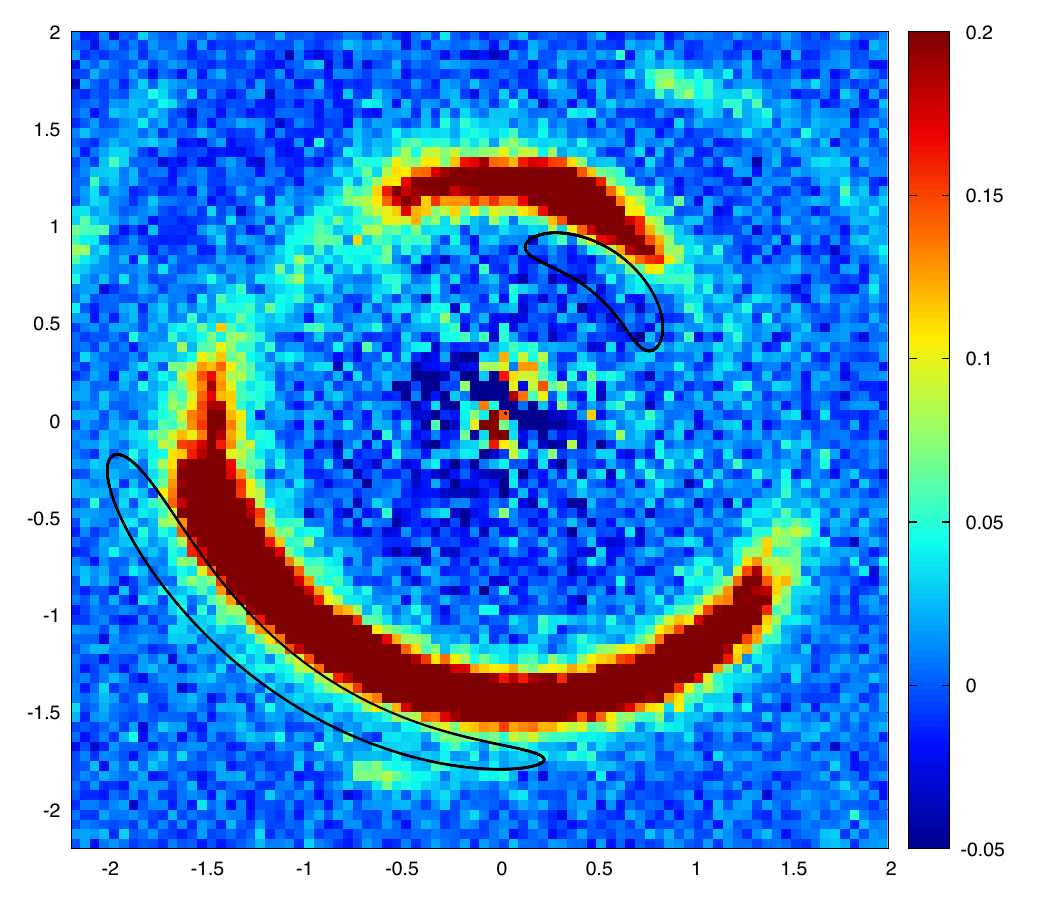}
		\label{fig:data_circ1}
	}
	\subfigure[$\alpha=1.1$]
	{
		\includegraphics[height=0.35\hsize,width=0.32\hsize]{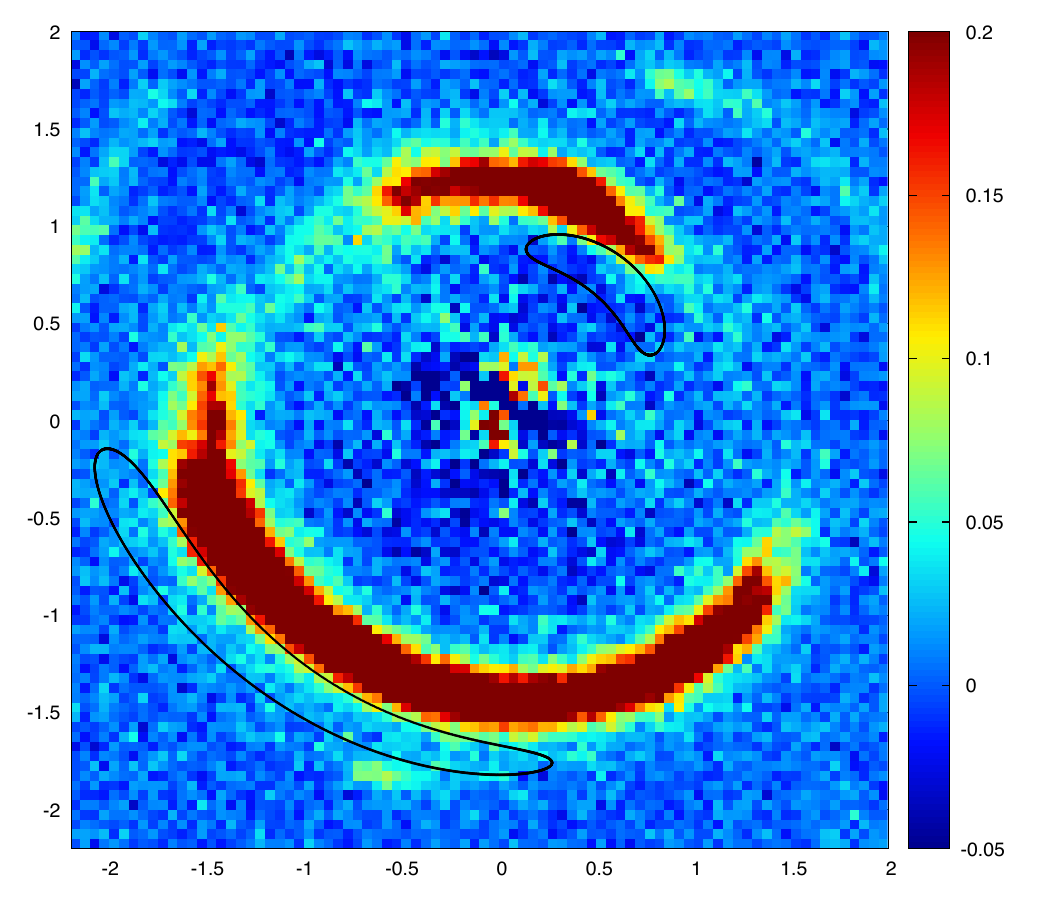}
		\label{fig:data_circ2}
	}
	\subfigure[$\alpha=1.3$]
	{
		\includegraphics[height=0.35\hsize,width=0.32\hsize]{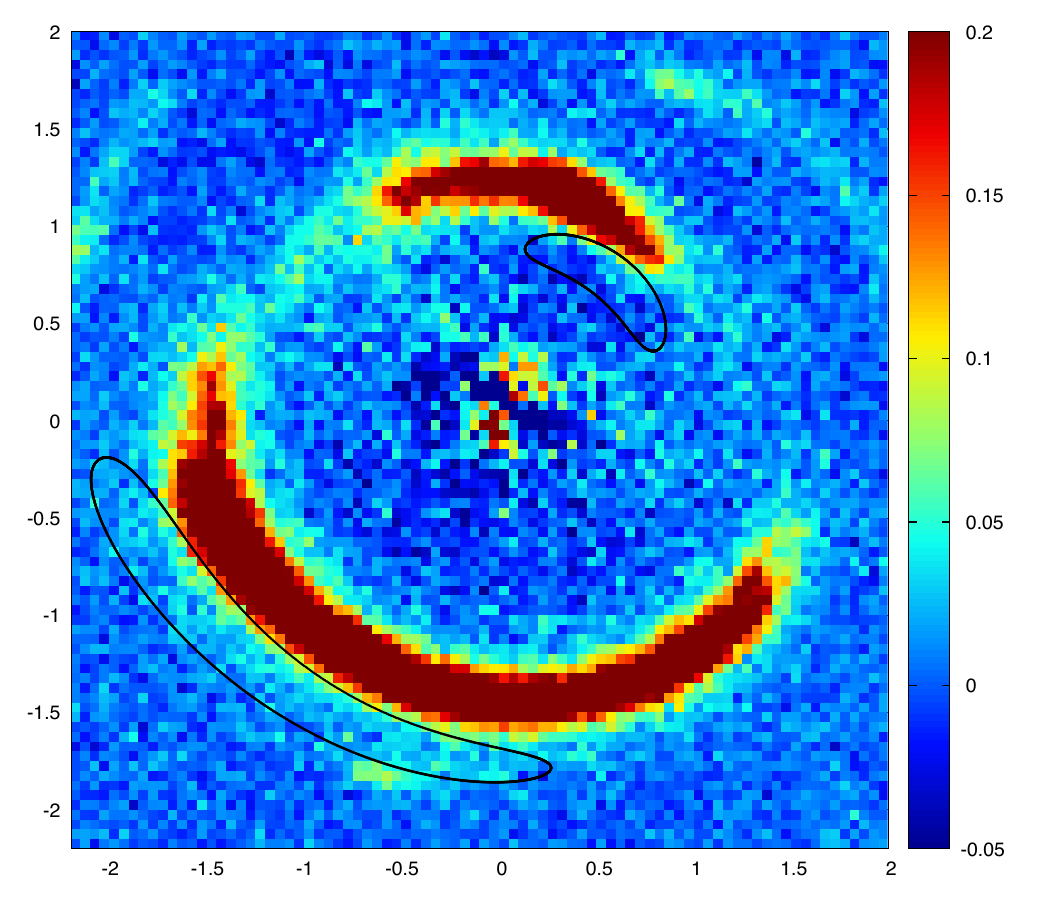}
		\label{fig:data_circ3}
	}
	\subfigure[residuals, $\alpha=0.96$]
	{
		\includegraphics[height=0.35\hsize,width=0.32\hsize]{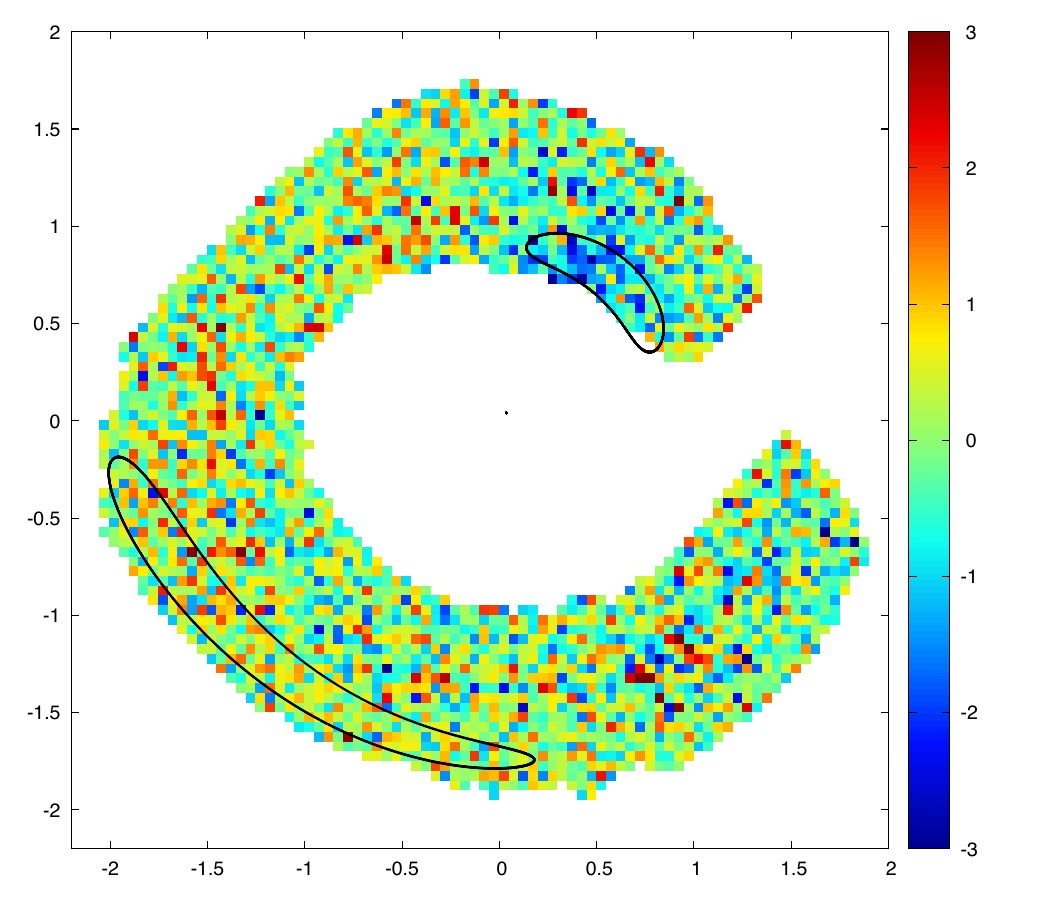}
		\label{fig:res_circ1}
	}
	\subfigure[residuals, $\alpha=1.1$]
	{
		\includegraphics[height=0.35\hsize,width=0.32\hsize]{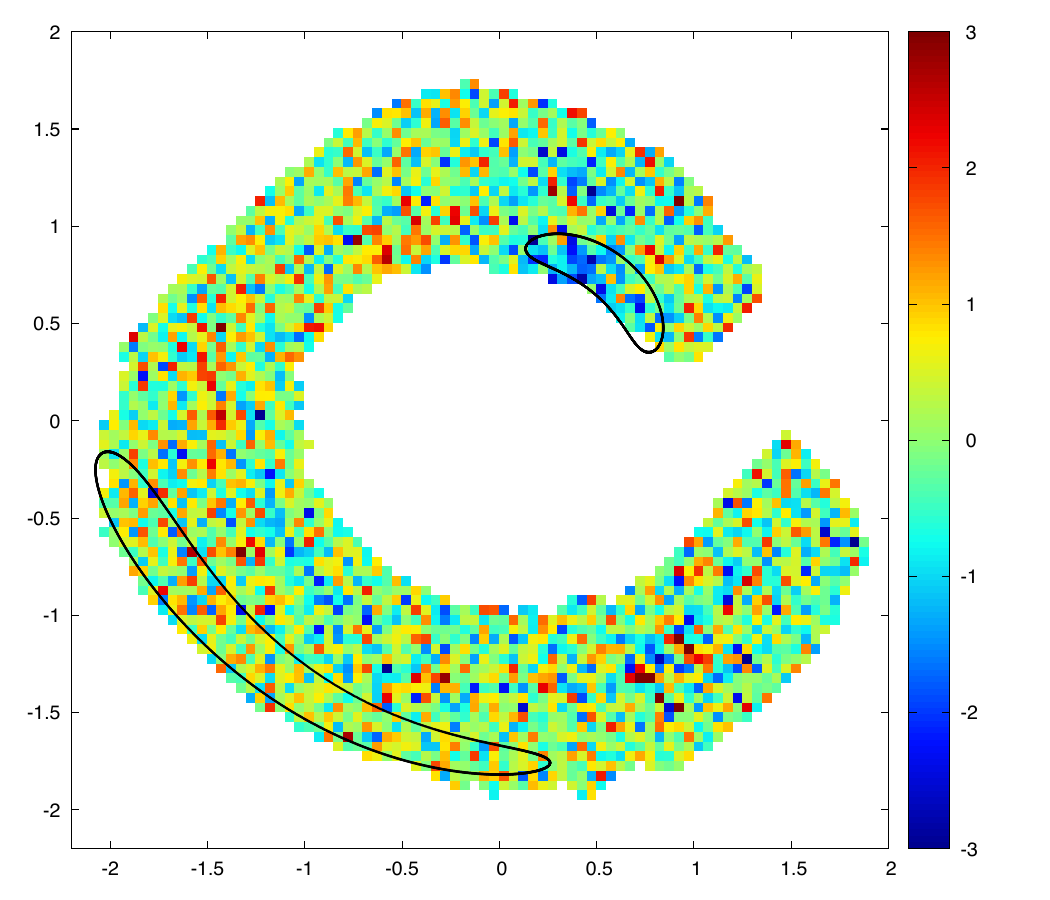}
		\label{fig:res_circ2}
	}
	\subfigure[residuals, $\alpha=1.3$]
	{
		\includegraphics[height=0.35\hsize,width=0.32\hsize]{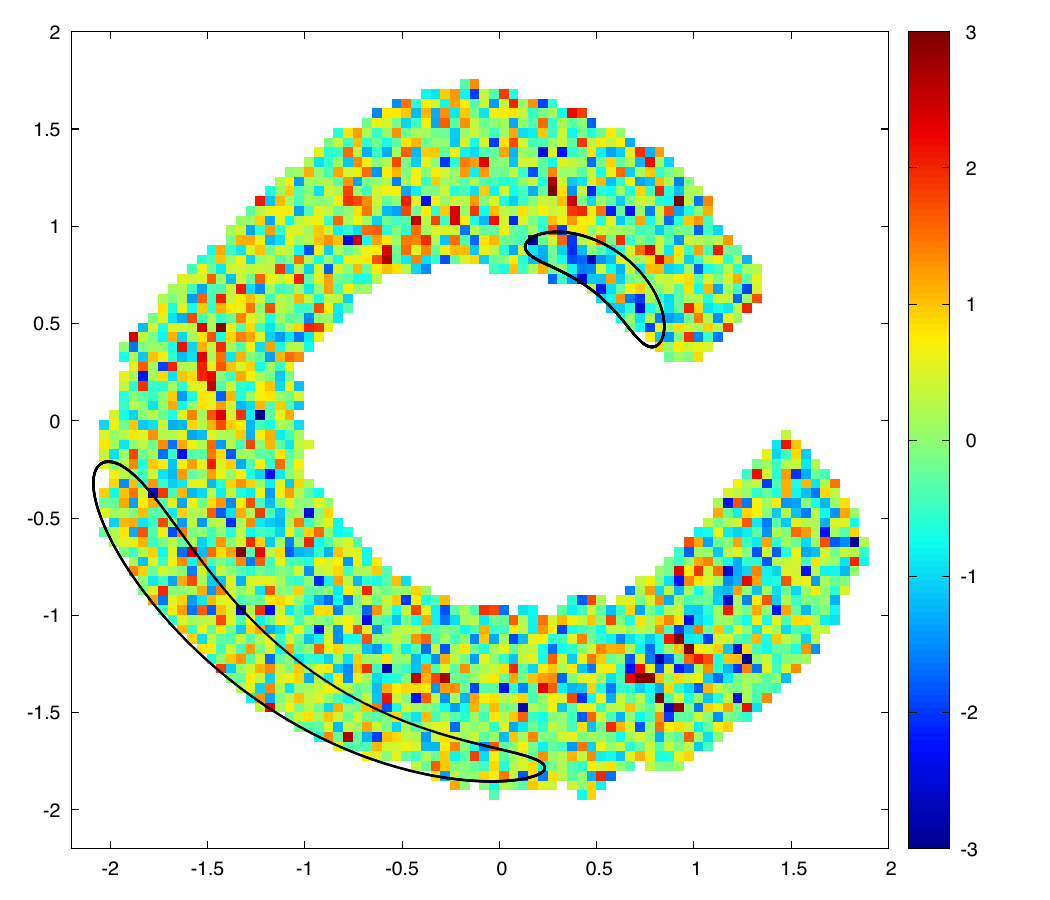}
		\label{fig:res_circ3}
	}
	\caption{Region below the upper arc of the lensed source $s1$ (smaller circled region in upper right) and its larger counterpart image using solutions with different density slopes for the host galaxy: $\alpha=0.96$, 1.1, and 1.3 respectively. Top row shows the data with the color bar capped at a low surface brightness, while bottom row shows the corresponding residuals (data minus model) from the model fits. Note that the smaller circled region is unexpectedly dark compared to its counterimage, but as the slope is increased, the counterimage is enlarged to include pixels with lower surface brightness, reducing the disparity between the two regions. For this figure we have enlarged the mask by two pixel lengths compared to the mask used in our modeling runs to show the greater extent of the region with negative residuals.}
\label{fig:varyslope}
\end{figure*}

\begin{figure}
	\centering
	\includegraphics[height=1.0\hsize,width=1.00\hsize]{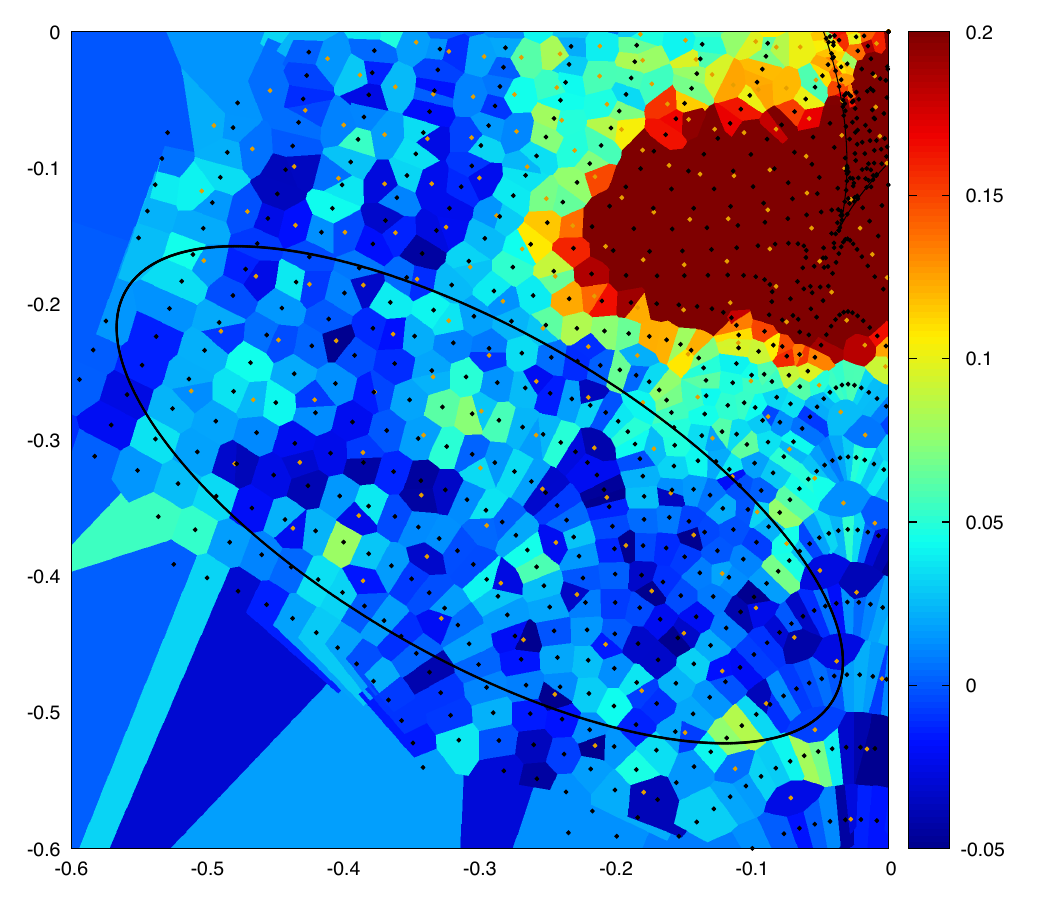}
	\label{checkerboard2}
	\caption{Inferred source pixels for the model that does not use supersampling, where we have zoomed in on the elliptical region whose lensed images lie in the circled regions in Figure \ref{fig:varyslope}, and we have capped the colorbar to provide greater contrast in the region circled by the ellipse. The ray-traced points are color-coded exactly the same as in Figure \ref{fig:bestfit_sourcegrid}, with orange points coming from the region of the upper lensed arc. Note that the checkerboard effect is again evident here, with pixels coming from ray-traced points near the upper arc preferring a lower surface brightness compared to other nearby pixels.}
\label{fig:checkerboard2}
\end{figure}

\begin{figure*}
	\centering
	\subfigure[trimmed mask]
	{
		\includegraphics[height=0.50\hsize,width=0.48\hsize]{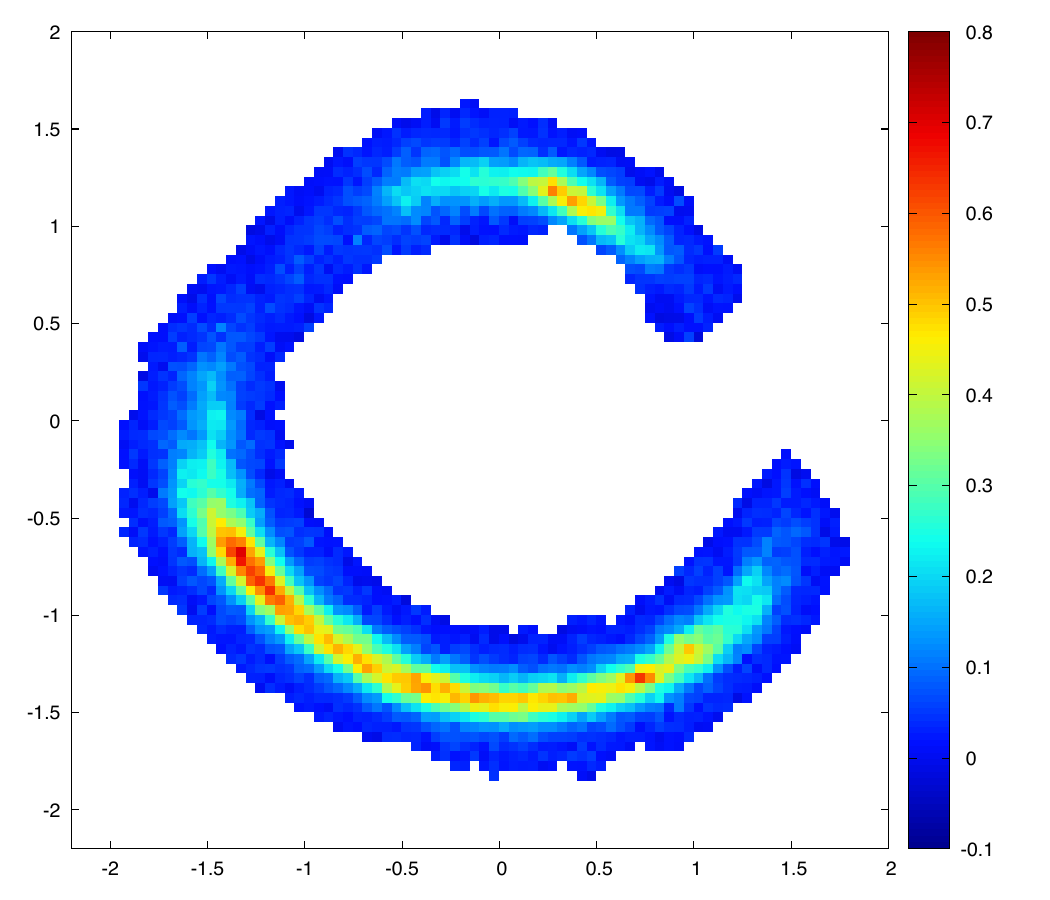}
		\label{fig:trimmed_mask}
	}
	\subfigure[inferred posterior probability in galaxy log-slope $\alpha$]
	{
		\includegraphics[height=0.48\hsize,width=0.48\hsize]{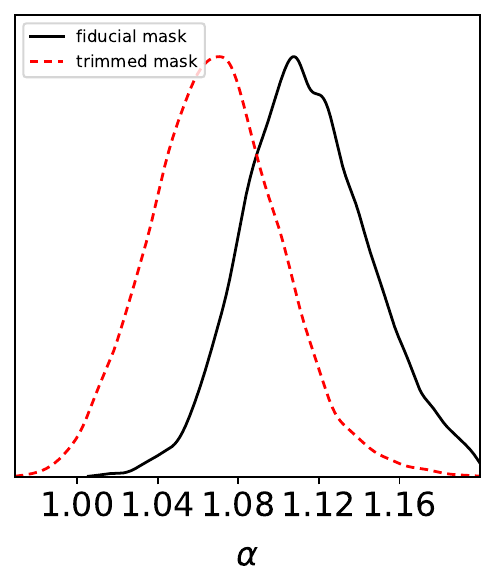}
		\label{fig:alpha_post_trimmed}
	}
	\caption{Effect on inferred host galaxy 2D log-slope $\alpha$ when a ``trimmed'' mask is used for the low-redshift source $s1$. (a) shows the data with the trimmed mask (compare to Figure \ref{fig:mask1}), where pixels under the upper arc have been removed. After rerunning the analysis, (b) shows the effect of the posterior on $\alpha$, where the median $\alpha$ has been reduced from 1.11 to 1.07, confirming that the region in the upper right of the mask is the principle source of bias in the galaxy slope.}
\label{fig:trimmed_mask_results}
\end{figure*}

\subsection{Constraint on the host galaxy's density slope}\label{sec:galaxyslope}

Before we discuss the effect of supersampling on the inferred subhalo properties, it is useful to first consider its effect on the host galaxy properties. The most striking difference, evident in Table \ref{tab:posterior_inferences}, is in the (negative) log-slope $\alpha$ of the projected density profile of the lens: when no supersampling is performed, the slope is consistent with isothermal (i.e. $\alpha=1$), in agreement with the results of B24 when both lensed sources $s1$ and $s2$ are modeled (note when comparing results that B24 uses the 3D log-slope $\gamma = \alpha+1$). However, when supersampling is used, the median inferred slope jumps higher by $\approx 0.1$, and in the model with a subhalo we find $\alpha = 1.11 \pm 0.03$. This is inconsistent with an isothermal slope at the 3$\sigma$ level, and thus is in tension with the results of kinematic modeling by \cite{turner2024} who infer $\alpha = 0.96\pm 0.02$ and the earlier modeling of \cite{sonnenfeld2012} who infer $\alpha = 0.98\pm0.02$ from a combination of lensing and kinematics. In the models with multipoles, this tension is weaker, with $\alpha = 1.07\pm0.04$, but a bias still appears to be present. It is worth exploring the source of this bias, especially as it may affect the subhalo constraints.

First, we should note that the tension would be much higher if not for the inclusion of the second source plane $s2$. In all prior models that include a subhalo but only model the $s1$ images, the median inferred slope is at least $\alpha \gnsim 1.2$. The highest evidence model in \cite{minor2021} yields $\alpha \approx 1.37$, while more recently, \cite{despali2024} find $\alpha \approx 1.6$ in their model with an NFW subhalo. We confirm this by rerunning our analyses and fitting only the source $s1$, the results of which are explored in detail in Appendix \ref{sec:1src_comp} (see Figure \ref{fig:alpha_post_1src} in particular for a comparison of posteriors in $\alpha$). The fact that the lensed arcs from $s1$ can be fairly well reproduced over a considerable range of slope values is likely a consequence of the source position transformation \citep{schneider2013} degeneracy, which is largely broken by the inclusion of multiple source planes. The bias appears to be effectively eliminated in non-supersampled models that fit both lensed sources, but is apparently more difficult to overcome when supersampling is used. To understand why this is the case, we choose points from the chain from our non-supersampled run (with subhalo included) with three different slopes $\alpha=0.96$, $\alpha=1.1$ and $\alpha=1.3$, turn supersampling on in each case, and reoptimize the model while keeping the slope constant. (Since there were no points in the chain with $\alpha=1.3$, we started from 1.1 and incrementally increased the slope to 1.3, reoptimizing each time, over three iterations.)

The resulting residuals are shown in Figure \ref{fig:varyslope}. Note that there is a prominent region in the upper right with negative residuals, implying the model is producing too much surface brightness in this region. This is a region of low signal, below the rightmost portion of the upper arc from $s1$, whose counterimage is a more magnified region below the lower arc; to visualize this, we generate an ellipse in the source plane whose lensed images map to the region in question (black curves). Since these regions differ in surface brightness, it is more advantageous for the reconstructed source to match the surface brightness of the larger counterimage since it comprises far more pixels. However, as the log-slope is increased, note that the counterimage is pushed further out to a region with lower surface brightness, reducing the difference in surface brightness and therefore resulting in smaller residuals. 

The question remains, why is this effect reduced in the non-supersampled fits? In Figure \ref{fig:checkerboard2} we plot the inferred source for the $\alpha=0.96$ solution obtained without using supersampling, and zero in on the region inside the ellipse whose lensed images correspond to the curves shown in Figure \ref{fig:varyslope}. As in Figure \ref{fig:bestfit_sourcegrid}, the ray traced points corresponding to the image pixels from the upper arc are colored in orange, while the points from the lower arc are colored in black. Note that again we find a ``checkerboard'' effect, with the pixels from the upper arc region preferring a lower surface brightness and vice versa; close inspection shows examples of ray-traced points that are very close together but whose pixels have noticeably different surface brightnesses. Since the gradient regularization allows for such fluctuations, and since \emph{all} the image pixels in the mask are being ray traced to create source pixels, the checkerboarding results in noise-level residuals in the lensed images. From this, one can predict that a stronger regularization scheme such as curvature regularization, should result in significant residuals since such fluctuations are suppressed. Indeed, this is evident in the fit in B24 that uses curvature regularization, where residuals can be seen (see their Figure E1, upper right panel) in the same region in question; such residuals are not evident in any of their other fits that use gradient regularization. For this reason, their fit using curvature regularization does indeed result in a higher density slope, especially when only one source plane $s1$ is used (they infer $\alpha \approx 1.4$ in that case).

We defer to Section \ref{sec:dark_spot_discussion} a discussion of the possible origin of the unexpected low surface brightness region identified here. However, to verify that the region we have identified is indeed driving the bias in the galaxy's density slope, we implement a ``trimmed'' mask where the low surface brightness region under the upper arc is removed from the mask (Figure \ref{fig:trimmed_mask}). After re-running our analysis with supersampling and a subhalo included, the inferred slope (plotted in Figure \ref{fig:alpha_post_trimmed}) shifts to $\alpha = 1.07 \pm 0.03$ (dashed red curve), which is 0.04 lower than its inferred value with the original mask (black curve). This is perhaps the best that could be expected from this approach, since residuals remain above the trimmed region where the signal is higher, but nevertheless reinforces our conclusion that a lower-than-expected surface brightness at and around the upper arc is tending to bias the density slope to a higher value.\footnote{It is possible that adding the third lensed source using the VLT/MUSE data from \cite{collett2020} would further help to reduce the bias in the slope. Based on the low resolution of the MUSE data and the relatively small extent of the arcs, we expect its impact to be relatively minor; however it is difficult to know for sure whether it would make a significant difference, since in B24 they do not compare solutions using only the two lensed sources versus using all three. Nevertheless, the third MUSE source could in principle provide a useful consistency check on whether the inferred slope is compatible with the data at a much more distant redshift ($z_{s3}=5.975$).}

\subsection{Modeling additional angular structure in the foreground galaxy}\label{sec:multipoles}
The foreground galaxy in J0946 shows a remarkable degree of isophote twist in the same radial range as the lensed arcs from the $s1$ source \citep{gavazzi2008}.  It is therefore natural to question whether unaccounted-for angular structure in the lensing galaxy could possibly mimic the presence of a substructure in this case.  In B24, they note that when additional angular structure is added to the lensing mass in the form of $m=3,4$ multipole terms, the fit performs marginally better compared to the subhalo-perturbed model, at the 1.5$\sigma$ confidence level (although they note that the highest Bayesian evidence is achieved when \emph{both} are added to the model). Here we check whether this result holds up when supersampling is performed. We include multipole terms exactly as in B24 and perform a nested sampling run, this time with 1400 live points to reduce the odds of missing the maximum a posteriori peak in the higher-dimensional parameter space. We find that, although the addition of multipoles is preferred over the un-perturbed model with an increase in log-evidence $\Delta\ln{\cal E} = 58.0$, significant residuals remain near the purported location of the substructure and its counterimage, and the ``checkerboard effect'' is still present in the source reconstruction. As a result, the subhalo-perturbed model is strongly preferred over the multipole-perturbed model  at the $12.8\sigma$ confidence level ($\Delta\ln{\cal E} = 84.52$). We conclude that additional angular structure (of the kind that would be expected from isodensity contour twist or ellipticity gradients) is very unlikely to mimic the effect of the localized substructure perturbation inferred in J0946, in agreement with the conclusion of \cite{vegetti2010} based on modeling pixellated corrections to the lensing potential.

We also perform modeling runs that include both multipoles and a subhalo, with and without supersampling. As in B24, this model produces the highest Bayesian evidence among all the models. The residuals and reconstructed sources from the best-fit point are shown (for the supersampled model) in the bottom row of Figure \ref{fig:bestfit_supersampled}, and the parameter inferences are listed in Table \ref{tab:posterior_inferences2}. Close inspection shows that the model with multipoles does indeed produce slightly smaller residuals in the vicinity of the subhalo and in its counterimage in the bottom right; however, in other regions, larger residuals are produced, and overall the best-fit $\chi^2$ is actually slightly higher for the model with multipoles compared to without. The reason this model is preferred by the Bayesian evidence, then, is due to its allowing a somewhat smoother source, particularly the low-redshift source $s1$. This seems to be a rather tenuous reason for preferring the model with multipoles, and it is a fair question whether a more sophisticated regularization scheme could result in a different evaluation by the Bayesian evidence, a point to which we will return in Section \ref{sec:source_prior_discussion}. 

Before moving on to discuss the subhalo constraints, we note here that in all of our fits, including the models with multipoles, the residuals for the $s2$ images are not quite reduced down to noise-level despite the low signal-to-noise of these images. This is most noticeable in the bottom-most image where 7-8 contiguous pixels have residuals in the same direction. We speculate this is most likely due to the limitations of the angular structure captured in our model: the first few multipoles cannot perfectly capture the isodensity contour twist and ellipticity gradient which is likely present in the lensing mass based on the observed isophotes. Perhaps another indication of this is the fact that the centroid of the lensing mass corresponding to $s1$ is typically offset from the centroid of its surface brightness in most of our fits by at least 1$\sigma$ (in our models with multipoles, the median offset is roughly $\Delta_{xc,s1}\approx -0.12\arcsec$, $\Delta_{yc_s1}\approx -0.1\arcsec$, though it is somewhat smaller in the models without multipoles). Note that this only affects the images of $s2$, so this could be an attempt to compensate for not capturing the angular structure of the lens perfectly well near the Einstein radius of the lensed images of $s2$. If a higher signal-to-noise exposure is modeled in the future (e.g. using \textit{JWST} observations), it may become necessary to move beyond the simple multipole-based model to a model that more directly captures the angular structure described above to achieve a good fit to both sets of lensed arcs.

\begin{figure*}
	\centering
	\subfigure[without multipoles]
	{
		\includegraphics[height=0.48\hsize,width=0.48\hsize]{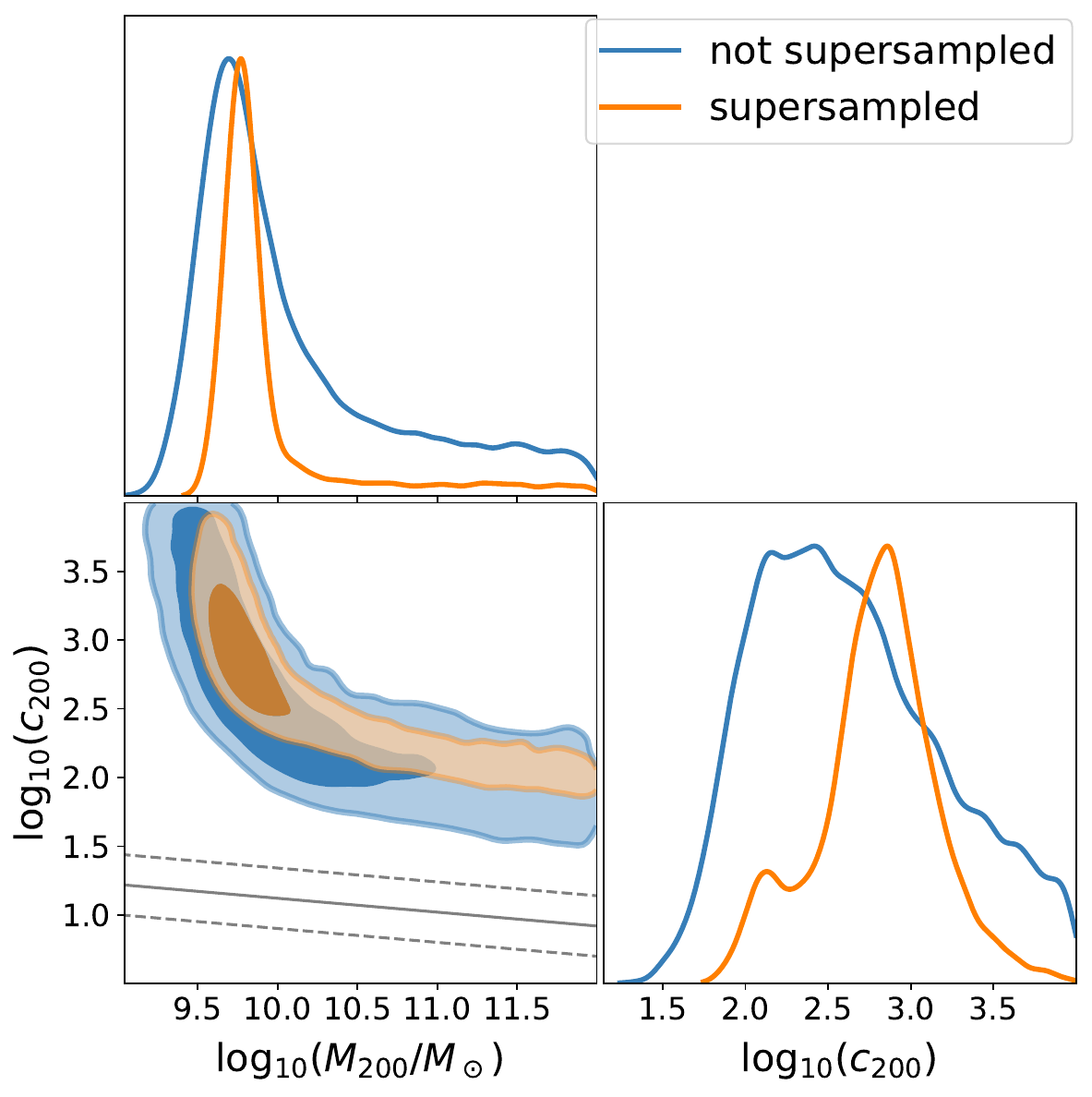}
		\label{fig:m200_vs_logc_nomult}
	}
	\subfigure[with multipoles]
	{
		\includegraphics[height=0.48\hsize,width=0.48\hsize]{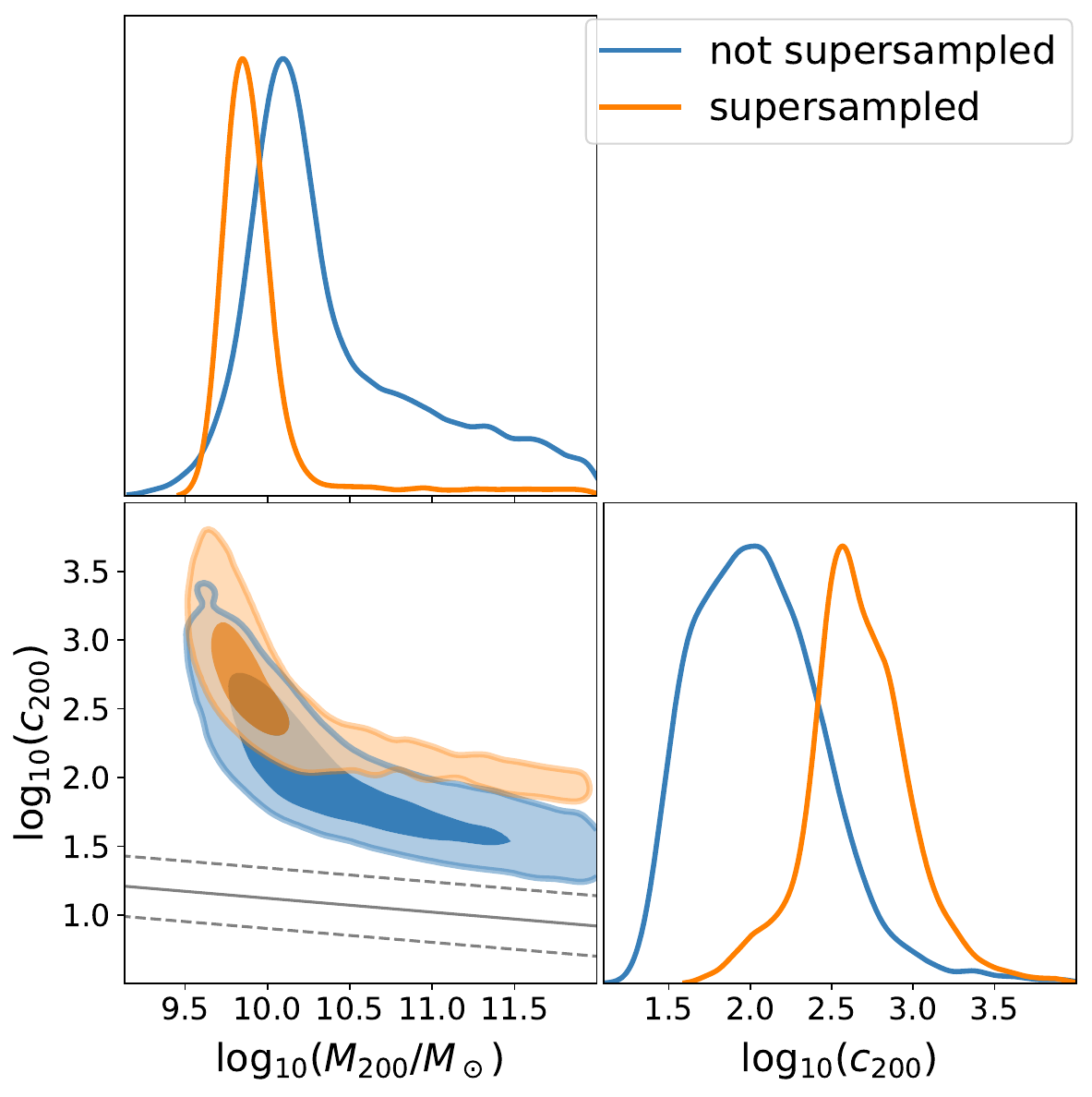}
		\label{fig:m200_vs_logc_mult}
	}
	\caption{Joint posteriors in $\log_{10}(m_{200})$ and $\log_{10}(c_{200})$, where $m_{200}$ and $c_{200}$ are defined as the mass and concentration the subhalo would have if it were a dark matter halo in the field, for models (a) without and (b) with multipoles included in the host galaxy model. Blue curves correspond to the models that do not use supersampling, while orange curves are the supersampled models; the contours enclose the 68\% and 95\% probability regions. Grey line shows the CDM concentration-mass relation for field halos from Dutton et al. (2014), with dashed lines showing $\pm 2\sigma$ scatter. Note that supersampling narrows the allowed parameter space and disfavors a subhalo with $c \lnsim 100$, even in the model with multipoles included.}
\label{fig:m200_vs_logc}
\end{figure*}

\begin{figure*}
	\centering
 	\subfigure[without multipoles]
	{
		\includegraphics[height=0.48\hsize,width=0.48\hsize]{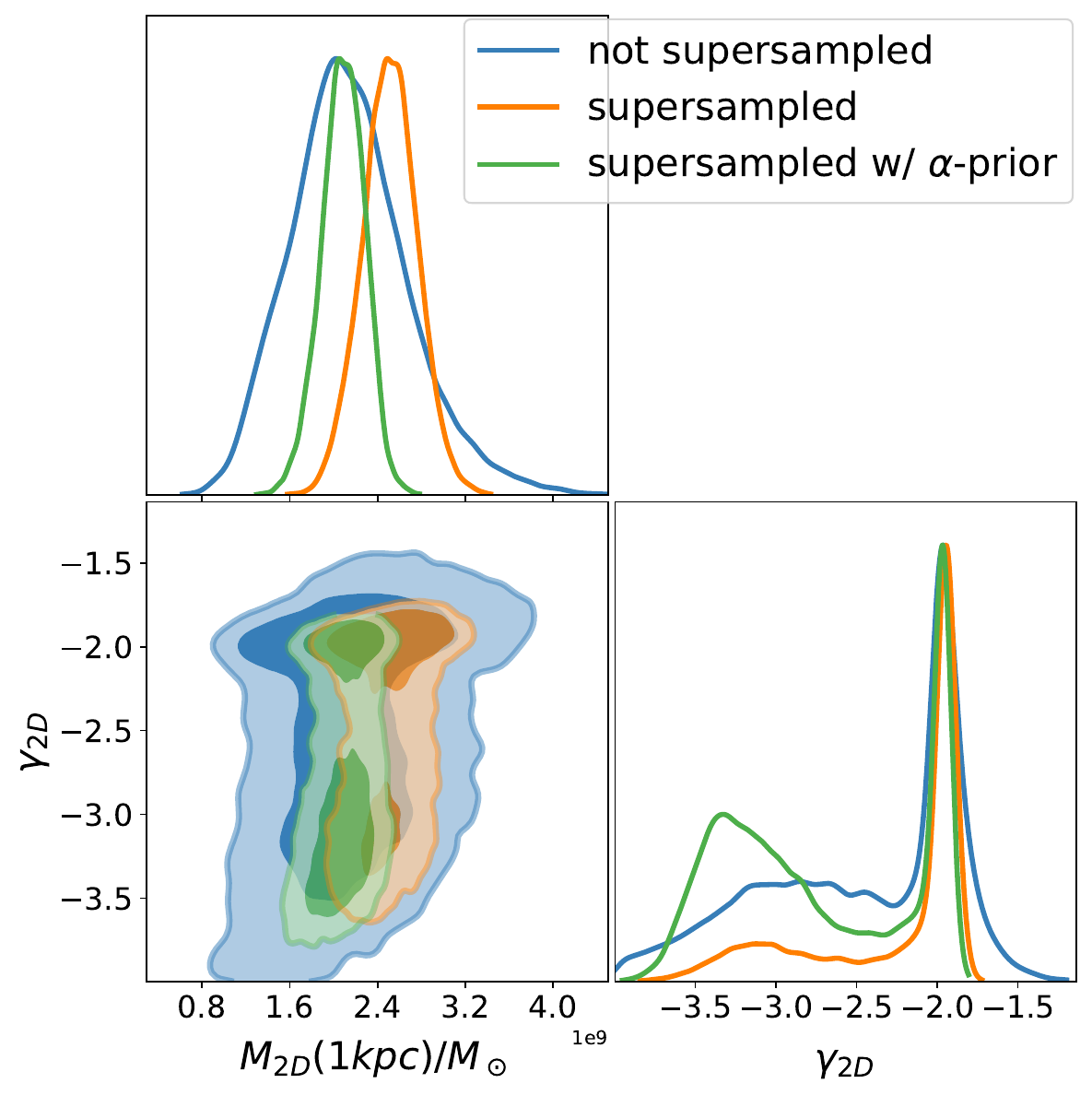}
		\label{fig:m1kpc_vs_logslope_nomult}
	}
	\subfigure[with multipoles]
	{
		\includegraphics[height=0.48\hsize,width=0.48\hsize]{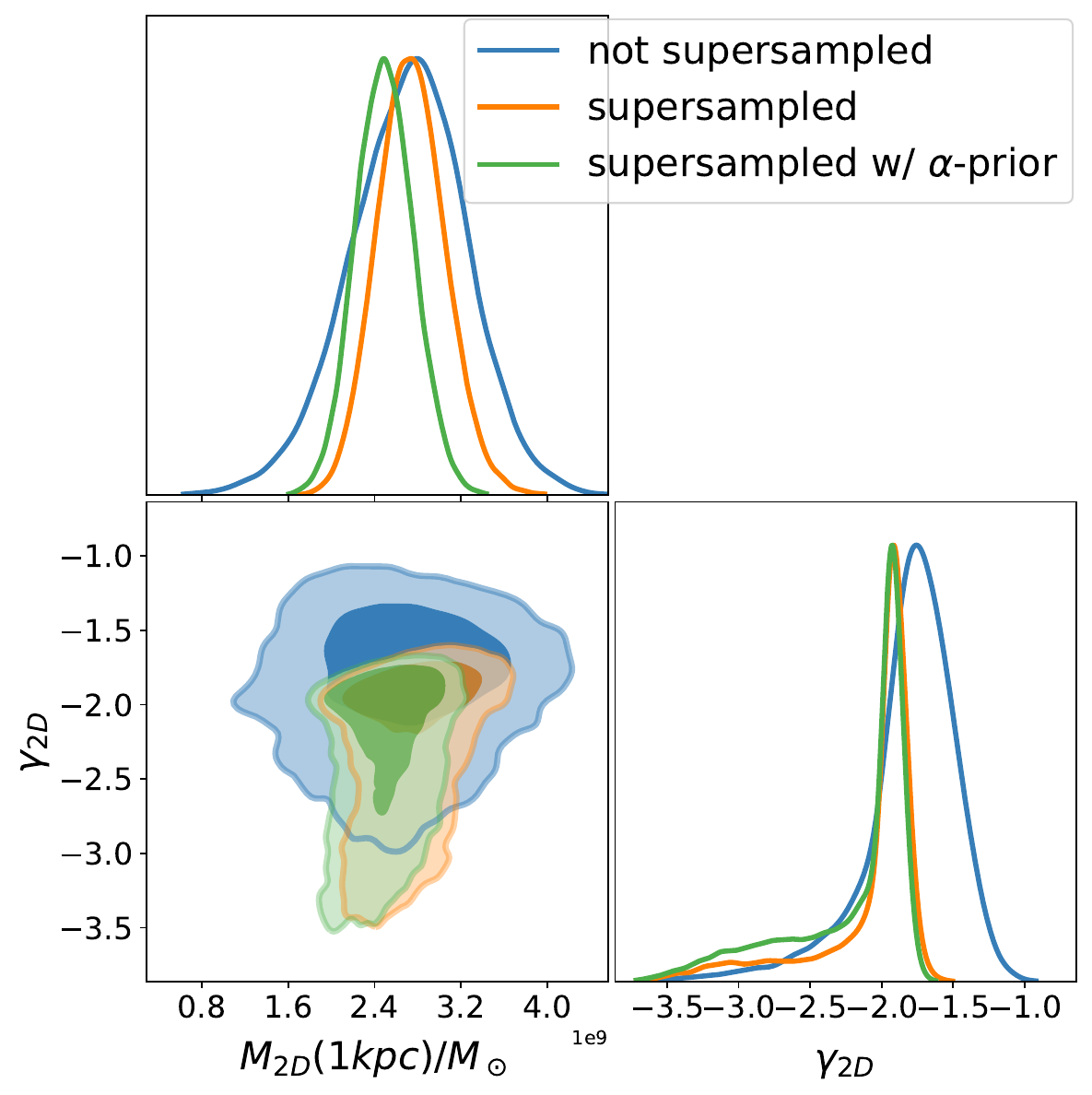}
		\label{fig:m1kpc_vs_logslope_mult}
	}
	\caption{Joint posteriors in the subhalo's projected mass within 1 kpc ($M_{2D}$(1kpc)) and the log-slope $\gamma_{2D}$ of the subhalo's projected density profile, for models (a) without and (b) with multipoles included in the host galaxy model. Blue curves correspond to models without supersampling, orange curves are the supersampled models, and green curves are the supersampled models that incorporate a prior on the host galaxy slope based on kinematics from Turner et al. 2024.}
\label{fig:m1kpc_vs_logslope}
\end{figure*}

\subsection{Constraints on the subhalo properties and comparison to CDM}\label{sec:subhalo_constraints}

To investigate how the subhalo constraints are affected by supersampling, we first plot the posterior in the inferred $m_{200}$ and concentration $c_{200}$ of the subhalo. (Recall that $m_{200}$ is the mass the subhalo \emph{would} have if it were a field halo without tidal stripping.) In Figure \ref{fig:m200_vs_logc_nomult} we plot the posteriors resulting from the models that do not include additional angular structure in the form of multipoles; the orange curves give the 95\% and 68\% contours for the supersampled model, whereas the blue curves give the corresponding contours without supersampling. We first note that the version without supersampling is quite similar to the corresponding posterior in B24, although our posterior includes a somewhat broader spread in concentrations, particularly going toward high concentrations; the reason for this is most likely due to improved constraints due to their inclusion of U-band data, which we do not include in this paper. However, when the data is modeled with supersampling, the posterior narrows considerably although it follows roughly the same degeneracy curve. In the posteriors for the models that include multipoles (Figure \ref{fig:m200_vs_logc_mult}), the difference is more dramatic: without supersampling, the posterior allows for subhalos that are significantly less concentrated ($c < 100$) as $m_{200}$ is increased beyond $10^{10}M_\odot$, in agreement with B24; however when supersampling is used, this parameter space is strongly disfavored, and instead the posterior requires $c \gnsim 100$ even for high $m_{200}$ values.

For a rough comparison to CDM expectations, we also show in Figure \ref{fig:m200_vs_logc} the median concentration-mass relation for field halos from \cite{dutton2014} (solid gray line) along with $\pm 2\sigma$ scatter (dashed lines; the scatter is $\sigma \approx 0.1$ dex). For our supersampled model, our median inferred $\log_{10}(c_{200})=2.8$; for points in the chain with this approximate value we find $\log_{10}(m_{200}/M_\odot)\approx 9.8$. From \cite{dutton2014}, the median expected log-concentration for field halos of this mass is $\bar c_{200}\approx$1.1. However, dark matter subhalos can be expected to be more concentrated than field halos of similar mass by as much as $2\sigma$ above the median (with a small fraction reaching as high as $3\sigma$; \citealt{minor2021b,moline2017}, hence $\log_{10}(c_{200}) \approx 1.3$ is a better point of comparison. Our median inferred log-concentration of the subhalo exceeds this value by $\approx 4\sigma_c$, where $\sigma_c$ is the posterior width of $\log_{10}(c_{200})$ to the left of the median. This is only a rough comparison, as we have not yet accounted for the impact of baryonic physics nor the expected stellar luminosity of the subhalos in question.

Before we delve deeper into the subhalo constraints and a more rigorous comparison with CDM simulations, it is useful to reframe the discussion in terms of quantities that are more directly constrained by the lensing data. In \cite{minor2017} it was shown that for perturbing subhalos, if the position of a subhalo is well-constrained, one can robustly infer the subhalo's projected mass enclosed within its perturbation radius $r_{\delta c}$, defined as the distance from the subhalo center to the point on the critical curve that is being perturbed the most, regardless of the density profile of the subhalo.  In the case of our models, this perturbation radius lies approximately in the range 1.25-1.3 kpc (the reason for the variation is due to the fact that the inferred position of the subhalo varies slightly, depending on the slope of the primary galaxy and whether multipoles are included in the model). For simplicity, in \cite{minor2021} the projected mass within 1 kpc was considered as the approximate robust mass, and we adopt the same quantity for consistency here. In addition, following that work, we consider the approximate log-slope of the subhalo's projected density profile $\gamma_{2D}$ near 1 kpc, defined as the average log-slope within the range (0.75 kpc,1.25 kpc). Thus for each point in our chains, we calculate $M_{2D}$(1kpc) and $\gamma_{2D}$ as derived parameters.

The resulting posteriors in $M_{2D}$(1kpc) and $\gamma_{2D}$ are plotted in Figure \ref{fig:m1kpc_vs_logslope}. Note that in contrast to $m_{200}$/$c_{200}$, there is no degeneracy between the two parameters, as expected since $M_{2D}$(1kpc) is expected to be well-constrained regardless of the subhalo's density profile. Note that in the models without multipoles (Figure \ref{fig:m1kpc_vs_logslope_nomult}), the allowed parameter space is reduced considerably when supersampling is used: in particular, the lower bound on $M_{2D}$(1kpc) is increased significantly, from $1.2\times10^9M_\odot$ to $2\times10^9M_\odot$ at the 95\% confidence level. In the models that include multipoles \ref{fig:m1kpc_vs_logslope_mult}, which are our highest evidence models, the difference is more dramatic: without supersampling, the subhalo's log-slope $\gamma_{2D}$ can be as shallow as $-1.3$ at the 95\% confidence level, whereas after supersampling the constraint on the slope is more stringent, with  $\gamma_{2D}$(1kpc)$ < -1.75$. It should also be noted that the inclusion of both lensed sources $s1$ and $s2$ plays a role here, since even in supersampled models, higher $M_{2D}$(1kpc) and shallower $\gamma_{2D}$ are allowed when only $s1$ is modeled (see Appendix \ref{sec:1src_comp} for a detailed comparison or single-source versus double-source models).

\begin{figure*}
	\centering
	\subfigure[without multipoles]
	{
		\includegraphics[height=0.48\hsize,width=0.48\hsize]{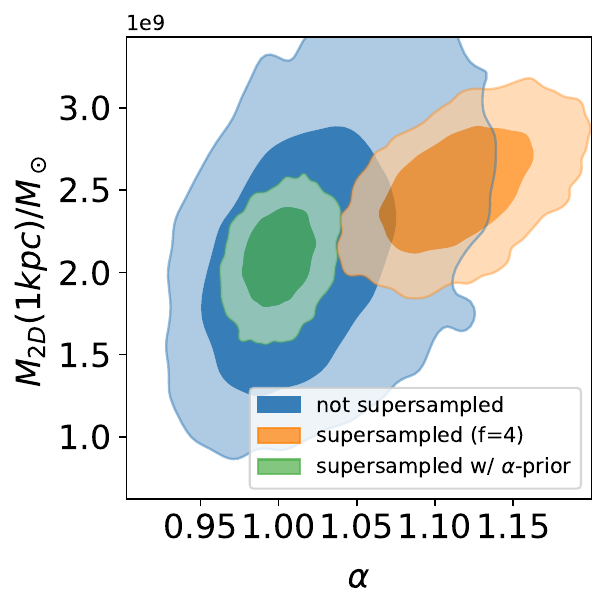}
		\label{fig:galaxy_slope_vs_m1kpc_nomult}
	}
	\subfigure[with multipoles]
	{
		\includegraphics[height=0.48\hsize,width=0.48\hsize]{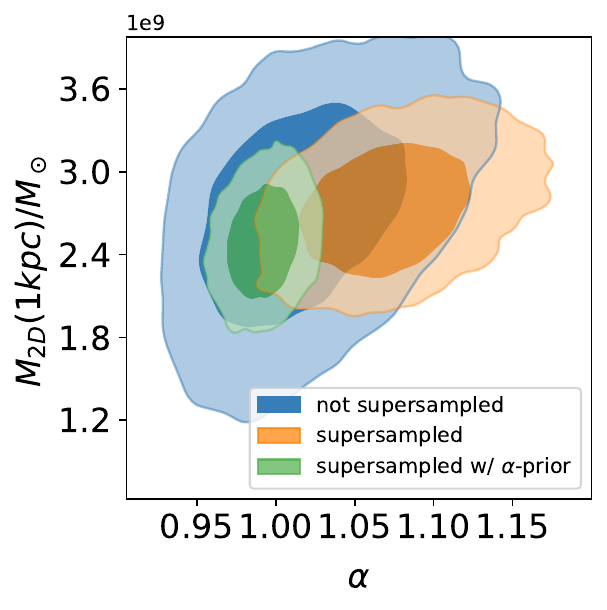}
		\label{fig:galaxy_slope_vs_m1kpc_mult}
	}
	\caption{Joint posteriors in the (negative) log-slope $\alpha$ of the host galaxy versus $M_{2D}$(1kpc), the subhalo's projected mass within 1 kpc, for models (a) without and (b) with multipoles included in the host galaxy model. Contours are defined the same as in Figure \ref{fig:m1kpc_vs_logslope}. Note that the smaller the subhalo mass $M_{2D}$(1kpc), the shallower the slope is required to be; the fit that included an $\alpha$-prior from kinematics (green contours) shows that for an isothermal slope $\alpha=1$, we have $M_{2D}$(1kpc$)\approx2\times10^9M_\odot$. This is roughly consistent with extrapolating from the correlation seen in our fiducial supersampled model (orange contours).}
\label{fig:galaxy_slope_vs_m1kpc}
\end{figure*}

Since we have seen in Section \ref{sec:galaxyslope} that our inferred slope of the primary galaxy's density profile is likely to be slightly biased, we do two additional modeling runs (with and without multipoles) where we implement a Gaussian prior in the galaxy slope $\alpha$, with $\bar\alpha=0.96$ and $\sigma_\alpha=0.02$, based on the constraint from kinematics in \cite{turner2024}. The $\alpha$-prior model without multipoles is plotted as the green contours in Figure \ref{fig:m1kpc_vs_logslope}, and parameter inferences are given in Table \ref{tab:posterior_inferences2}. Note from the model with multipoles (Figure \ref{fig:m1kpc_vs_logslope_mult}), the constraints on have been tightened significantly to $M_{2D}$(1kpc) $= (2.5\pm0.5)\times10^9M_\odot$, with the lower bound now $M_{2D}$(1kpc) $>2.0\times10^9M_\odot$. In the model without multipoles (Figure \ref{fig:m1kpc_vs_logslope_nomult}), the change is more pronounced: we infer $M_{2D}$(1kpc) $= (2.1\pm0.2)\times10^9M_\odot$, with the lower bound now $M_{2D}$(1kpc) $>1.7\times10^9M_\odot$. By comparison, the subhalo's density slope $\gamma_{2D}$ does not change up to the two significant figures we have quoted here.

To better understand the tightened constraints under this prior, in Figure \ref{fig:galaxy_slope_vs_m1kpc_nomult} we plot the joint posterior in $M_{2D}$(1kpc) versus the galaxy log-slope $\alpha$. Note that in the supersampled model that doesn't use the $\alpha$-prior (orange curves), there is a clear correlation between the galaxy log-slope versus $M_{2D}$(1kpc). The reason for this is twofold: first, the robust quantity identified in \cite{minor2021} is actually the mass within $r_{\delta c}$ divided by the host galaxy log-slope. Thus, if the subhalo and host galaxy positions were fixed, we would expect the correlation in Figure \ref{fig:galaxy_slope_vs_m1kpc_nomult} to follow the line of constant $M_{2D}$(1kpc)$/\alpha$. In fact the correlation is a bit steeper than this, because the subhalo position, host galaxy position and Einstein radius all change slightly as $\alpha$ changes. Extrapolating down to the isothermal case $\alpha=1$, we can expect a median inferred $M_{2D}$(1kpc)~$ \approx 2.5\times10^9M_\odot$ and $\approx 2.0\times10^9M_\odot$ with and without multipoles respectively, and this is approximately what we find when the $\alpha$-prior is implemented. 

\begin{figure*}
	\centering
	\subfigure[$1.0\times10^9M_\odot$ residuals, s.s. on]
	{
		\includegraphics[height=0.35\hsize,width=0.32\hsize]{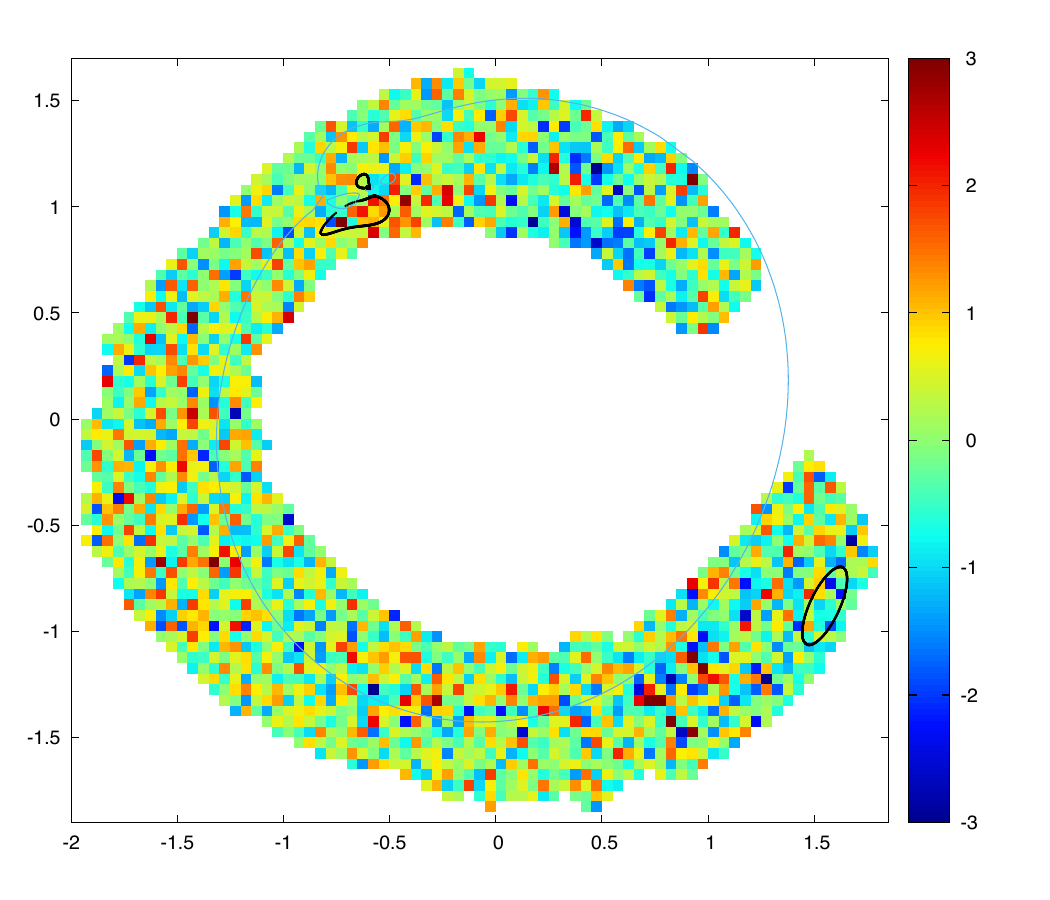}
		\label{fig:res_m1e9}
	}
         \subfigure[$1.0\times10^9M_\odot$, unconvolved image]
	{
		\includegraphics[height=0.35\hsize,width=0.32\hsize]{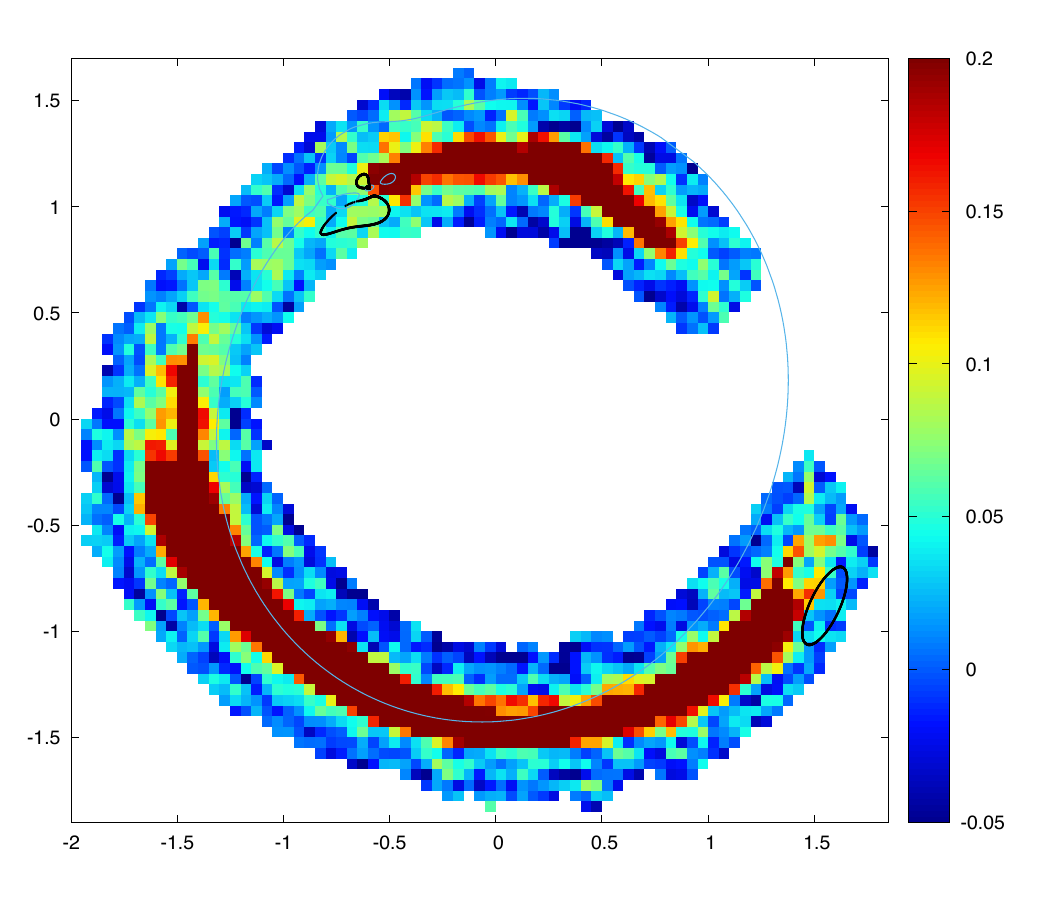}
		\label{fig:img_nopsf_m1e9}
	}
	\subfigure[$1.0\times10^9M_\odot$, reconstructed source]
	{
		\includegraphics[height=0.35\hsize,width=0.32\hsize]{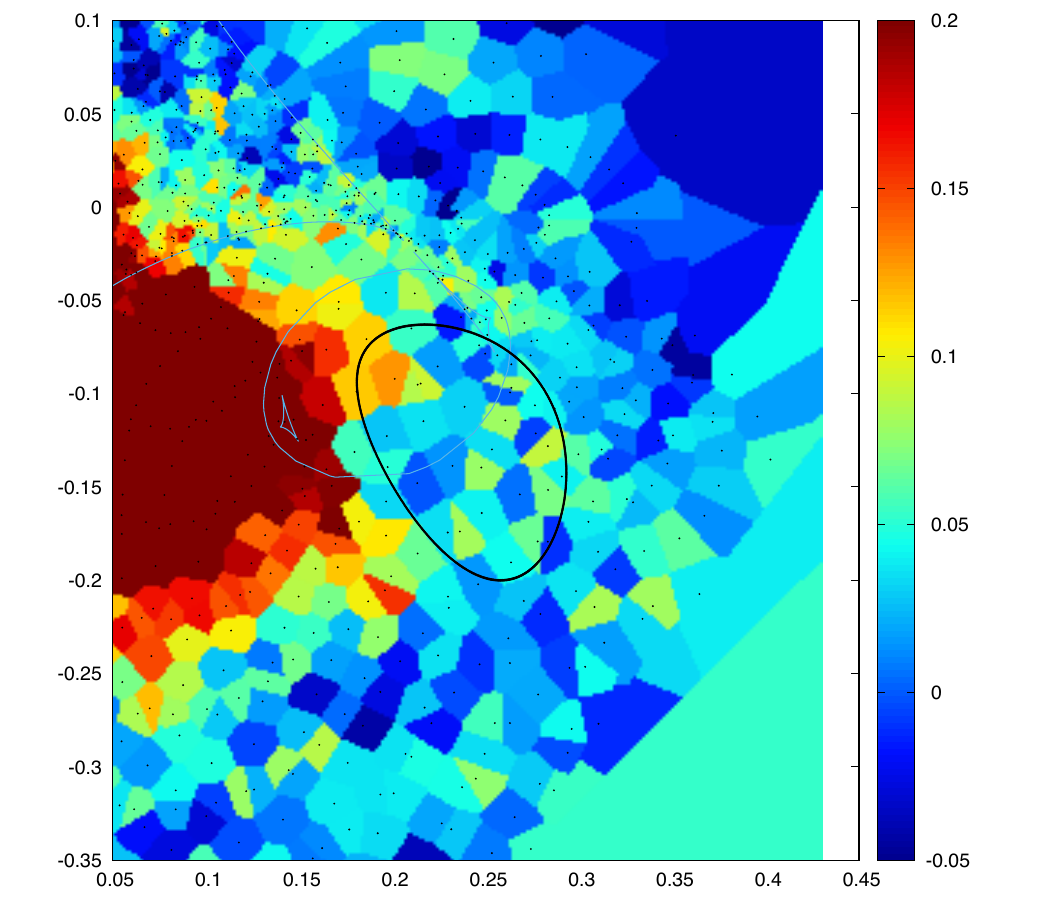}
		\label{fig:src_m1e9}
	}
	\subfigure[$2.8\times10^9M_\odot$ residuals, s.s. on]
	{
		\includegraphics[height=0.35\hsize,width=0.32\hsize]{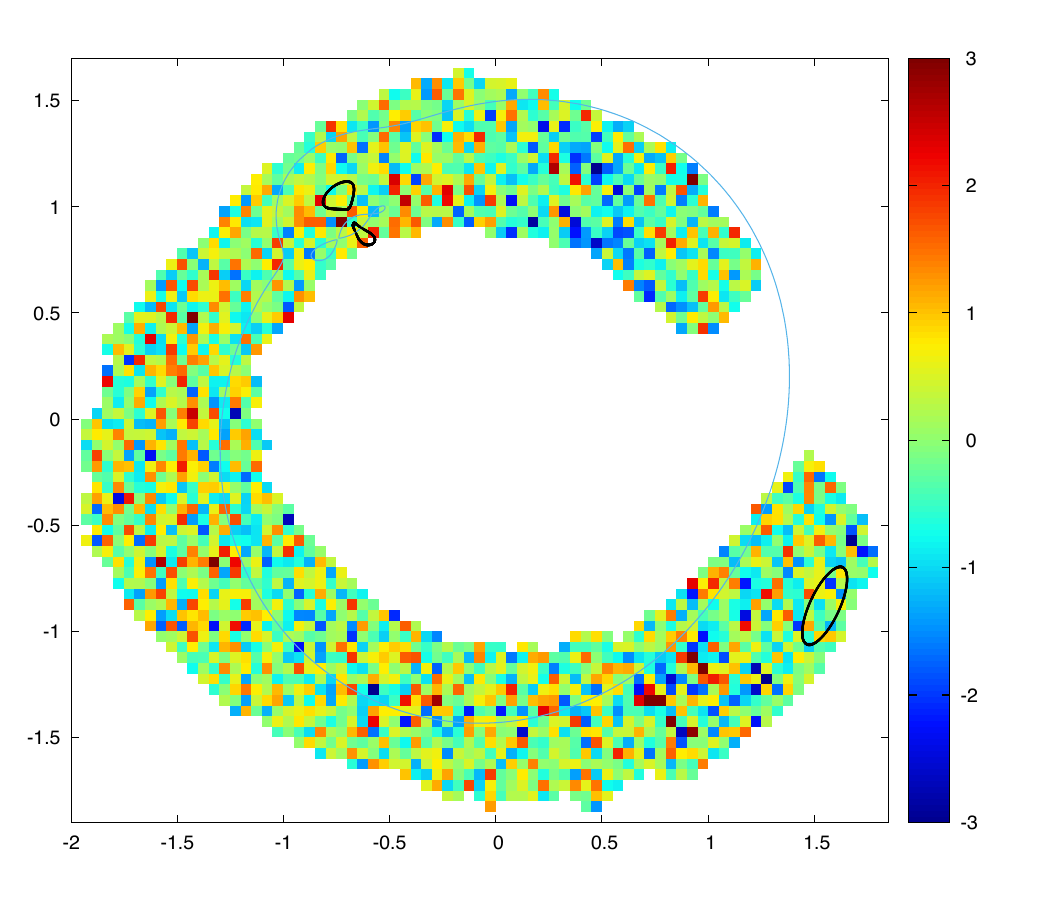}
		\label{fig:res_m2.8e9}
	}
 	\subfigure[$2.8\times10^9M_\odot$, unconvolved image]
	{
		\includegraphics[height=0.35\hsize,width=0.32\hsize]{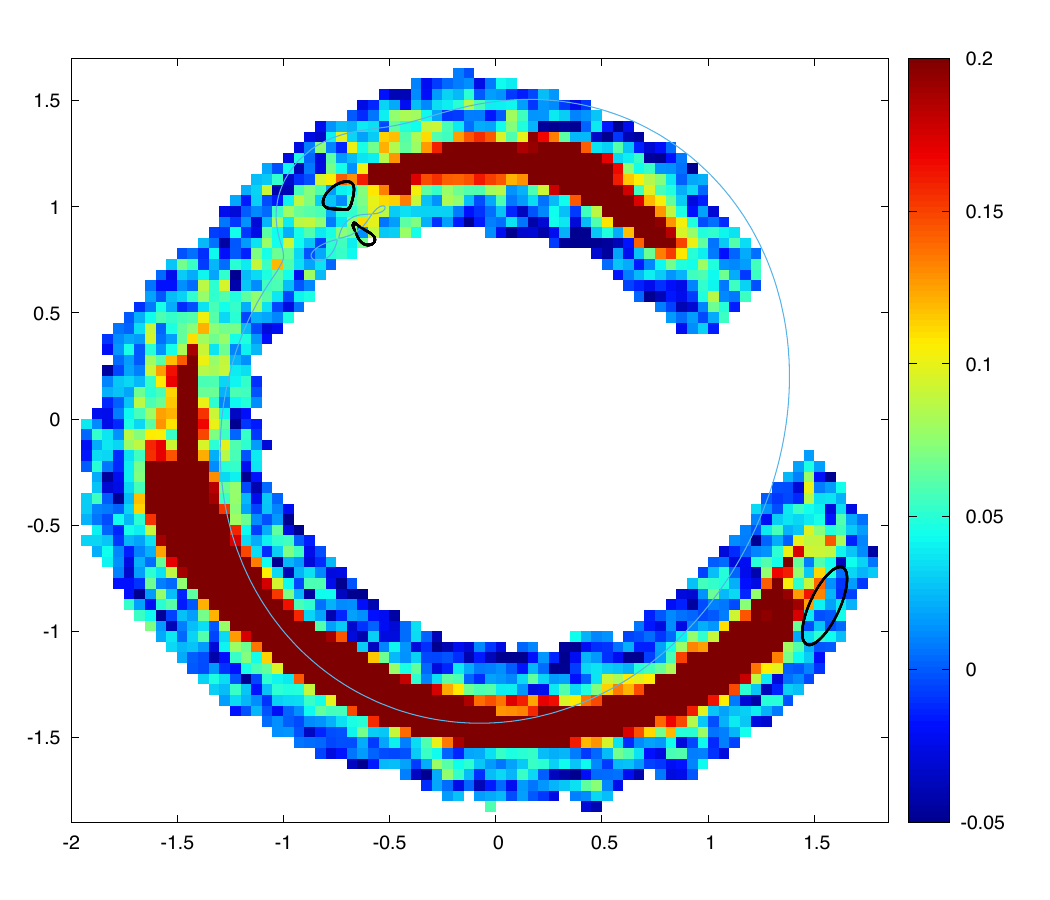}
		\label{fig:img_nopsf_m2.8e9}
	}
 	\subfigure[$2.8\times10^9M_\odot$, reconstructed source]
	{
		\includegraphics[height=0.35\hsize,width=0.32\hsize]{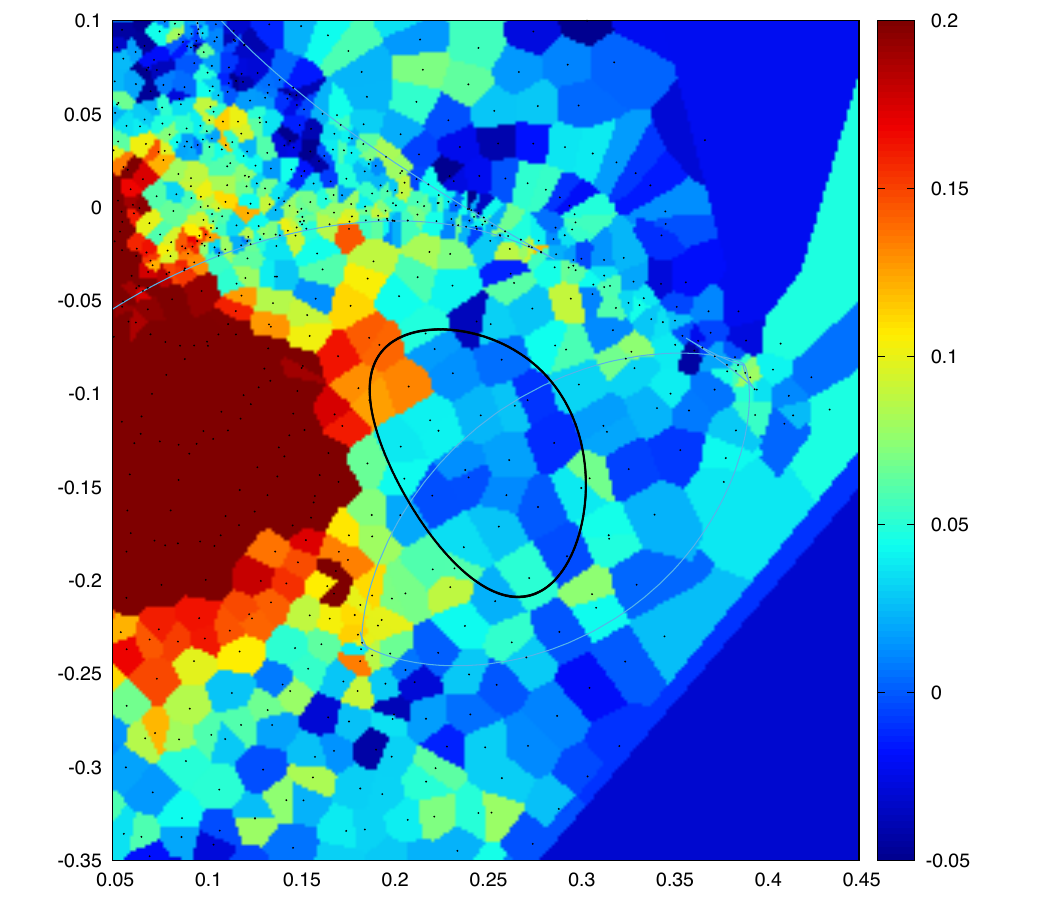}
		\label{fig:src_m2.8e9}
	}
	\caption{Solutions using two different points from the chain generated by fitting to the data without supersampling: one with subhalo mass $M_{2D}$(1kpc$)=1.0\times10^9M_\odot$ (top row) and the other with $M_{2D}$(1kpc$)=2.8\times10^9M_\odot$ (bottom row). Left panels show the normalized residuals that occur when supersampling is turned on and the regularization parameter is reoptimized; middle panels show the lensed images produced by the model (without supersampling) before it is convolved with the PSF; right panels shows the corresponding reconstructed source pixels, where we have capped the colorbar at 0.2 to better show the contrast in the outer regions. We have circled a region along with its corresponding counterimage, where significant residuals appear for the lower-mass substructure (panel (a)); note the disparity in surface brightness between these two regions required to fit the data (panel (b)). By contrast, for the more massive subhalo the surface brightness of the two regions before convolving with the PSF is more consistent (panel (e)), allowing for a better fit when supersampling is used (panel (d)).}
\label{fig:m1kpc_comp}
\end{figure*}

\subsubsection{Comparison to subhalo candidates from Illustris TNG100 and TNG50 simulations}

We will explore the reason why supersampling alters the subhalo constraints in the following section; here, we compare our new constraints to expectations from CDM. In \cite{minor2021} the authors identified over 3000 subhalos of galaxies in the Illustris TNG100-1 cosmological simulation, which includes star formation and stellar and AGN feedback physics \citep{nelson2015}; candidate subhalos were chosen whose total mass and stellar mass approximately matched what we observe for the host galaxy in J0946 \citep{auger2010}. For each of these candidate subhalos, they calculated $M_{2D}$(1kpc) and $\gamma_{2D}$ along 1000 different lines of sight and found that for subhalos with $M_{2D}$(1kpc) > $2\times 10^9M_\odot$, only one subhalo had $\gamma_{2D} < -1.5$ for the top 10\% highest density lines of sight. This one subhalo is disfavored, however, because its stellar mass $\sim 2\times10^9 M_\odot$ would almost certainly have a visible stellar light component, given their conservative upper bound on the subhalo's V-band stellar luminosity $L_V \gnsim 10^8L_\odot$. More recently, \cite{despali2024} have repeated the analysis with the Illustris TNG50 simulation \citep{nelson2019}, which has better sub-kpc resolution, and found several candidate subhalos that satisfy $\gamma_{2D} < -1.5$; however all of them have stellar lumonisities $L_V$ well above $10^8L_\odot$ (assuming a Salpeter-like IMF) and therefore would likely be visible. The number of candidate subhalos with $\gamma_{2D} < -1.5$ increases somewhat if we broaden our range of subhalo $M_{2D}$(1kpc) down to $1\times 10^9M_\odot$; in this case they find a few subhalo candidates with $L_V \approx 1\times10^8M_\odot$, which could in principle evade detection if the stellar mass-to-light ratio were high enough. Alternatively, if we consider shallower slopes, e.g. within the range $\gamma_{2D} \in (-1.5,-1.2)$, they find dozens of subhalos in this range, although all of them have $L_V > 10^8L_\odot$.

In view of these findings, there are two ways the subhalo could be made more consistent with CDM: either the subhalo's projected mass $M_{2D}$(1kpc) must be reduced well below $2\times 10^9 M_\odot$, or else the slope $\gamma_{2D}$ must be made shallower than $-1.5$ (most likely a combination of both would be required). In our non-supersampled models, both of these possibilities seem to be realized to some extent: in the model without multipoles (blue curve in Figure \ref{fig:m1kpc_vs_logslope_nomult}), $M_{2D}$(1kpc) is allowed to be as low as roughly $1\times 10^9M_\odot$, while in the model with multipoles (blue curve in Figure \ref{fig:m1kpc_vs_logslope_mult}), the slope is allowed to be as shallow as $-1.28$. However, our hopes are dashed when we rerun the modeling with supersampling, as both solutions are disfavored: we now have $M_{2D}$(1kpc)$ > 2.0\times10^9M_\odot$ at 95\% CL (in the model where the $\alpha$-prior from \cite{turner2024} is used, this is reduced to $M_{2D}$(1kpc$ > 1.7\times10^9M_\odot$), and $\gamma_{2D}$(1kpc)$ < -1.75$ at 95\% CL for all models. Thus, it appears that the perturbing substructure remains a strong outlier in CDM under the assumption of a dark matter subhalo.

Finally, we note that the $\alpha$-prior solution has a pronounced mode with a rather steep slope at 1 kpc, $\gamma_{2D} < -2.5$, which is also present to some degree in the other solutions in Figure \ref{fig:m1kpc_vs_logslope_nomult}. One might naively expect that the ``steep slope'' end of the posterior is dominated by solutions with extremely high concentration $c_{200}$, but this is not the case: a subhalo whose scale radius is well within 1 kpc has a slope  that asymptotes to $\gamma_{2D} = -2$ since this is the outer 2D log-slope of an unstripped NFW halo. Thus, any log-slope steeper than -2 must be generated by tidal stripping. It is important to recognize that the specific peak of the posterior around $\gamma_{2D} = -2$ is partly a consequence of the subhalo model itself, and not because it is a significantly better fit at $\gamma_{2D}=-2$ compared to, e.g. -2.5 or -3. However, a significant number of solutions in the mode with $\gamma_{2D} < -2.5$ include subhalos with $m_{200} \gnsim 10^{11}M_\odot$ that have extreme tidal stripping, with $r_t \lnsim r_{max}$, the radius at maximum circular velocity. It is unclear how viable such solutions could be in CDM, as subhalos that have undergone such severe tidal stripping may be disrupted almost entirely (although a small cusp is likely to remain; cf. \citealt{errani2020}). In \cite{despali2024} many of the subhalo candidates have infall masses greater than $10^{11}M_\odot$, but very few of them have slopes steeper than -2; those who do have a high stellar mass fraction $M_*/M_{DM} \gnsim 0.5$, indicating these steep slopes occur through baryon cooling and adiabatic contraction, rather than primarily through severe tidal stripping. In any case, such candidates are ruled out by having an expected luminosity beyond the observed limit $L_V > 10^8L_\odot$.

\subsubsection{Why are the subhalo constraints altered by supersampling?}\label{sec:subhalo_discussion}

Now it remains to be explained why the subhalo constraints are altered by supersampling. First we consider a solution with subhalo mass $M_{2D}$(1kpc$)\approx 1.0\times10^9M_\odot$, which fell within the 95\% confidence region when supersampling was not used, but was excluded in the fits that used supersampling. To do this, we choose the best-fit point in the chain that did not use supersampling where $M_{2D}$ lies within the range (0.98,1.02)$\times10^9M_\odot$, yielding a solution with $M_{2D}$(1kpc$)=1.0\times 10^9M_\odot$. This solution produces noise-level residuals without supersampling, but when supersampling is turned on, after reoptimizing the regularization parameter we find significant positive residuals in the vicinity in the subhalo (Figure \ref{fig:res_m1e9}), and negative residuals in its counterimage in the lower right. (Note, the residuals in the upper right are due to the systematic identified in Section \ref{sec:galaxyslope} and are present on some level regardless of the subhalo parameters, so we ignore these residuals in the discussion here.) We focus specifically on the region surrounded by the ellipse shown in the lower-right; the counterimage of the ellipse near the subhalo encircles a portion (though by no means all) of the residuals in question. If we plot the images from the model without supersampling and without convolving with the PSF, we obtain the image in Figure \ref{fig:img_nopsf_m1e9}. Note that the surface brightness within the elliptical region is noticeably darker than that of its counterimage in the upper left, as is required by the data. Since these two circled regions are lensed from the same region of the source, a checkerboard pattern of pixel surface brightnesses is required to give them difference surface brightnesses, and indeed this is what we see in the reconstructed source (Figure \ref{fig:src_m1e9}). However, this solution works because we are only ray tracing the pixel centers; when supersampling is used, the subpixels overlap neighboring source pixels, such that the image pixels are now required to have consistent surface brightnesses, worsening the fit. In addition, we find that significant residuals in these regions remain even after reoptimizing the lens parameters.

For comparison, we now choose a point with $M_{2D}$(1kpc$)=2.8\times10^9M_\odot$, which is allowed in both the supersampled and non-supersampled runs. Note that the counterimage of the elliptical region (Figure \ref{fig:img_nopsf_m2.8e9}) now has a very different shape, covering a smaller area, and the surface brightnesses are now more consistent. Although the brightness within the teardrop-shaped region in the upper left appears somewhat lower than in the data  (compare e.g. Figure \ref{fig:data_circ1}), convolving with the PSF allows it to be consistent with the data, since neighboring pixels are significantly brighter. As a result, there is no checkerboard pattern required in the corresponding region of the source (Figure \ref{fig:src_m2.8e9}), and residuals are significantly reduced compared to the $M_{2D}$(1kpc$)=1.0\times10^9M_\odot$ solution when supersampling is turned on (Figure \ref{fig:res_m2.8e9}), despite the fact that we have not reoptimized the lens parameters which would doubtless produce a better fit.

Other regions of subhalo parameter space where the supersampling runs do not overlap with the non-supersampling runs follow in a similar vein, although the precise regions where residuals occur can vary somewhat. The essential point is that when supersampling is used, the subhalo's parameter space is restricted by the requirement that when image pixels overlap when ray-traced to the source plane, they must have consistent surface brightnesses. This consistency is supposed to be encouraged by the source regularization prior when supersampling is not used, but as we have seen, this does not always prevent the ``checkerboard'' effect of source pixels when the suppression of residuals is significant enough. The result is that we obtain reconstructed sources that have very delicately chosen fluctuations to effectively hide the inconsistent brightnesses of the image pixels. Such implausible solutions are eliminated when supersampling of the image pixels performed, narrowing the allowed parameter space.

\subsection{Dependence on the number of pixel splittings}\label{sec:num_splittings}

For our main results, we used a supersampling factor $N_{\rm sp}=4$ (i.e. $4\times 4$ splitting of image pixels), which makes the ray tracing more computationally intensive than without supersampling. We now check whether a smaller number of splittings can achieve the same result by redoing the fit with $N_{\rm sp}=2$. In this case, because the number of subpixels is relatively small, we do not use a clustering algorithm to define the source pixel locations, but rather choose a simpler approach: for each image pixel, we choose a subpixel whose ray-traced source position will define a source pixel. Hence, for that particular subpixel, no interpolation will be needed to determine the surface brightness, whereas the other three ray-traced subpixels will require interpolation with neighboring source pixels. In this way, we save time not only by having fewer ray-traced points and more sparse matrices, but also because the clustering algorithm is not required to define the source pixels. To reduce gridding effects as much as possible, the subpixel that is used to define a source pixel is rotated as we move to neighboring image pixels (the pattern is: upper-left subpixel, followed by lower-right, then upper-right, then lower-left, then repeat). 

The results are quite similar to that of the $N_{\rm sp}=4$ case. The fit without a subhalo is again rather poor, and the change in log-evidence when a subhalo is included in the modeling is $\Delta \ln{\cal E} = 118.5$, equivalent to a $15.2\sigma$ detection. The detection significance is therefore only slightly reduced compared to the $N_{\rm sp}=4$ model, but still very high. Interestingly, the parameters of the main lensing galaxy are slightly better constrained in the $N_{\rm sp}=4$ case, although the subhalo parameter constraints are nearly identical. In view of this, it seems likely that even $N_{\rm sp}=2$ splitting is sufficient for establishing detections in most cases, although the parameter constraints might be slightly degraded compared to a higher supersampling factor $N_{\rm sp}$.

It should be noted that even in the absence of supersampling, the checkerboard effect can be diminished if the number of source pixels is chosen to be smaller than the number of image pixels $N_d$ within the mask. For example, if we choose $N_s = N_d/2$, then the surface brightness values for half of the image pixels are determined by ray tracing to the source plane and interpolating in nearby source pixels, which reduces the checkerboarding effect and produces significant residuals if a subhalo is not included. If we redo the analysis using $N_d/2$ source pixels and no supersampling, we find a subhalo detection significance of $\sim 9\sigma$; this is not nearly as high compared to when supersampling is used, but more significant nonetheless.

\section{Discussion}\label{sec:discussion}

\subsection{Origin of the ``dark'' spot biasing the host galaxy slope in supersampled fits}\label{sec:dark_spot_discussion}

Now that we have identified the unexpectedly low surface brightness around the upper arc as a culprit in biasing the density slope, the question remains as to its origin. One possibility is dust extinction in the region around the upper arc. The presence of a dust lane through the galaxy center is evident in the \textit{HST} U-band (F336W) image \citep{sonnenfeld2012}; however it is not straightforward to establish the presence of dust in the region of concern because the foreground light is much dimmer in the U-band. In the fits of B24 that use curvature regularization (and hence do not exhibit the checkerboard effect), residuals in the region of concern are significantly higher in the I-band compared to the U-band, which would seem to cast doubt on the dust hypothesis. However the source is considerably less extended in the U-band, so it is unclear whether dust extinction would be as noticeable given how dark the source is in this region. \textit{JWST} observations should easily settle the question of whether dust extinction is present near the lensed images, since dust extinction should be significantly diminished in the longer wavelength near-IR bands of the NIRCam instrument; in addition, continuum dust emission may be detectable in the near-infrared with the MIRI instrument, possibly allowing for a useful dust-correction.

Another possibility for the apparent dark spot would be an imperfect subtraction of the foreground galaxy, whose isophotes show a remarkable degree of twist. The fact that the foreground subtraction can make a difference is evidenced in the fits of \cite{vegetti2010} who tried out two different foreground subtractions: one using the original B-spline method of \cite{bolton2008}, and the other using an elliptical S\'ersic profile. They infer a log-slope that is nearly 0.1 higher for the S\'ersic-subtracted case ($\alpha \approx 2.2$ versus 2.3 for B-spline and S\'ersic, respectively). However, the B-spline foreground modeling can capture ellipticity gradients and isophote twist fairly well, so it is unclear whether a more flexible foreground model will improve matters. Perhaps the masking of the lensed arcs during the foreground fitting is partly to blame, as some lensed surface brightness around the masked areas remain during this procedure; if that is the case, simultaneous modeling of the lensed arcs and the foreground galaxy (again, with a flexible model that can capture twist) may overcome this issue, but would be computationally quite expensive. It is even conceivable that dust and foreground subtraction are intertwined here, in that the central dust lane could be biasing the foreground model and thereby compromising the fit in regions further from the center. In the latter case, \textit{JWST} observations would again allow us to circumvent this issue.

A third possibility is that there is additional structure in the lensing potential that is not being captured by our models. In our fit that include multipoles, a lower slope is indeed allowed; however when we compare residuals for points in the chain that have $\alpha\approx 1$, the multipole-perturbed model achieves only marginal improvement in the low-signal region under the upper arc, although there is a greater improvement in the higher-signal region of the upper arc itself. This raises the question whether there might be an additional perturbing substructure near the region in question. However, such a substructure was not obvious in the modeling of \cite{vegetti2010} who employed pixellated corrections to the lensing potential to identify the original substructure in the first place. If an additional perturber is present, it would likely need to have a mass smaller than $\sim$ few$\times10^8M_\odot$ to have not been evident in their analysis \citep{vegetti2009}. It might be a useful exercise to add pixellated corrections to the potential while also including supersampling and both lensed sources (and perhaps multiple bands as well), but this is beyond the scope of this work.

The systematic we have identified here highlights the utility of modeling data in multiple bands, particularly including bands where the foreground surface brightness is relatively small compared to the lensed images. In B24 they include U-band data (more precisely, the \textit{HST} F336W band) where foreground light is negligible in the modeling, and its impact is particularly noticeable in their fits that use curvature regularization: in this case, the inferred slope $\alpha$ is reduced by 0.13 when the U-band data is included in the modeling. The payoff in including multiple bands is not only to tighten constraints, but also to protect against biases of the kind we have identified here. We leave the modeling of both bands in the context of supersampling to future work. In the meantime, we have mitigated the bias by imposing a prior on the density slope from kinematic measurements \citep{turner2024}, the results of which were discussed in Section \ref{sec:subhalo_constraints}.

\subsection{Could the type of source prior affect the subhalo constraints?}
\label{sec:source_prior_discussion}

Although the source prior does not significantly impact the detection significance, it could still in principle have an impact on the inferred subhalo properties. In our analyses that use supersampling, we find that the inferred subhalo properties are quite similar between the two fits that use gradient versus curvature regularization. However, we saw in Section \ref{sec:multipoles} that the model that includes multipoles and a subhalo was preferred only because of the greater smoothness of the source, rather than an overall reduction in residuals (although there was some improvement in the vicinity of the subhalo and its counterimage), leaving us to wonder whether a more sophisticated regularization scheme might affect the subhalo constraints. In either case, there are significant residuals in the brightest region of the lower lensed arc due to the presence of a cusp in the inferred source galaxy discussed in Section \ref{sec:results_no_ss}, raising the possibility that this may lead to bias in the inferred subhalo properties. It is possible that different regularizations, e.g. using a Matern kernel as in the method of \cite{vernardos2022} or using a luminosity-weighted regularization \citep{nightingale2018} that can fit the central cusp better may result in somewhat different inferred subhalo parameters. We leave a wider investigation of how different regularization priors affect the inferred subhalo parameters to future work.

\section{Conclusions}\label{sec:conclusions}

We have modeled the gravitational lens SDSSJ0946+1006, reconstructing both lensed sources that give rise to extended arcs and including a perturbing dark substructure. We find that when supersampling of image pixels is performed, wherein each pixel is split into subpixels which are ray traced to the source plane and averaged, the detection significance of the substructure increases dramatically to $\sim 17\sigma$, far greater compared to when supersampling is not used. This is consistent with the results of \cite{nightingale2024}, who modeled the low-redshift source and also found a high detection significance for the substructure using supersampling. The key distinction is that supersampling enforces the requirement that image pixels that overlap when ray traced to the source plane have consistent surface brightnesses. This consistency is \emph{encouraged} by the prior on the source galaxy smoothness, but is not always realized without supersampling, depending on the type of source regularization used and the resulting change in the quality of fit. In contrast, the high detection significance of the substructure attained with supersampling is not sensitive to the assumed regularization prior on the source galaxies.

In addition to the detection significance, we find that the combination of supersampling and modeling both lensed sources tightens the constraints on the substructure's central mass and density (Figure \ref{fig:m1kpc_vs_logslope}). Without including any prior from kinematic measurements, in our highest evidence model we constrain the subhalo's projected mass within 1kpc, $M_{2D}$(1kpc$)=(2.7\pm0.6)\times10^9M_\odot$ and its projected density log-slope over the interval (0.75kpc,1.25kpc) is constrained to be $\gamma_{2D} = -1.9^{+0.2}_{-1.3}$ at the 95\% confidence level. We also find that the subhalo's effective concentration $c_{200} > 100$ at 95\% confidence. If we impose a prior on the host galaxy's density log-slope from kinematics \citep{turner2024}, the constraints are tightened to $M_{2D}$(1kpc$)=(2.5\pm0.5)\times10^9M_\odot$ and $\gamma_{2D} = -2.0^{+0.2}_{-1.3}$. This subhalo remains an outlier in CDM by comparison to analogue subhalos in the Illustris TNG100 \citep{minor2021} and TNG50 \citep{despali2024} simulations. The constraints may be tightened further by including observations in additional \textit{HST} bands, e.g. U-band (F336W) observations, as well as including the faint third lensed source detected by VLT/MUSE observations \citep{collett2020} as is done in \cite{ballard2024}, which we leave to future work.

Although noise-level residuals can be achieved without a substructure when supersampling is neglected, this is a mirage: it requires the source galaxy to be quite noisy, to allow for the part of the lensed arc near the purported location of the substructure to have a higher surface brightness compared to its counterpart image (Figure \ref{fig:bestfit_sourcegrid}), even though this noisiness is not evident in the individual lensed images themselves. In contrast, when supersampling of the image pixels is performed---even the most modest level of supersampling (using $2\times 2$ pixel splittings, without supersampling the PSF)---this problem is overcome and a good fit cannot be achieved without including a substructure in the model (Figure \ref{fig:bestfit_supersampled}). This highlights the importance of using supersampling when modeling gravitationally lensed arcs, either during or after model fitting, to more robustly detect perturbations to the smooth lens model due to substructure.

Finally, supersampling has allowed us to identify a region of unexpectedly low surface brightness that has biased the the host galaxy's density profile to be steeper than isothermal in prior studies, and is likely biasing our supersampled models to a lesser extent. This bias has been minimized by the inclusion of both source planes corresponding to the two sets of extended lensed arcs in the model, as well as including a prior on the host galaxy slope from kinematics. The bias may be minimized further by including images in multiple bands and modeling the third lensed source at $z_{s3}\approx6$ identified by VLT/MUSE observations (as is done in \citealt{ballard2024} without supersampling). We speculate that this relatively dark spot may be a result of either dust extinction, an imperfect foreground subtraction, or a combination of these two effects. In either case, follow-up observations via the James Webb Space Telescope (\textit{JWST}) may help eliminate this bias by providing relatively dust-free images and higher signal-to-noise data to facilitate a more detailed model for the foreground galaxy light.

The high detection significance inferred here places the existence of the substructure in SDSSJ0946+1006 on firmer ground, and should motivate deeper follow-up observations. In addition to an improved signal-to-noise of the lensed images in bands with negligible dust extinction, imaging with \textit{JWST} would either detect a stellar light signal from the substructure, or else significantly lower the upper bound on its stellar luminosity. In either case, the result would shed light on the apparent tension with $\Lambda$CDM in addition to strengthening constraints on cosmological parameters from this unique system.

\section*{Acknowledgements}

The author would like to thank Manoj Kaplinghat, Daniel Ballard, Wolfgang Enzi and Giulia Despali for insightful discussions, and Daniel Ballard and Tom Collett for sharing their foreground subtraction and PSF fits. He would also like to thanks Giorgos Vernardos, James Chan, and Aymeric Galan and the anonymous referee for giving valuable feedback on the manuscript.
QM was supported by the generosity of Eric and Wendy Schmidt by recommendation of the Schmidt Futures program.
QM gratefully acknowledges a grant of computer time from ACCESS allocation TG-AST130007.

\section*{Data availability}
No new data were generated or analysed in support of this research.

\begin{table*}
\centering
\begin{tabular}{|l|l|l|l|l|l|}
\hline
\textbf{Parameter} & \textbf{Prior} & \multicolumn{4}{c|}{\textbf{Posterior inference (median with 95\% credible interval)}} \\
\hline
& & \multicolumn{2}{c|}{without supersampling} & \multicolumn{2}{c|}{with $N_{\rm sp}=4$ supersampling} \\
\hline

$R_{e}$(arcsec) & $\mathcal{U}(1.1,1.8)$ & $1.400_{-0.002}^{+0.003}$ & $1.393_{-0.009}^{+0.005}$ &  $1.397_{-0.002}^{+0.002}$ & $1.391_{-0.004}^{+0.004}$  \\
$\alpha$ & $\mathcal{U}(0.7,1.6)$ & $0.96_{-0.05}^{+0.08}$ &  $1.02_{-0.07}^{+0.10}$ &  $1.11_{-0.08}^{+0.08}$ &  $1.11_{-0.06}^{+0.07}$ \\
$e_{1} \equiv (1-q)\cos(2\theta)$ & $\mathcal{U}(-0.2,0.2)$ & $-0.0061_{-0.0186}^{+0.0198}$ &  $-0.0034_{-0.0200}^{+0.0198}$ &  $0.017_{-0.016}^{+0.016}$ &  $0.012_{-0.013}^{+0.013}$ \\
$e_{2} \equiv (1-q)\sin(2\theta)$ & $\mathcal{U}(-0.2,0.2)$ & $-0.036_{-0.016}^{+0.018}$ &  $-0.0038_{-0.0231}^{+0.0331}$ &  $-0.064_{-0.018}^{+0.017}$ &  $0.0071_{-0.0196}^{+0.0148}$ \\
$x_{c}$(arcsec) & $\mathcal{U}(-0.1,0.1)$ & $0.028_{-0.008}^{+0.006}$ &  $0.039_{-0.008}^{+0.010}$ &  $0.012_{-0.004}^{+0.004}$ &  $0.039_{-0.005}^{+0.005}$ \\
$y_{c}$(arcsec) & $\mathcal{U}(-0.1,0.1)$ & $0.044_{-0.006}^{+0.007}$ &  $0.041_{-0.009}^{+0.008}$ &  $0.040_{-0.004}^{+0.004}$ &  $0.033_{-0.005}^{+0.005}$ \\
$\Gamma_{1}$ & $\mathcal{U}(-0.15,0.15)$ & $0.061_{-0.008}^{+0.008}$ &  $0.069_{-0.009}^{+0.009}$ &  $0.073_{-0.008}^{+0.007}$ &  $0.079_{-0.006}^{+0.005}$ \\
$\Gamma_{2}$ & $\mathcal{U}(-0.15,0.15)$ & $-0.070_{-0.007}^{+0.006}$ &  $-0.065_{-0.008}^{+0.008}$ &  $-0.086_{-0.008}^{+0.008}$ &  $-0.070_{-0.006}^{+0.005}$ \\
\hline
$\log_{10}(m_{200}/M_\odot)$ & $\mathcal{U}(8,12)$ & ... &  $9.88_{-0.49}^{+1.91}$ &  ... &  $9.80_{-0.20}^{+1.86}$ \\
$\log_{10}(c_{200})$ & $\mathcal{U}(-1,4)$ & ... &  $2.57_{-0.82}^{+1.27}$ &  ... &  $2.81_{-0.79}^{+0.71}$ \\
$\log_{10}(r_{t}/$kpc$)$ & $\mathcal{U}(-2,3)$ & ... &  $0.33_{-1.32}^{+2.52}$ &  ... &  $1.14_{-1.92}^{+1.77}$ \\
$x_{c,{\rm sub}}$(arcsec) & $\mathcal{U}(-1.2,-0.2)$ & ... &  $-0.64_{-0.07}^{+0.08}$ &  ... &  $-0.64_{-0.02}^{+0.04}$ \\
$y_{c,{\rm sub}}$(arcsec) & $\mathcal{U}(0.7,1.3)$ & ... &  $0.98_{-0.10}^{+0.14}$ &  ... &  $1.00_{-0.03}^{+0.05}$ \\
\hline
$R_{e,s1}$(arcsec) & $\mathcal{U}(0.01,0.5)$ & $0.13_{-0.04}^{+0.07}$ &  $0.17_{-0.05}^{+0.08}$ &  $0.24_{-0.06}^{+0.05}$ &  $0.24_{-0.04}^{+0.05}$ \\
$\Delta x_{c,s1}$(arcsec) & $\mathcal{N}(0,0.1)$ & $-0.10_{-0.05}^{+0.08}$ &  $-0.096_{-0.053}^{+0.066}$ &  $0.052_{-0.097}^{+0.096}$ &  $-0.079_{-0.035}^{+0.042}$ \\
$\Delta y_{c,s1}$(arcsec) & $\mathcal{N}(0,0.1)$ & $-0.11_{-0.09}^{+0.11}$ &  $-0.091_{-0.088}^{+0.087}$ &  $0.12_{-0.14}^{+0.17}$ &  $-0.030_{-0.062}^{+0.059}$ \\
\hline
$M_{2D}$(1kpc)$(10^9M_\odot)$ & ... & ... &  $2.11_{-0.90}^{+1.23}$ &  ... &  $2.51_{-0.50}^{+0.50}$ \\
$\gamma_{2D}$(0.75, 1.25 kpc) & ... & ... &  $-2.28_{-1.48}^{+0.62}$ &  ... &  $-1.98_{-1.45}^{+0.12}$ \\
$m_{sub,tot}(10^9M_\odot)$ & ... & ... &  $3.18_{-1.82}^{+8.82}$ &  ... &  $4.64_{-2.31}^{+5.70}$ \\
\hline
$\ln\mathcal{E}$ & ... & $-2349.4$ &  $-2333.1$ &  $-2460.7$ &  $-2318.1$ \\
\hline
\end{tabular}
\caption{Parameter inferences for models with versus without a subhalo, and with versus without supersampling, along with the log-evidence for each model (final row). The median parameter values are quoted along with error bars denoting the 95\% credible interval in that parameter. The parameters with $s1$ in the subscript describe the lensing mass of the source galaxy $s1$, which is modeled with an isothermal sphere; $\Delta x_{c,s1}$ and $\Delta y_{c,s1}$ give the offset in the position of the lensing mass from the centroid of the reconstructed source $s1$, and $R_{e,s1}$ gives its Einstein radius with respect to the higher source redshift $z_{s2}$. The final three parameters are derived parameters related to subhalo properties: $M_{2D}$(1kpc) is the projected subhalo mass within 1 kpc, $\gamma_{2D}$(0.75,1.25kpc) is the average projected density slope within an interval around 1kpc; and $m_{sub,tot}$ is the total subhalo mass.}
\label{tab:posterior_inferences}
\end{table*}

\begin{table*}
\centering
\begin{tabular}{|l|l|l|l|l|l|}
\hline
\textbf{Parameter} & \textbf{Prior} & \multicolumn{4}{c|}{\textbf{Posterior inference (median with 95\% credible interval)}} \\
\hline
& & multipoles, & multipoles & $\alpha$-prior, & $\alpha$-prior, \\
& & no supersampling & $(N_{\rm sp}=4)$ & no multipoles & multipoles \\
\hline
$R_{e}$(arcsec) & (1.1,1.8) & $1.389_{-0.011}^{+0.008}$ &  $1.391_{-0.006}^{+0.005}$ &  $1.396_{-0.004}^{+0.003}$ &  $1.394_{-0.005}^{+0.004}$ \\
$\alpha$ & $(0.7,1.6)^*$ & $1.01_{-0.07}^{+0.09}$ &  $1.07_{-0.06}^{+0.08}$ &  $1.00_{-0.03}^{+0.03}$ &  $0.99_{-0.03}^{+0.03}$ \\
$e_{1} \equiv (1-q)\cos(2\theta)$ & $(-0.2,0.2)$ & $-0.0034_{-0.0208}^{+0.0220}$ &  $0.017_{-0.014}^{+0.014}$ &  $0.010_{-0.013}^{+0.011}$ &  $0.014_{-0.013}^{+0.013}$ \\
$e_{2} \equiv (1-q)\sin(2\theta)$ & $(-0.2,0.2)$ & $0.040_{-0.030}^{+0.029}$ &  $0.021_{-0.020}^{+0.019}$ &  $0.00085_{-0.01726}^{+0.01236}$ &  $0.019_{-0.018}^{+0.016}$ \\
$x_{c}$(arcsec) & $(-0.1,0.1)$ & $0.039_{-0.011}^{+0.011}$ &  $0.041_{-0.006}^{+0.007}$ &  $0.037_{-0.005}^{+0.004}$ &  $0.041_{-0.006}^{+0.006}$ \\
$y_{c}$(arcsec) & $(-0.1,0.1)$ & $0.026_{-0.011}^{+0.011}$ &  $0.029_{-0.006}^{+0.006}$ &  $0.035_{-0.004}^{+0.005}$ &  $0.030_{-0.005}^{+0.005}$ \\
$\Gamma_{1}$ & $(-0.15,0.15)$ & $0.071_{-0.009}^{+0.010}$ &  $0.080_{-0.006}^{+0.007}$ &  $0.072_{-0.005}^{+0.005}$ &  $0.074_{-0.005}^{+0.005}$ \\
$\Gamma_{2}$ & $(-0.15,0.15)$ & $-0.057_{-0.009}^{+0.008}$ &  $-0.063_{-0.008}^{+0.007}$ &  $-0.064_{-0.005}^{+0.004}$ &  $-0.057_{-0.006}^{+0.005}$ \\
\hline
$\log_{10}(m_{200}/M_\odot)$ & $(8,12)$ & $10.22_{-0.56}^{+1.55}$ &  $9.88_{-0.22}^{+1.53}$ &  $9.83_{-0.31}^{+1.97}$ &  $9.84_{-0.22}^{+1.80}$ \\
$\log_{10}(c_{200})$ & $(-1,4)$ & $2.03_{-0.59}^{+0.94}$ &  $2.65_{-0.64}^{+0.69}$ &  $2.70_{-0.70}^{+0.89}$ &  $2.64_{-0.73}^{+0.70}$ \\
$\log_{10}(r_{t}/$kpc$)$ & $(-2,3)$ & $0.68_{-0.98}^{+2.20}$ &  $1.28_{-1.90}^{+1.63}$ &  $0.002_{-0.952}^{+2.784}$ &  $0.88_{-1.57}^{+2.01}$ \\
$x_{c,{\rm sub}}$(arcsec) & $(-1.2,-0.2)$ & $-0.64_{-0.06}^{+0.06}$ &  $-0.64_{-0.02}^{+0.03}$ &  $-0.64_{-0.02}^{+0.04}$ &  $-0.64_{-0.03}^{+0.03}$ \\
$y_{c,{\rm sub}}$(arcsec) & $(0.7,1.3)$ & $1.00_{-0.09}^{+0.14}$ &  $0.98_{-0.04}^{+0.04}$ &  $1.00_{-0.03}^{+0.06}$ &  $0.99_{-0.04}^{+0.04}$ \\
\hline
$R_{e,s1}$(arcsec) & $(0.01,0.5)$ & $0.17_{-0.05}^{+0.07}$ &  $0.21_{-0.05}^{+0.06}$ &  $0.16_{-0.03}^{+0.03}$ &  $0.15_{-0.02}^{+0.02}$ \\
$\Delta x_{c,s1}$(arcsec) & $\mathcal{N}(0,0.1)$ & $-0.11_{-0.05}^{+0.06}$ &  $-0.12_{-0.04}^{+0.05}$ &  $-0.11_{-0.03}^{+0.04}$ &  $-0.14_{-0.03}^{+0.03}$ \\
$\Delta y_{c,s1}$(arcsec) & $\mathcal{N}(0,0.1)$ & $-0.13_{-0.09}^{+0.09}$ &  $-0.097_{-0.072}^{+0.074}$ &  $-0.11_{-0.06}^{+0.06}$ &  $-0.16_{-0.06}^{+0.06}$ \\
\hline
\hline
$A_{3}$ & $(-0.05,0.05)$ & $0.0017_{-0.0083}^{+0.0100}$ &  $-0.0008_{-0.0064}^{+0.0061}$ &  ... &  $-0.0016_{-0.0051}^{+0.0051}$ \\
$B_{3}$ & $(-0.05,0.05)$ & $0.022_{-0.008}^{+0.007}$ &  $0.0025_{-0.0068}^{+0.0057}$ &  ... &  $0.0021_{-0.0053}^{+0.0045}$ \\
$A_{4}$ & $(-0.05,0.05)$ & $0.022_{-0.011}^{+0.012}$ &  $0.016_{-0.007}^{+0.006}$ &  ... &  $0.016_{-0.005}^{+0.005}$ \\
$B_{4}$ & $(-0.05,0.05)$ & $-0.019_{-0.015}^{+0.014}$ &  $-0.016_{-0.008}^{+0.009}$ &  ... &  $-0.016_{-0.007}^{+0.008}$ \\
\hline
$M_{2D}$(1kpc)$(10^9M_\odot)$ & ... & $2.73_{-1.14}^{+1.01}$ &  $2.73_{-0.58}^{+0.64}$ &  $2.09_{-0.42}^{+0.36}$ &  $2.49_{-0.51}^{+0.54}$ \\
$\gamma_{2D}$(0.75, 1.25 kpc) & ... & $-1.76_{-0.98}^{+0.47}$ &  $-1.94_{-1.30}^{+0.19}$ &  $-2.85_{-0.77}^{+0.93}$ &  $-1.98_{-1.32}^{+0.18}$ \\
$m_{sub}(10^9M_\odot)$ & ... & $7.27_{-4.70}^{+18.26}$ &  $6.05_{-3.42}^{+8.19}$ &  $2.56_{-0.75}^{+4.27}$ &  $4.56_{-2.21}^{+7.10}$ \\
\hline
$\ln\mathcal{E}$ & ... & $-2316.3$ & $-2305.9$  & $-2324.7$  & $-2308.6$  \\
\hline
\end{tabular}
\caption{Parameter inferences for models that include multipoles, with and without supersampling (columns 3 and 4), and models that include a prior on the host galaxy 2D log-slope $\alpha$ from kinematics, with and without multipoles (columns 5 and 6). Note that both models that used the $\alpha$-prior also used $N_{\rm sp}=4$ supersampling. Parameter definitions are the same as in Table \ref{tab:posterior_inferences}, except with the additional parameters for the $m=3,4$ multipole amplitudes $A_3$,$B_3$,$A_4$,$B_4$. $^*$The $\alpha$-prior models use a Gaussian prior $\alpha \in \mathcal{N}(0.96,0.02)$ from kinematic modeling \citep{turner2024}.}
\label{tab:posterior_inferences2}
\end{table*}

\bibliography{jackpotss}{}

\begin{thebibliography}{}
\expandafter\ifx\csname natexlab\endcsname\relax\def\natexlab#1{#1}\fi
\providecommand{\url}[1]{\href{#1}{#1}}
\providecommand{\dodoi}[1]{doi:~\href{http://doi.org/#1}{\nolinkurl{#1}}}
\providecommand{\doeprint}[1]{\href{http://ascl.net/#1}{\nolinkurl{http://ascl.net/#1}}}
\providecommand{\doarXiv}[1]{\href{https://arxiv.org/abs/#1}{\nolinkurl{https://arxiv.org/abs/#1}}}

\bibitem[{{Andrade} {et~al.}(2022){Andrade}, {Fuson}, {Gad-Nasr}, {Kong},
  {Minor}, {Roberts}, \& {Kaplinghat}}]{andrade2022}
{Andrade}, K.~E., {Fuson}, J., {Gad-Nasr}, S., {et~al.} 2022, \mnras, 510, 54,
  \dodoi{10.1093/mnras/stab3241}

\bibitem[{{Andrade} {et~al.}(2019){Andrade}, {Minor}, {Nierenberg}, \&
  {Kaplinghat}}]{andrade2019}
{Andrade}, K.~E., {Minor}, Q., {Nierenberg}, A., \& {Kaplinghat}, M. 2019,
  \mnras, 487, 1905, \dodoi{10.1093/mnras/stz1360}

\bibitem[{{Auger} {et~al.}(2010){Auger}, {Treu}, {Bolton}, {Gavazzi},
  {Koopmans}, {Marshall}, {Moustakas}, \& {Burles}}]{auger2010}
{Auger}, M.~W., {Treu}, T., {Bolton}, A.~S., {et~al.} 2010, \apj, 724, 511,
  \dodoi{10.1088/0004-637X/724/1/511}

\bibitem[{{Ballard} {et~al.}(2024){Ballard}, {Enzi}, {Collett}, {Turner}, \&
  {Smith}}]{ballard2024}
{Ballard}, D.~J., {Enzi}, W. J.~R., {Collett}, T.~E., {Turner}, H.~C., \&
  {Smith}, R.~J. 2024, \mnras, 528, 7564, \dodoi{10.1093/mnras/stae514}

\bibitem[{{Baltz} {et~al.}(2009){Baltz}, {Marshall}, \& {Oguri}}]{baltz2009}
{Baltz}, E.~A., {Marshall}, P., \& {Oguri}, M. 2009, \jcap, 1, 015,
  \dodoi{10.1088/1475-7516/2009/01/015}

\bibitem[{{Birrer} {et~al.}(2015){Birrer}, {Amara}, \&
  {Refregier}}]{birrer2015}
{Birrer}, S., {Amara}, A., \& {Refregier}, A. 2015, \apj, 813, 102,
  \dodoi{10.1088/0004-637X/813/2/102}

\bibitem[{{Bolton} {et~al.}(2008){Bolton}, {Burles}, {Koopmans}, {Treu},
  {Gavazzi}, {Moustakas}, {Wayth}, \& {Schlegel}}]{bolton2008}
{Bolton}, A.~S., {Burles}, S., {Koopmans}, L. V.~E., {et~al.} 2008, \apj, 682,
  964, \dodoi{10.1086/589327}

\bibitem[{{Collett} \& {Auger}(2014)}]{collett2014}
{Collett}, T.~E., \& {Auger}, M.~W. 2014, \mnras, 443, 969,
  \dodoi{10.1093/mnras/stu1190}

\bibitem[{{Collett} \& {Smith}(2020)}]{collett2020}
{Collett}, T.~E., \& {Smith}, R.~J. 2020, \mnras, 497, 1654,
  \dodoi{10.1093/mnras/staa1804}

\bibitem[{{Curtin} {et~al.}(2023){Curtin}, {Edel}, {Shrit}, {Agrawal}, {Basak},
  {Balamuta}, {Birmingham}, {Dutt}, {Eddelbuettel}, {Garg}, {Jaiswal},
  {Kaushik}, {Kim}, {Mukherjee}, {Sai}, {Sharma}, {Parihar}, {Swain}, \&
  {Sanderson}}]{curtin2023}
{Curtin}, R., {Edel}, M., {Shrit}, O., {et~al.} 2023, The Journal of Open
  Source Software, 8, 5026, \dodoi{10.21105/joss.05026}

\bibitem[{{Despali} {et~al.}(2024){Despali}, {Heinze}, {Fassnacht}, {Vegetti},
  {Spingola}, \& {Klessen}}]{despali2024}
{Despali}, G., {Heinze}, F.~M., {Fassnacht}, C.~D., {et~al.} 2024, arXiv
  e-prints, arXiv:2407.12910, \dodoi{10.48550/arXiv.2407.12910}

\bibitem[{{Dutton} \& {Macci{\`o}}(2014)}]{dutton2014}
{Dutton}, A.~A., \& {Macci{\`o}}, A.~V. 2014, \mnras, 441, 3359,
  \dodoi{10.1093/mnras/stu742}

\bibitem[{{Errani} \& {Pe{\~n}arrubia}(2020)}]{errani2020}
{Errani}, R., \& {Pe{\~n}arrubia}, J. 2020, \mnras, 491, 4591,
  \dodoi{10.1093/mnras/stz3349}

\bibitem[{{Etherington} {et~al.}(2024){Etherington}, {Nightingale}, {Massey},
  {Tam}, {Cao}, {Niemiec}, {He}, {Robertson}, {Li}, {Amvrosiadis}, {Cole},
  {Diego}, {Frenk}, {Frye}, {Harvey}, {Jauzac}, {Koekemoer}, {Lagattuta},
  {Lange}, {Limousin}, {Mahler}, {Sirks}, \& {Steinhardt}}]{etherington2024}
{Etherington}, A., {Nightingale}, J.~W., {Massey}, R., {et~al.} 2024, \mnras,
  \dodoi{10.1093/mnras/stae1375}

\bibitem[{{Feroz} {et~al.}(2009){Feroz}, {Hobson}, \& {Bridges}}]{feroz2009}
{Feroz}, F., {Hobson}, M.~P., \& {Bridges}, M. 2009, \mnras, 398, 1601,
  \dodoi{10.1111/j.1365-2966.2009.14548.x}

\bibitem[{{Galan} {et~al.}(2024){Galan}, {Vernardos}, {Minor}, {Sluse}, {Van de
  Vyvere}, \& {Gomer}}]{galan2024}
{Galan}, A., {Vernardos}, G., {Minor}, Q., {et~al.} 2024, arXiv e-prints,
  arXiv:2406.08484, \dodoi{10.48550/arXiv.2406.08484}

\bibitem[{{Gavazzi} {et~al.}(2008){Gavazzi}, {Treu}, {Koopmans}, {Bolton},
  {Moustakas}, {Burles}, \& {Marshall}}]{gavazzi2008}
{Gavazzi}, R., {Treu}, T., {Koopmans}, L. V.~E., {et~al.} 2008, \apj, 677,
  1046, \dodoi{10.1086/529541}

\bibitem[{{Goss} \& {Wu}(1999)}]{goss1999}
{Goss}, M., \& {Wu}, K. 1999, Tech.rep., HP Labs, 12

\bibitem[{{Hezaveh} {et~al.}(2016){Hezaveh}, {Dalal}, {Marrone}, {Mao},
  {Morningstar}, {Wen}, {Blandford}, {Carlstrom}, {Fassnacht}, {Holder},
  {Kemball}, {Marshall}, {Murray}, {Perreault Levasseur}, {Vieira}, \&
  {Wechsler}}]{hezaveh2016}
{Hezaveh}, Y.~D., {Dalal}, N., {Marrone}, D.~P., {et~al.} 2016, \apj, 823, 37,
  \dodoi{10.3847/0004-637X/823/1/37}

\bibitem[{{Minor} {et~al.}(2021{\natexlab{a}}){Minor}, {Gad-Nasr},
  {Kaplinghat}, \& {Vegetti}}]{minor2021}
{Minor}, Q., {Gad-Nasr}, S., {Kaplinghat}, M., \& {Vegetti}, S.
  2021{\natexlab{a}}, \mnras, 507, 1662, \dodoi{10.1093/mnras/stab2247}

\bibitem[{{Minor} {et~al.}(2021{\natexlab{b}}){Minor}, {Kaplinghat}, {Chan}, \&
  {Simon}}]{minor2021b}
{Minor}, Q., {Kaplinghat}, M., {Chan}, T.~H., \& {Simon}, E.
  2021{\natexlab{b}}, \mnras, 507, 1202, \dodoi{10.1093/mnras/stab2209}

\bibitem[{{Minor} \& {Kaplinghat}(2008)}]{minor2008}
{Minor}, Q.~E., \& {Kaplinghat}, M. 2008, \mnras, 391, 653,
  \dodoi{10.1111/j.1365-2966.2008.13777.x}

\bibitem[{{Minor} {et~al.}(2017){Minor}, {Kaplinghat}, \& {Li}}]{minor2017}
{Minor}, Q.~E., {Kaplinghat}, M., \& {Li}, N. 2017, \apj, 845, 118,
  \dodoi{10.3847/1538-4357/aa7fee}

\bibitem[{{Molin{\'e}} {et~al.}(2017){Molin{\'e}}, {S{\'a}nchez-Conde},
  {Palomares-Ruiz}, \& {Prada}}]{moline2017}
{Molin{\'e}}, {\'A}., {S{\'a}nchez-Conde}, M.~A., {Palomares-Ruiz}, S., \&
  {Prada}, F. 2017, \mnras, 466, 4974, \dodoi{10.1093/mnras/stx026}

\bibitem[{{Nelson} {et~al.}(2015){Nelson}, {Pillepich}, {Genel},
  {Vogelsberger}, {Springel}, {Torrey}, {Rodriguez-Gomez}, {Sijacki}, {Snyder},
  {Griffen}, {Marinacci}, {Blecha}, {Sales}, {Xu}, \& {Hernquist}}]{nelson2015}
{Nelson}, D., {Pillepich}, A., {Genel}, S., {et~al.} 2015, Astronomy and
  Computing, 13, 12, \dodoi{10.1016/j.ascom.2015.09.003}

\bibitem[{{Nelson} {et~al.}(2019){Nelson}, {Springel}, {Pillepich},
  {Rodriguez-Gomez}, {Torrey}, {Genel}, {Vogelsberger}, {Pakmor}, {Marinacci},
  {Weinberger}, {Kelley}, {Lovell}, {Diemer}, \& {Hernquist}}]{nelson2019}
{Nelson}, D., {Springel}, V., {Pillepich}, A., {et~al.} 2019, Computational
  Astrophysics and Cosmology, 6, 2, \dodoi{10.1186/s40668-019-0028-x}

\bibitem[{{Nierenberg} {et~al.}(2014){Nierenberg}, {Treu}, {Wright},
  {Fassnacht}, \& {Auger}}]{nierenberg2014}
{Nierenberg}, A.~M., {Treu}, T., {Wright}, S.~A., {Fassnacht}, C.~D., \&
  {Auger}, M.~W. 2014, \mnras, 442, 2434, \dodoi{10.1093/mnras/stu862}

\bibitem[{{Nightingale} \& {Dye}(2015)}]{nightingale2015}
{Nightingale}, J.~W., \& {Dye}, S. 2015, \mnras, 452, 2940,
  \dodoi{10.1093/mnras/stv1455}

\bibitem[{{Nightingale} {et~al.}(2018){Nightingale}, {Dye}, \&
  {Massey}}]{nightingale2018}
{Nightingale}, J.~W., {Dye}, S., \& {Massey}, R.~J. 2018, \mnras, 478, 4738,
  \dodoi{10.1093/mnras/sty1264}

\bibitem[{{Nightingale} {et~al.}(2024){Nightingale}, {He}, {Cao},
  {Amvrosiadis}, {Etherington}, {Frenk}, {Hayes}, {Robertson}, {Cole}, {Lange},
  {Li}, \& {Massey}}]{nightingale2024}
{Nightingale}, J.~W., {He}, Q., {Cao}, X., {et~al.} 2024, \mnras, 527, 10480,
  \dodoi{10.1093/mnras/stad3694}

\bibitem[{{Petkova} {et~al.}(2014){Petkova}, {Metcalf}, \&
  {Giocoli}}]{petkova2014}
{Petkova}, M., {Metcalf}, R.~B., \& {Giocoli}, C. 2014, \mnras, 445, 1954,
  \dodoi{10.1093/mnras/stu1860}

\bibitem[{{Schneider} {et~al.}(1992){Schneider}, {Ehlers}, \&
  {Falco}}]{schneider1992}
{Schneider}, P., {Ehlers}, J., \& {Falco}, E.~E. 1992, {Gravitational Lenses},
  \dodoi{10.1007/978-3-662-03758-4}

\bibitem[{{Schneider} \& {Sluse}(2013)}]{schneider2013}
{Schneider}, P., \& {Sluse}, D. 2013, \aap, 559, A37,
  \dodoi{10.1051/0004-6361/201321882}

\bibitem[{Sibson(1981)}]{sibson1981}
Sibson, R. 1981, Interpreting Multivariate Data, ed. V.~Barnett (New York: John
  Wiley and Sons)

\bibitem[{{Smith} \& {Collett}(2021)}]{smith2021}
{Smith}, R.~J., \& {Collett}, T.~E. 2021, \mnras, 505, 2136,
  \dodoi{10.1093/mnras/stab1399}

\bibitem[{{Sonnenfeld} {et~al.}(2012){Sonnenfeld}, {Treu}, {Gavazzi},
  {Marshall}, {Auger}, {Suyu}, {Koopmans}, \& {Bolton}}]{sonnenfeld2012}
{Sonnenfeld}, A., {Treu}, T., {Gavazzi}, R., {et~al.} 2012, \apj, 752, 163,
  \dodoi{10.1088/0004-637X/752/2/163}

\bibitem[{{Suyu} {et~al.}(2006){Suyu}, {Marshall}, {Hobson}, \&
  {Blandford}}]{suyu2006}
{Suyu}, S.~H., {Marshall}, P.~J., {Hobson}, M.~P., \& {Blandford}, R.~D. 2006,
  \mnras, 371, 983, \dodoi{10.1111/j.1365-2966.2006.10733.x}

\bibitem[{{Tessore} \& {Metcalf}(2015)}]{tessore2015}
{Tessore}, N., \& {Metcalf}, R.~B. 2015, \aap, 580, A79,
  \dodoi{10.1051/0004-6361/201526773}

\bibitem[{{Turner} {et~al.}(2024){Turner}, {Smith}, \& {Collett}}]{turner2024}
{Turner}, H.~C., {Smith}, R.~J., \& {Collett}, T.~E. 2024, \mnras, 528, 3559,
  \dodoi{10.1093/mnras/stae263}

\bibitem[{{Vegetti} \& {Koopmans}(2009)}]{vegetti2009}
{Vegetti}, S., \& {Koopmans}, L.~V.~E. 2009, \mnras, 392, 945,
  \dodoi{10.1111/j.1365-2966.2008.14005.x}

\bibitem[{{Vegetti} {et~al.}(2010){Vegetti}, {Koopmans}, {Bolton}, {Treu}, \&
  {Gavazzi}}]{vegetti2010}
{Vegetti}, S., {Koopmans}, L.~V.~E., {Bolton}, A., {Treu}, T., \& {Gavazzi}, R.
  2010, \mnras, 408, 1969, \dodoi{10.1111/j.1365-2966.2010.16865.x}

\bibitem[{{Vegetti} {et~al.}(2012){Vegetti}, {Lagattuta}, {McKean}, {Auger},
  {Fassnacht}, \& {Koopmans}}]{vegetti2012}
{Vegetti}, S., {Lagattuta}, D.~J., {McKean}, J.~P., {et~al.} 2012, \nat, 481,
  341, \dodoi{10.1038/nature10669}

\bibitem[{{Vernardos} \& {Koopmans}(2022)}]{vernardos2022}
{Vernardos}, G., \& {Koopmans}, L.~V.~E. 2022, \mnras, 516, 1347,
  \dodoi{10.1093/mnras/stac1924}

\end{thebibliography}
\bibliographystyle{aasjournal}

\appendix
\setcounter{section}{0}
\section{Comparison with models that include only the low-redshift source $s1$}\label{sec:1src_comp}
\begin{figure*}
	\centering
 	\subfigure[without multipoles]
	{
		\includegraphics[height=0.48\hsize,width=0.48\hsize]{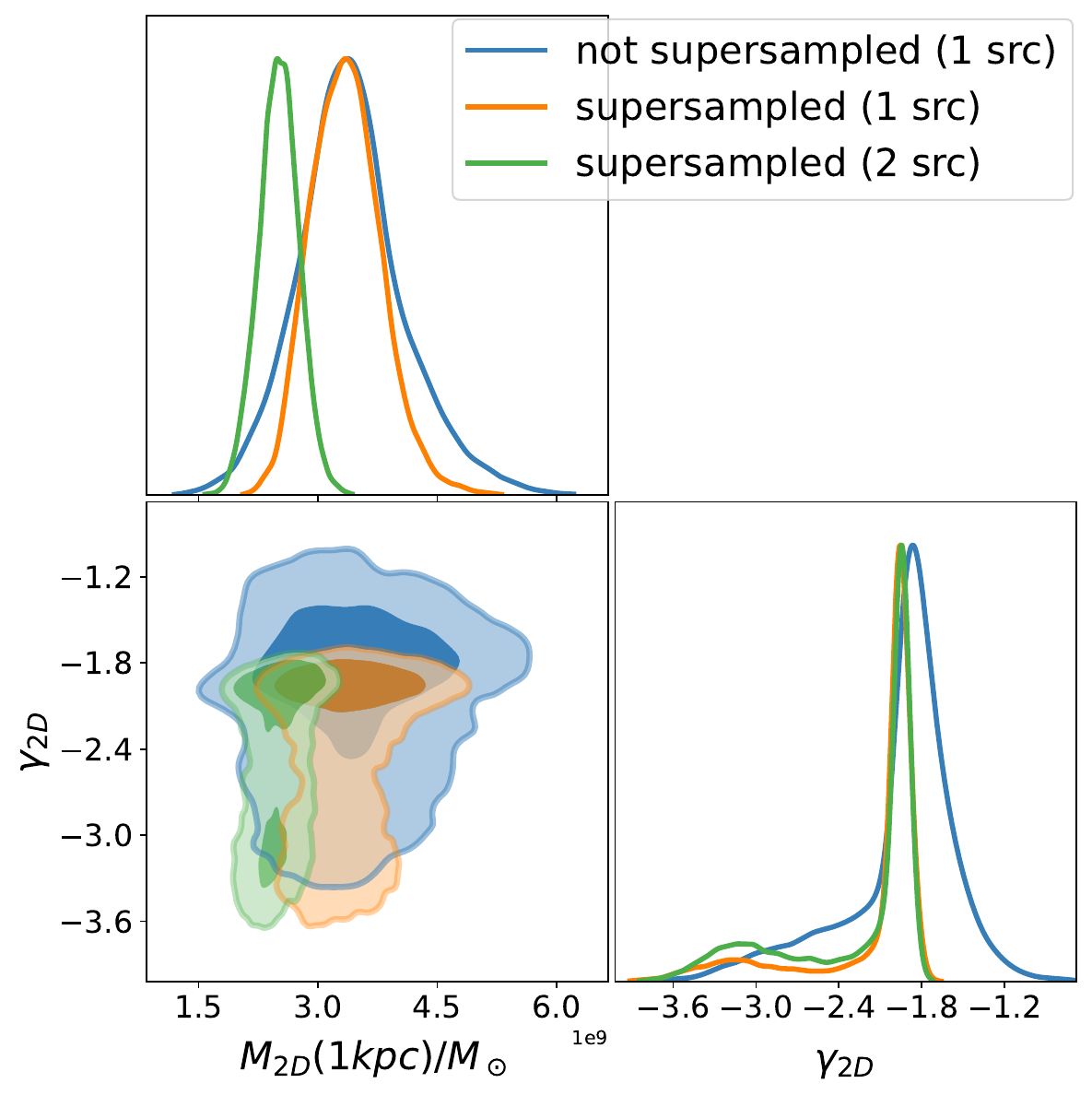}
		\label{fig:m1kpc_vs_logslope_1src_nomult}
	}
	\subfigure[with multipoles]
	{
		\includegraphics[height=0.48\hsize,width=0.48\hsize]{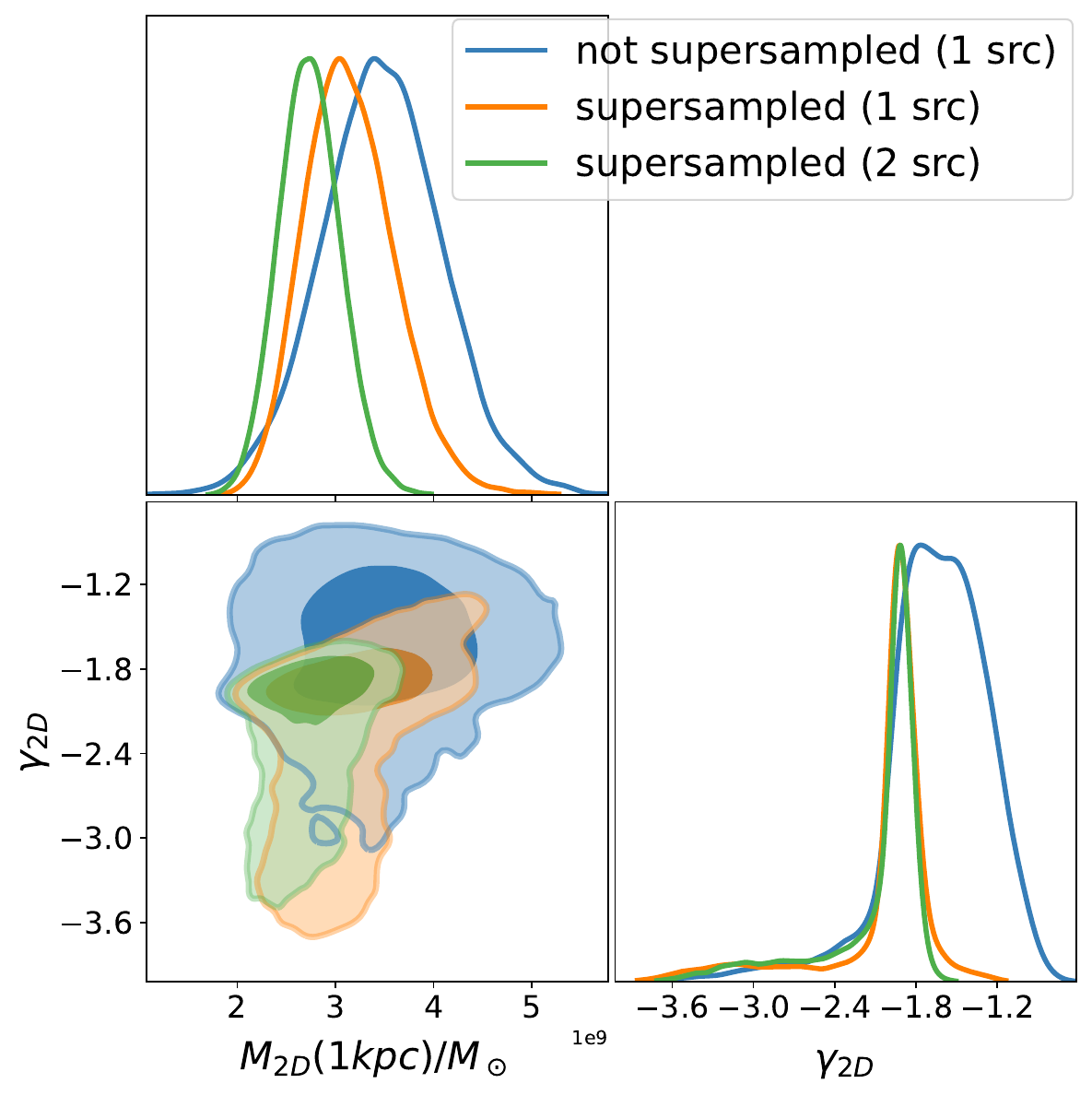}
		\label{fig:m1kpc_vs_logslope_1src_mult}
	}
	\caption{Joint posteriors in the subhalo's projected mass within 1 kpc ($M_{2D}$(1kpc)) and the log-slope $\gamma_{2D}$ of the subhalo's projected density profile, for models (a) without and (b) with multipoles included in the host galaxy model. Blue curves correspond to single-source models without supersampling, orange curves are single-source supersampled models, and green curves are the supersampled models that model both lensed sources (note these are identical to the orange curves in Figure \ref{fig:m1kpc_vs_logslope}).}
\label{fig:m1kpc_vs_logslope_1src}
\end{figure*}

\begin{figure}
	\centering
	\includegraphics[height=0.5\hsize,width=0.50\hsize]{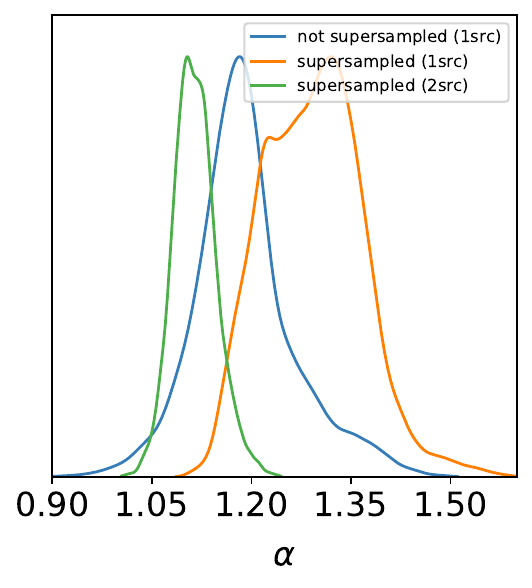}
	\caption{Posterior in the host galaxy (negative) 2D log-slope $\alpha$. Blue and orange curves correspond to single-source models that are not supersampled and supersampled, respectively, while green curve corresponds to the supersampled model that includes both lensed sources.}
\label{fig:alpha_post_1src}
\end{figure}

With the exception of B24, all prior studies that modeled the subhalo in J0946 modeled only the low-redshift source \citep{vegetti2010,minor2021,nightingale2024,despali2024}. Hence to provide a point of comparison to prior work, we repeat our analyses that included a subhalo in this paper, but this time only modeling the low-redshift source. The resulting constraints on the subhalo projected mass $M_{2D}$(1kpc) and density log-sloper $\gamma_{2D}$ are shown in Figure \ref{fig:m1kpc_vs_logslope_1src}. Note that when only the low-redshift source is modeled, higher values of $M_{2D}$(1kpc) are allowed, even when supersampling is performed. When supersampling is not performed, however, $\gamma_{2D}$ is allowed to be as shallow as $-1.2$, or even a bit shallower in the case with multipoles. These solutions are entirely compatible with those of \cite{minor2021} and \cite{despali2024} (hereafter M21 and D24) who both find solutions with $m_{2D}$(1kpc) as high as $4\times10^9M_\odot$ and slopes $\gamma_{2D}$ as shallow as $-1.2$. In Figure \ref{fig:alpha_post_1src} we plot posteriors in the host galaxy log-slope $\alpha$ for single-source versus double source models. Note that the single-source models allow significantly steeper slopes, up to 1.5, or even up to 1.6 in the supersampled case.

The solutions of M21 and D24 are in fact biased toward even higher $\alpha$ values than our posteriors show; for example, in their highest evidence model, M21 infer $\alpha > 1.32$ at the 95\% confidence level, while in D24 $\alpha$ can run as steep as 1.6. The reason for this may be due to the fact that both papers use a larger mask than we have used here, including more of the problematic region that gives rise to a bias in the host galaxy slope $\alpha$ as discussed in Section \ref{sec:galaxyslope}. Furthermore, in our unsupersampled models we see that solutions with $\alpha > 1.3$ require the subhalo log-slope $\gamma_{2D}$ to be shallower than $\approx$ -2, which is consistent with the solutions of D24. Note that M21 did use $N_{\rm sp}=2$ supersampling, but in their fits that used multipoles they inferred a solution with $\gamma_{2D} \approx -1.3$. Indeed, we see in Figure \ref{fig:m1kpc_vs_logslope_1src_mult} that the single-source model with supersampling does allow for a solution with such a shallow slope and relatively high $M_{2D}$(1kpc) $\approx 4\times10^9M_\odot$; this branch of the posterior requires $\alpha \gnsim 1.2$. This mass is a bit higher than in M21, but nevertheless close to  their solution. The reason the solution in M21 with multipoles does not allow for slopes as steep as -2 is likely because they infer $\alpha > 1.32$ at 95\% CL due to the enlarged mask, and perhaps also because of the less flexible source model. In any event, the solutions in M21 and D24 that have $\gamma_{2D} > -1.7$ are no longer viable when both sources are modeled and supersampling is performed.

\section{Effect of the source pixellation in supersampled models}\label{sec:pixellation}
In all the models that use supersampling in this work, we have set the number of source pixels $N_s$ to be equal to half the number of data pixels within the mask, i.e. $N_s = N_d/2$ for both sources $s1$ and $s2$. This is done to lessen the computational cost, but is in contrast with the models without supersampling for which we use $N_s = N_d$. It is worth investigating whether there is any substantial change in our results if we use twice the number of source pixels in our supersampled runs, so that $N_s = N_d$.

\begin{figure*}
	\centering
 	\subfigure[$N_s=N_d/2$]
	{
		\includegraphics[height=0.48\hsize,width=0.48\hsize]{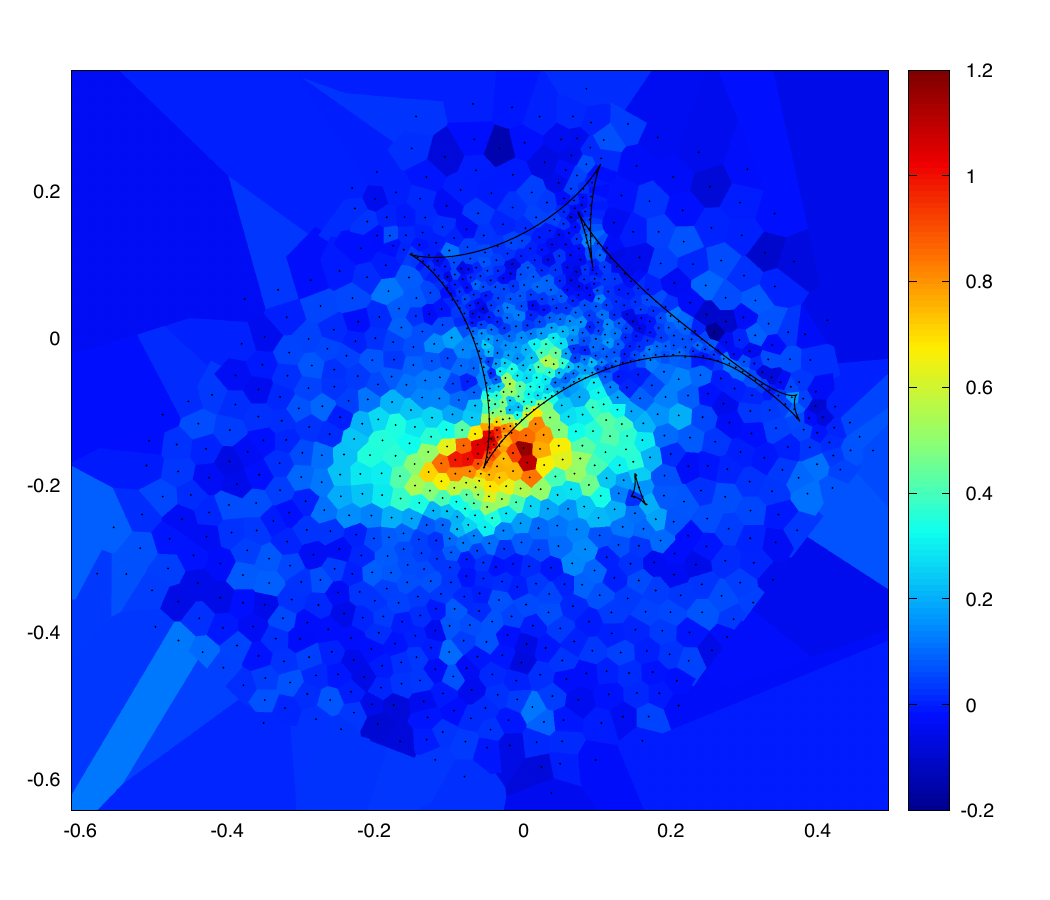}
		\label{fig:src1_pix}
	}
	\subfigure[$N_s=N_d$]
	{
		\includegraphics[height=0.48\hsize,width=0.48\hsize]{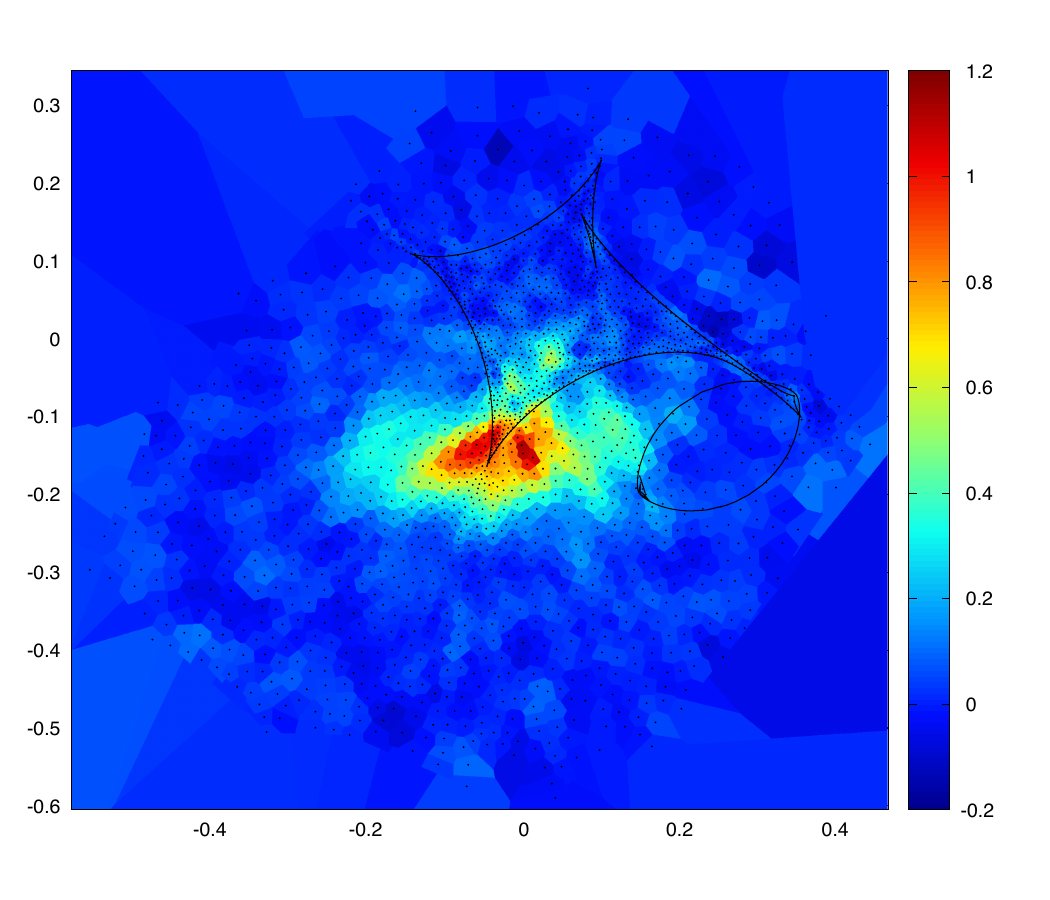}
		\label{fig:src1_morepix}
	}
	\caption{Reconstructed source $s1$ in supersampled models with two different pixellations: (a) the number of source pixels is equal to half the number of data pixels; (b) the number of source pixels is equal to the number of data pixels. Black points show the centroids of the Voronoi pixels, while black curves are the caustics. To make the pixellation evident, here we do not interpolate to find the surface brightness at each point, unlike the reconstructions shown in Figures \ref{fig:bestfit_nosup}-\ref{fig:bestfit_supersampled}. Note that the source pixel centroids do not correspond to any specific ray-traced image pixels, but rather are determined by a clustering algorithm after ray tracing all the subpixels, as discussed in Section \ref{sec:src_reconstruction}.}
\label{fig:src_pixcomp}
\end{figure*}

\begin{figure}
	\centering
	\includegraphics[height=0.5\hsize,width=0.50\hsize]{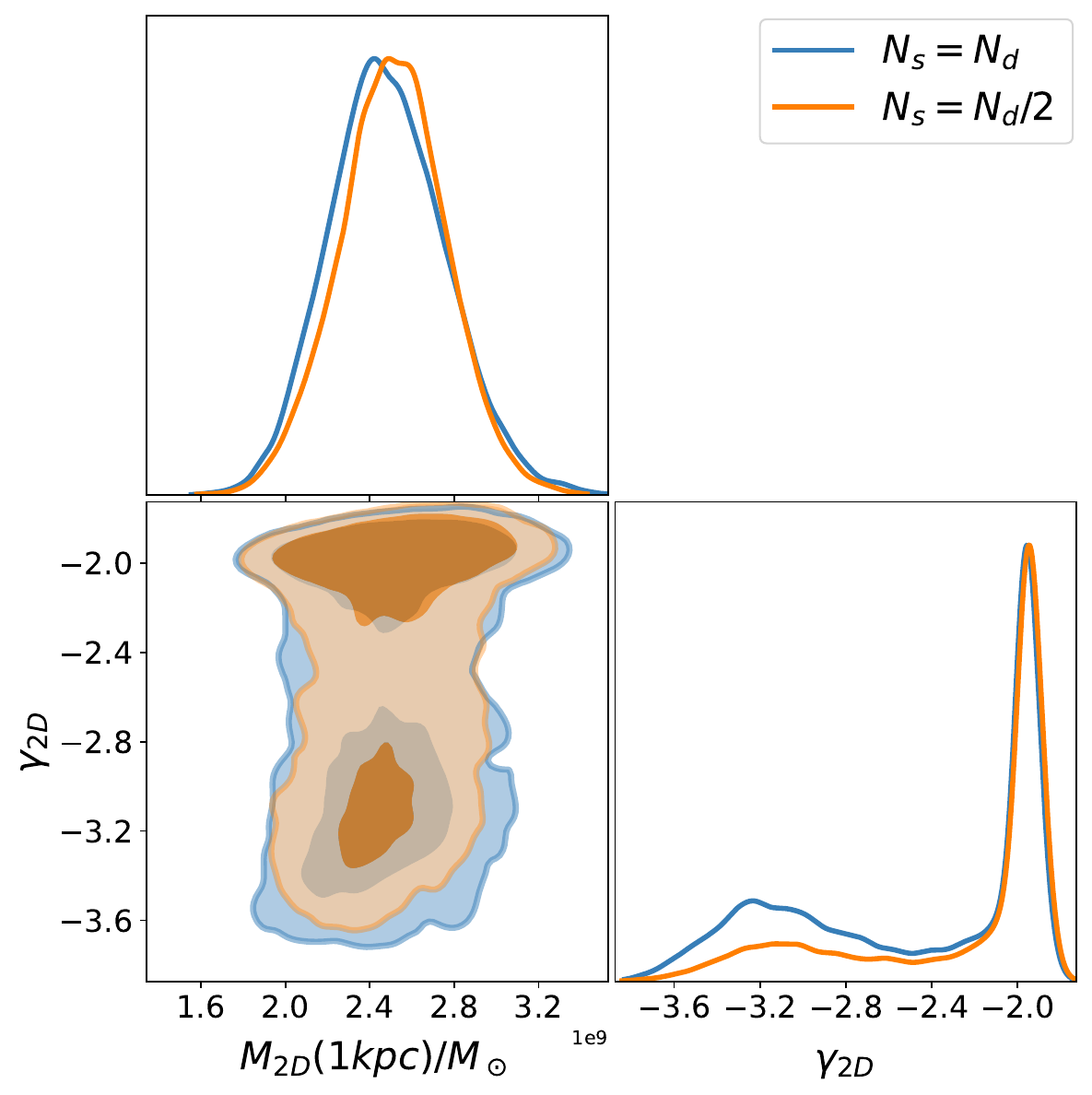}
	\caption{Joint posteriors in the subhalo's projected mass within 1 kpc ($M_{2D}$(1kpc)) and the log-slope $\gamma_{2D}$ of the subhalo's projected density profile. Blue curve corresponds to the model that sets the number of source pixels equal to the number of data pixels within the mask ($N_s=N_d$), while orange curve is the model that sets $N_s$ to half the number of data pixels ($N_s=N_d/2$). }
 
\label{fig:m1kpc_vs_logslope_pix}
\end{figure}

To check this, we rerun our supersampled fit that includes a subhalo using $N_s=N_d$, but without including multipoles in the model. The resulting source reconstruction of $s1$ from the best-fit model is shown in Figure \ref{fig:src1_morepix}, which we contrast with our fiducial model in Figure \ref{fig:src1_pix}. Here we do not use interpolation when displaying the reconstructed sources, so the Voronoi pixellation is evident. Note that in these supersampled models, the pixel centroids do not correspond to any specific ray-traced image pixel, but are rather determined by a K-means clustering algorithm from all the ray-traced subpixels as in \cite{nightingale2015} (see Section \ref{sec:src_reconstruction} for details). Comparing the two reconstructions, we see that there is indeed improved source resolution, particularly in the central regions where the surface brightness peaks. Unsurprisingly, there is at least a modest reduction in residuals, as the $\chi^2$/pixel improves from 0.91 to 0.89. Moving on to the posterior inferences, we find that there is a slight reduction in the error bars on certain parameters of the host galaxy due to the extra constraining power of the additional source pixels: for example, the host galaxy log-slope inference changes from $\alpha = 1.113_{-0.075}^{+0.072}$ to $\alpha = 1.113_{-0.056}^{0.067}$. However, the parameter inferences regarding the subhalo are remarkably consistent. In Figure \ref{fig:m1kpc_vs_logslope_pix} we plot posteriors in the $M_{2D}$(1kpc) and $\gamma_{2D}$ of the subhalo for both models, and find they very nearly overlap, with the higher resolution model finding slightly higher probability for steeper slopes ($\gamma \lnsim -2.5$). Thus, we conclude that the subhalo results from our models (whose parameter inferences are listed in Tables \ref{tab:posterior_inferences} and \ref{tab:posterior_inferences2}) are unlikely to change significantly if we double the number of source pixels.

We should note for completeness's sake, that it is possible in principle for the ``checkerboard'' effect (demonstrated most clearly in Section \ref{sec:checkerboarding}) to arise even in supersampled models. One obvious (and very unlikely) scenario would be if the modeler chooses the number of source pixels to be equal to or greater than the total number of subpixels (this would be  $N_s = N_{\rm sp}^2N_d$). In this case, there would be enough source pixels that they could in principle reproduce the checkerboard pattern, except this time on the level of the subpixels. A more likely scenario where this might occur would be if the source pixellation is allowed to become more adaptive. In \cite{nightingale2018} this is achieved by allowing the source pixels to cluster more in regions where there is greater surface brightness via a weighted K-means algorithm. This should only be a problem if there is a danger of checkerboarding occuring in a high-luminosity region, but is a potential pitfall to be aware of. A more dangerous prescription would be if one allows the source pixels to cluster adaptively around a point (or points) in such a way that is not tied to luminosity, but rather optimized based on what the Bayesian evidence prefers. In this scenario, it would be entirely possible for source pixels to cluster more in a region where it can achieve checkerboarding on the level of the ray-traced subpixels, so that i.e. a supersampled model that doesn't include a subhalo might still achieve good residuals. In adaptive clustering methods such as these, it would be wise to impose a limit on how densely source pixels can cluster to make sure the checkerboarding phenomenon does not reestablish itself on an even finer scale.

%\label{lastpage}
\end{document}